\RequirePackage{fix-cm}
\RequirePackage{fixltx2e}
\documentclass[11pt,aip, preprint]{revtex4-1}
\usepackage{lmodern}
\usepackage[T1]{fontenc}
\usepackage[utf8]{inputenc}
\usepackage{geometry}
\usepackage{color}
\usepackage{amsmath}
\usepackage{amssymb}
\usepackage{graphicx}
\PassOptionsToPackage{version=3}{mhchem}
\usepackage{mhchem}
\usepackage{microtype}
\usepackage[unicode=true,
 bookmarks=false,
 breaklinks=true,pdfborder={0 0 0},pdfborderstyle={},backref=false,colorlinks=true]
 {hyperref}
\hypersetup{pdftitle={Comparison of methods},
 citecolor =blue, linkcolor = blue,  urlcolor  = blue}

\makeatletter

\providecommand{\tabularnewline}{\\}

\usepackage{datetime}
\usepackage{refstyle}
\usepackage{longtable}
\usepackage{url}
\usepackage[title,toc,page,header]{appendix} 

\makeatother

\begin{document}

\title{Probabilistic performance estimators for computational chemistry
methods: Systematic Improvement Probability and Ranking Probability
Matrix. II. Applications}

\author{Pascal PERNOT}

\affiliation{Institut de Chimie Physique, UMR8000, CNRS, Université Paris-Saclay,
91405 Orsay, France}
\email{Pascal.Pernot@universite-paris-saclay.fr
}

\author{Andreas SAVIN }

\affiliation{Laboratoire de Chimie Théorique, CNRS and UPMC Université Paris 06,
Sorbonne Universités, 75252 Paris, France}
\email{Andreas.Savin@lct.jussieu.fr}

\begin{abstract}
In the first part of this study (Paper I), we introduced the systematic
improvement probability (SIP) as a tool to assess the level of improvement
on absolute errors to be expected when switching between two computational
chemistry methods. We developed also two indicators based on robust
statistics to address the uncertainty of ranking in computational
chemistry benchmarks: $P_{inv}$, the inversion probability between
two values of a statistic, and $\mathbf{P}_{r}$, the ranking probability
matrix. In this second part, these indicators are applied to nine
data sets extracted from the recent benchmarking literature. We illustrate
also how the correlation between the error sets might contain useful
information on the benchmark dataset quality, notably when experimental
data are used as reference.
\end{abstract}
\maketitle

\section{Introduction}

In Paper\,I \citep{Pernot2020}, we considered the uncertainty sources
impacting the values of benchmarking statistics (scores) and we presented
tools to estimate the uncertainty on statistics and to compare them.
We briefly summarize them here.

First, one compares system-by-system the absolute errors of two methods
$M_{i}$ and $M_{j}$. The systematic improvement probability ($\mathrm{SIP}{}_{i,j}$)
is defined as the fraction of systems for which $M_{i}$ has smaller
absolute errors than $M_{j}$. A SIP matrix can be built for a set
of methods, enabling to detect the methods with the best performances
in terms of absolute errors. A mean gain ($\mathrm{MG}{}_{i,j}$;
a negative value) is estimated, providing the expected decrease of
absolute errors when using $M_{i}$ instead of $M_{j}$. The mean
loss ($\mathrm{ML}{}_{i,j}$) is defined accordingly. The MUE difference
between both methods can be expressed as a combination of SIP, MG
and ML, illustrating the balance between gains and losses when switching
between two methods. 

Then, one compares statistics, taking into account their uncertainty
and correlation. For comparison of pairs of values, one uses $P_{inv}$
which gives the probability that the sign of the observed difference
is the opposite of the true one, considering the use of limited size
datasets. We have shown in Paper\,I \citep{Pernot2020} that $P_{inv}\simeq p_{g}/2$,
where $p_{g}$ is the $p$-value for the test of the equality of the
two values. This is tested in the first example below, as well as
the comparison to a $p$-value of the test ignoring correlations,
$p_{unc}$. To compare the statistics for a set of several methods,
we use the ranking probability matrix $\mathbf{P}_{r}$, which gives
the probability for each method to have any rank, considering the
limited size of the data set.

To avoid hypotheses on the errors distributions, bootstrap-based sampling
methods were used for the estimation of statistics uncertainty, $p$-values,
$P_{inv}$ and $\mathbf{P}_{r}$. The algorithms are detailed in Paper\,I
\citep{Pernot2020}, and some specific choices have been made regarding
the statistics: based on the recommendations of Wilcox and Erceg-Hurn
\citep{Wilcox2012}, quantiles are estimated by the Harrell and Davis
method \citep{Harrell1982}, and correlation coefficients are estimated
by the Spearman method (rank correlation), unless stated otherwise.

In the following, these methods are illustrated and validated on several
datasets taken from the recent benchmarking literature and covering
a wide range of dataset sizes and properties. In the next section,
the datasets are introduced and treated sequentially with a common
framework. This is followed by a global discussion covering the topics
of both papers, and a general conclusion.

\section{Applications\label{sec:Examples}}

Nine datasets have been extracted from the recent benchmarking literature.
Our selection is mostly based on the coverage of a representative
range of properties, dataset sizes (between a few tens to a few thousands)
and reference type (experimental or calculated)(Table\,\ref{tab:Case-studies}).
Besides such selection criteria, a major quality of the datasets is
their \emph{availability}, and their authors have to be praised for
that. Through these various examples, our intent is not to validate
or invalidate the original studies, but only to illustrate the properties
and interest of our proposed tools.

All cases are treated with a common framework: an introduction ; the
analysis of the correlation matrices for error sets and statistics
(MUE and $Q_{95}$); the analysis of the MUE and $Q_{95}$ statistics
and their inversion probabilities; the analysis of the SIP statistics;
and finally, the ranking probability matrices.

\begin{table}[!t]
\noindent \begin{centering}
\begin{tabular}{llcccl}
\hline 
Case & Property & $N$ & $K$ & Ref.$^{1}$ & Source\tabularnewline
\hline 
PER2018 & Intensive atomization energies & 222 & 9 & exp & \citep{Pernot2018}\tabularnewline
BOR2019 & Band gaps & 471 & 15 & exp & \citep{Borlido2019}\tabularnewline
NAR2019 & Enthalpies of formation & 469 & 4 & cal & \citep{Narayanan2019}\tabularnewline
CAL2019 & London Dispersion Corrections & 41 & 10$\times$3 & cal & \citep{Caldeweyher2019}\tabularnewline
JEN2018 & Non-covalent interaction energies & 66 & 6 & cal & \citep{Jensen2018}\tabularnewline
DAS2019 & Dielectric Constants & 23 & 6 & exp & \citep{Das2019}\tabularnewline
THA2015 & Polarizability & 135 & 7 & exp & \citep{Thakkar2015}\tabularnewline
WU2015 & Polarizability & 145 & 7 & cal & \citep{Wu2015b}\tabularnewline
ZAS2019 & Effective atomization energies & 6211 & 3 & cal & \citep{Zaspel2019}\tabularnewline
\hline 
\end{tabular}
\par\end{centering}
\caption{\label{tab:Case-studies}Case studies: $N$ is the number of systems
in the dataset and $K$ is the number of compared methods.\protect \\
$^{1}$ Nature of the reference data: experimental (exp) or calculated
(cal).}
\end{table}

\subsection{PER2018\label{subsec:Pernot2018}}

We consider here the intensive atomization energies \citep{Perdew2016}
estimated with 9 DFAs on the G3/99 dataset \citep{Curtiss2000}, and
extracted from a recent article by Pernot and Savin \citep{Pernot2018,Pernot2019}.
This medium-sized dataset ($N=222$) presents several non-normal error
distributions, and was used to illustrate the interest for benchmarks
of using $Q_{95}$ as a complement to the MUE, and to illustrate our
former definition of $P_{inv}$. Here we focus on the correlations
and their impact on the comparison of statistics. 
\begin{figure}[!t]
\noindent \begin{centering}
\begin{tabular}{ccc}
\includegraphics[width=0.32\columnwidth]{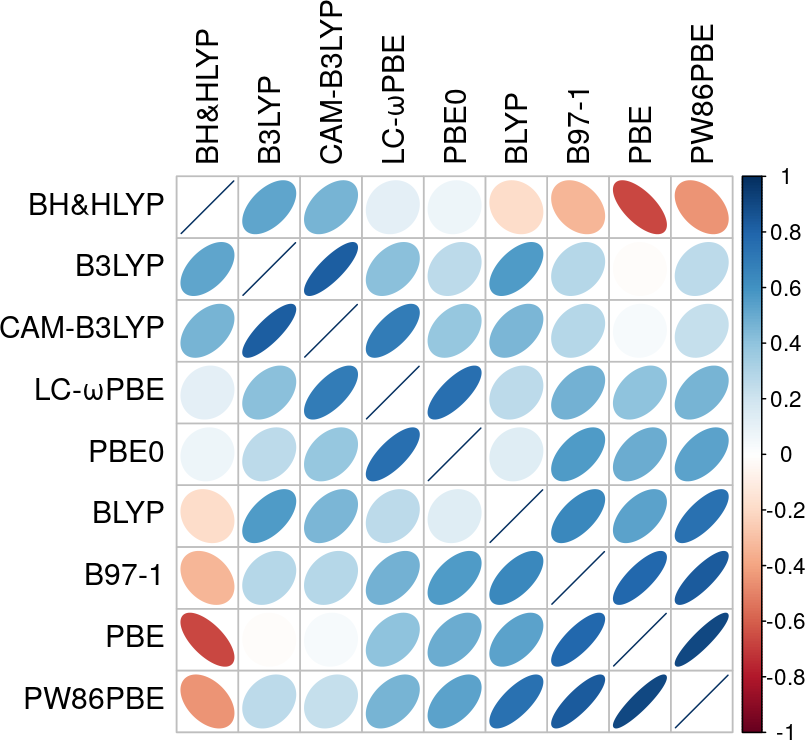} & 
~\includegraphics[width=0.32\columnwidth]{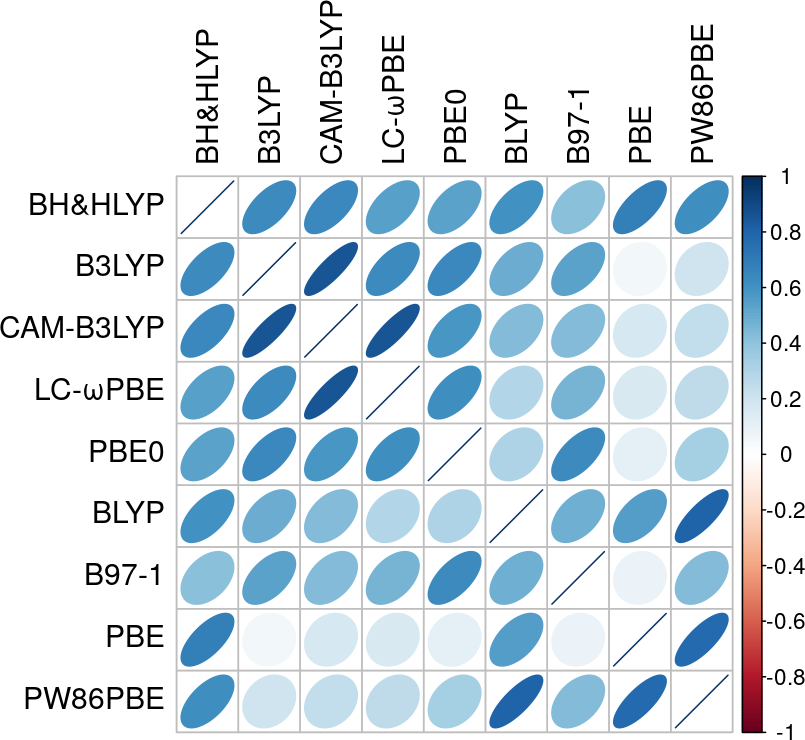} & 
~\includegraphics[width=0.32\columnwidth]{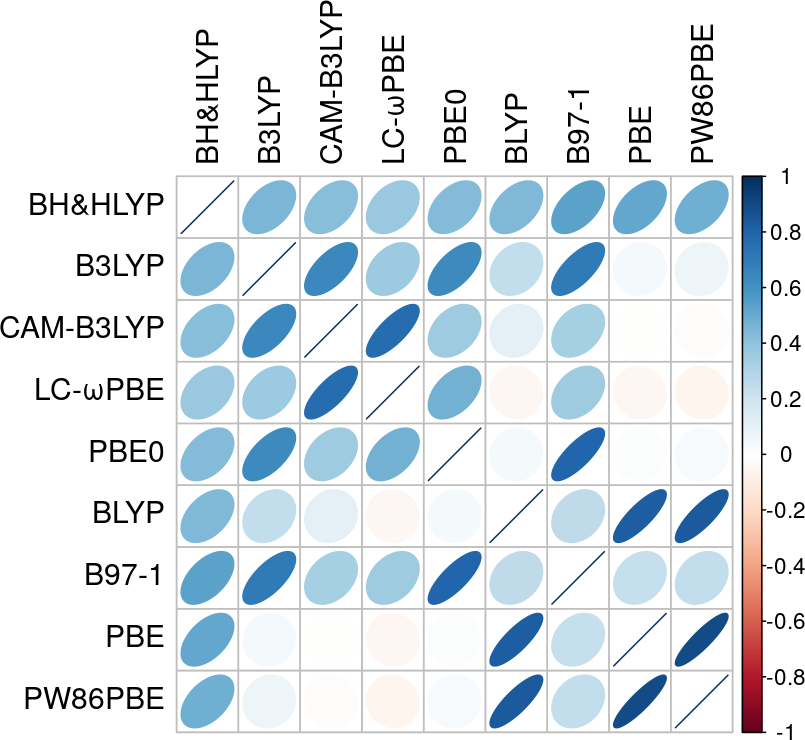}\tabularnewline
\multicolumn{3}{c}{\includegraphics[clip,width=0.99\columnwidth]{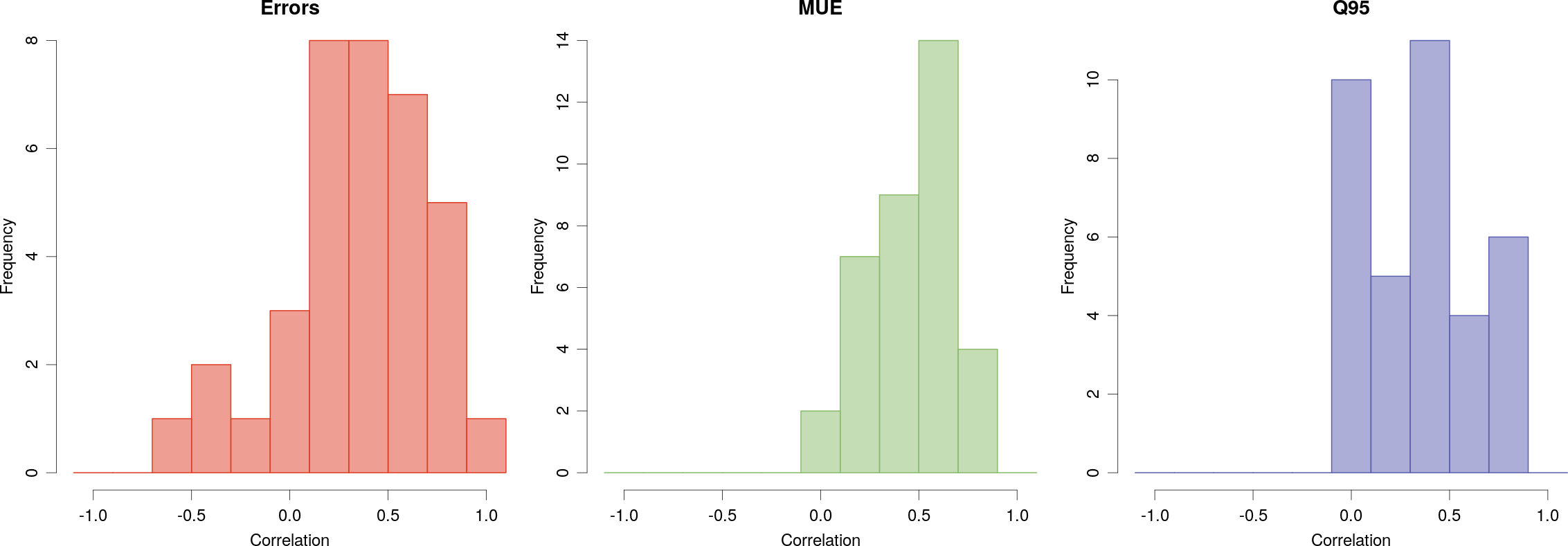}}\tabularnewline
\end{tabular}
\par\end{centering}
\noindent \centering{}\caption{\label{fig:pernot1}Case PER2018 - correlations: (top) rank correlation
matrices between errors sets, MUE and $Q_{95}$; (bottom) histogram
of non-diagonal elements of the corresponding correlation matrices.
The methods are ordered by a clustering of the errors correlation
matrix by the complete linkage method \citep{Defays1977} implemented
in the \texttt{R} function\texttt{ hclust} \citep{CiteR}.}
 
\end{figure}

\paragraph{Correlations.}

The correlation matrices between the error sets and their statistics
are represented in Fig.\,\ref{fig:pernot1}, along with histograms
of their non-diagonal elements. The errors sets are positively correlated,
with a wide distribution of correlation coefficients, except for pairs
involving BH\&HLYP, which presents negative correlations with four
other methods. When considering the scores, all correlations are positive
or null. Globally, the correlations are weaker for $Q_{95}$ than
for the MUE, except for a few pairs. The maximum of the histograms
shifts from 0.6 for MUE to 0 for $Q_{95}$, but large correlation
values are nevertheless still observed for $Q_{95}$. These observations
confirm the main trends from the numerical study of correlation transfer
in Paper\,I \citep{Pernot2020}. 
\begin{table}[!t]
\begin{centering}
{\small{}}%
\begin{tabular}{llr@{\extracolsep{0pt}.}lr@{\extracolsep{0pt}.}lr@{\extracolsep{0pt}.}lr@{\extracolsep{0pt}.}lr@{\extracolsep{0pt}.}lr@{\extracolsep{0pt}.}lr@{\extracolsep{0pt}.}lr@{\extracolsep{0pt}.}lr@{\extracolsep{0pt}.}lr@{\extracolsep{0pt}.}lr@{\extracolsep{0pt}.}lr@{\extracolsep{0pt}.}lr@{\extracolsep{0pt}.}lr@{\extracolsep{0pt}.}l}
\hline 
Methods &  & \multicolumn{2}{c}{MUE } & \multicolumn{2}{c}{$p_{unc}$} & \multicolumn{2}{c}{$p_{g}$} & \multicolumn{2}{c}{$P_{inv}$} & \multicolumn{2}{c}{} & \multicolumn{2}{c}{$Q_{95}$} & \multicolumn{2}{c}{$p_{unc}$} & \multicolumn{2}{c}{$p_{g}$} & \multicolumn{2}{c}{$P_{inv}$} & \multicolumn{2}{c}{} & \multicolumn{2}{c}{MSIP } & \multicolumn{2}{c}{SIP } & \multicolumn{2}{c}{MG } & \multicolumn{2}{c}{ML }\tabularnewline
 &  & \multicolumn{2}{c}{kcal/mol} & \multicolumn{2}{c}{} & \multicolumn{2}{c}{} & \multicolumn{2}{c}{} & \multicolumn{2}{c}{} & \multicolumn{2}{c}{kcal/mol} & \multicolumn{2}{c}{} & \multicolumn{2}{c}{} & \multicolumn{2}{c}{} & \multicolumn{2}{c}{} & \multicolumn{2}{c}{} & \multicolumn{2}{c}{} & \multicolumn{2}{c}{kcal/mol} & \multicolumn{2}{c}{kcal/mol}\tabularnewline
\cline{1-1} \cline{3-10} \cline{13-20} \cline{23-30} 
B3LYP  &  & 1&18(9)  & 0&00  & 0&00  & 0&00  & \multicolumn{2}{c}{} & 4&5(5)  & 0&00  & 0&00  & 0&00  & \multicolumn{2}{c}{} & 0&57(3)  & 0&53(3)  & -1&05(10)  & 0&48(5) \tabularnewline
B97-1  &  & \textbf{0}&\textbf{85(5) } & \multicolumn{2}{c}{- } & \multicolumn{2}{c}{-} & \multicolumn{2}{c}{-} & \multicolumn{2}{c}{} & \textbf{2}&\textbf{7(4) } & \multicolumn{2}{c}{- } & \multicolumn{2}{c}{- } & \multicolumn{2}{c}{-} & \multicolumn{2}{c}{} & 0&61(3)  & \multicolumn{2}{c}{-} & \multicolumn{2}{c}{-} & \multicolumn{2}{c}{-}\tabularnewline
BH\&HLYP  &  & 4&8(2)  & 0&00  & 0&00  & 0&00  & \multicolumn{2}{c}{} & 11&7(6)  & 0&00  & 0&00  & 0&00  & \multicolumn{2}{c}{} & 0&06(1)  & 0&95(2)  & -4&3(2)  & 0&8(2) \tabularnewline
BLYP  &  & 1&6(1)  & 0&00  & 0&00  & 0&00  & \multicolumn{2}{c}{} & 5&3(6)  & 0&00  & 0&00  & 0&00  & \multicolumn{2}{c}{} & 0&43(3)  & 0&77(3)  & -1&2(1)  & 0&6(1) \tabularnewline
CAM-B3LYP  &  & \textbf{0}&\textbf{90(9) } & 0&64  & 0&57  & 0&29  & \multicolumn{2}{c}{} & 4&1(4)  & 0&00  & 0&00  & 0&00  & \multicolumn{2}{c}{} & \textbf{0}&\textbf{74(3) } & 0&33(3)  & -1&3(2)  & 0&59(4) \tabularnewline
LC-$\omega$PBE  &  & 1&09(10)  & 0&03  & 0&00  & 0&00  & \multicolumn{2}{c}{} & 4&3(5)  & 0&01  & 0&00  & 0&00  & \multicolumn{2}{c}{} & 0&65(3)  & 0&43(3)  & -1&1(1)  & 0&44(3) \tabularnewline
PBE  &  & 2&8(2)  & 0&00  & 0&00  & 0&00  & \multicolumn{2}{c}{} & 8&1(8)  & 0&00  & 0&00  & 0&00  & \multicolumn{2}{c}{} & 0&30(2)  & 0&81(3)  & -2&6(2)  & 0&8(1) \tabularnewline
PBE0  &  & \textbf{0}&\textbf{92(7) } & 0&44  & 0&24  & 0&12  & \multicolumn{2}{c}{} & 3&3(5)  & 0&33  & 0&02  & 0&01  & \multicolumn{2}{c}{} & 0&66(3)  & 0&50(3)  & -0&74(7)  & 0&61(4) \tabularnewline
PW86PBE  &  & 1&6(1)  & 0&00  & 0&00  & 0&00  & \multicolumn{2}{c}{} & 6&1(9)  & 0&00  & 0&00  & 0&00  & \multicolumn{2}{c}{} & 0&49(3)  & 0&59(3)  & -1&6(2)  & 0&43(6) \tabularnewline
\hline 
\end{tabular}{\small\par}
\par\end{centering}
\noindent \centering{}\caption{\label{tab:pernot}Case PER2018 - absolute error statistics: $p$-values,
inversion probabilities and SIP statistics for comparison with the
DFA of smallest MUE (B97-1). The best scores and the values for which
$p_{g}>0.05$ are in boldface. The SIP, MG and ML columns correspond
to the B97-1 row of the corresponding matrices. Uncertainty is presented
in parenthesis notation.}
\end{table}

\paragraph{Statistics.}

The statistics are reported in Table\,\ref{tab:pernot}. Note that,
due to the use of a different quantile estimation algorithm, the values
of $Q_{95}$ have changed slightly from the values reported in the
original article \citep{Pernot2018}. 

There is a group of three methods (B97-1, CAM-B3LYP and PBE0) with
small MUE values. Considering the $p_{g}$ values, one cannot reject
the hypothesis that the observed differences are due to the limited
size of the datasets. Note that the same conclusion would have been
reached when ignoring correlation ($p_{unc}$), as the neglect of
correlation increases the $p$-values, but no other one reaches the
0.05 threshold. However, the $p_{unc}$ value for LC-$\omega$PBE
reaches 0.03, not far from the threshold. Consistently, the MUE inversion
probability $P_{inv}$ computed in the reference article \citep{Pernot2019},
included LC-$\omega$PBE in the group of methods with a sizable risk
of inversion. As demonstrated in\,Eq.\,31 of Paper\,I \citep{Pernot2020},
the revised version of $P_{inv}$ accounting for correlations is now
practically equal to $p_{g}/2$, which rejects LC-$\omega$PBE as
a contender for the head group. When picking B97-1 instead of CAM-B3LYP
based on the MUE, there is a 29\,\% chance to be wrong, \emph{i.e.},
that the MUE of CAM-B3LYP is indeed smaller than B97-1 due to the
dataset size. This risks falls to 12\,\% for PBE0.

\begin{figure}[!tb]
\noindent \centering{}\includegraphics[width=0.32\textwidth]{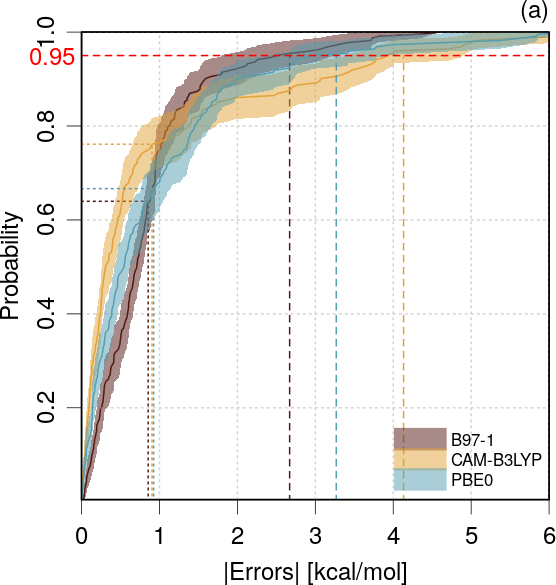}~\includegraphics[width=0.32\textwidth]{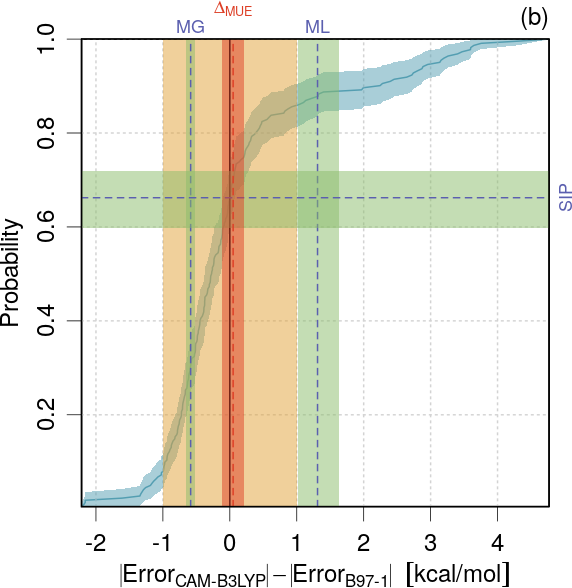}~\includegraphics[width=0.32\textwidth]{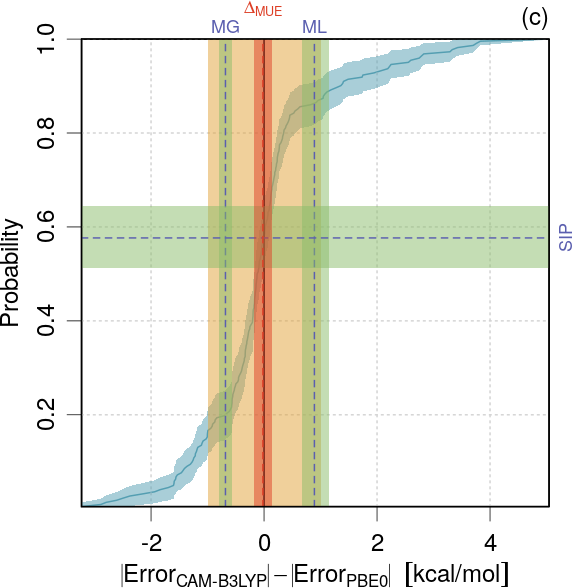}\caption{\label{fig:pernot1-1}Case PER2018 - absolute errors statistics: (a)
ECDF and statistics of absolute errors. The MUE values are depicted
by vertical dotted lines, and the $Q_{95}$ values by vertical dashed
lines; (b-c) ECDF and statistics of the difference of absolute errors.
The green- and red-shaded bands represent 95\,\% confidence intervals
for the reported statistics (SIP: systematic improvement probability;
MG: mean gain; ML: mean loss, $\Delta_{MUE}$: MUE difference). The
orange bar depicts the chemical accuracy (1\,kcal/mol). It is a visual
aid to evaluate the pertinence of the observed differences.}
\end{figure}

The situation is different for $Q_{95}$, where the neglect of correlation
would lead to the conclusion that PBE0 (3.3(5)\,kcal/mol) is not
significantly distinct from B97-1 (2.7(4)\,kcal/mol; $p_{unc}=0.33$)
whereas the correct value is given by $p_{g}=0.02$. In this example,
$Q_{95}$ can help us to rank the three best methods, for which the
MUE is not discriminant. This is linked to the presence of different
tails in the absolute errors distributions (cf. Fig.\,\ref{fig:pernot1-1}(a)).

This example illustrates and confirms the relations between $p_{unc}$,
$p_{g}$ and $P_{inv}$ expressed in Paper\,I \citep{Pernot2020},
Section\,II.D.3. In the following examples, only $P_{inv}$ is reported
to alleviate the results tables.

\paragraph{SIP analysis.}

\begin{figure}[t]
\noindent \begin{centering}
\includegraphics[width=0.45\textwidth]{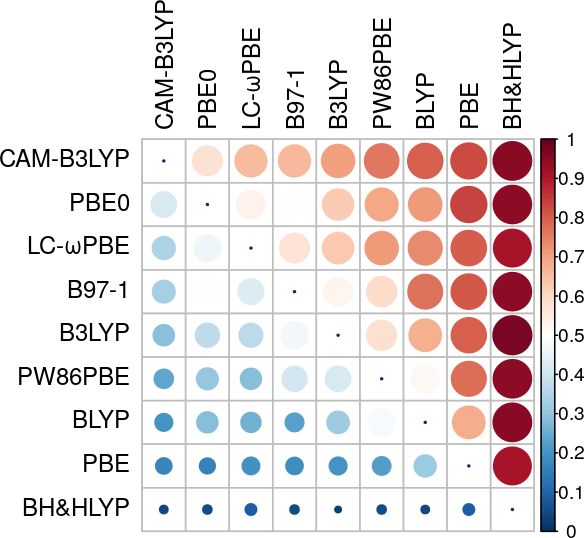}
\par\end{centering}
\noindent \centering{}\caption{\label{fig:SIPMAT-example}Case PER2018: SIP matrix. A line with a
majority of red patches signals a method with good SIP performances.
The SIP value is color-coded and the area of a disk is proportional
to the corresponding value. The methods are ordered by decreasing
value of MSIP. }
\end{figure}

The SIP analysis brings another view on the head trio (B97-1, CAM-B3LYP
and PBE0), as the method with the highest MSIP is CAM-B3LYP. One can
see on the SIP matrix in Fig.\,\ref{fig:SIPMAT-example}, that indeed,
the row for CAM-B3LYP is fully reddish, when those for B97-1 and PBE0
present also blue and white patches. We note also that B97-1 provides
a nearly full improvement over BH\&HLYP (SIP = 0.95(2)). 

The ECDF of the difference of absolute errors for CAM-B3LYP and B97-1
helps to understand the contradiction between the MUE and MSIP ranks
(Fig.\,\ref{fig:pernot1-1}(b)). The MUE difference for this pair
is statistically not significant ($p_{g}=0.57$), the SIP value for
CAM-B3LYP over B97-1 is 0.67 (1-0.33), the mean gain -0.6 kcal/mol
and the mean loss 1.3 kcal/mol, due to the heavy tail in the CAM-B3LYP
error distribution (these numbers correspond to the reciprocal comparison
of the one presented in Table\,\ref{tab:pernot}). So by switching
from B97-1 to CAM-B3LYP, one would have to accept a 33\,\% risk to
degrade the intensive atomization energies by 1.3\,kcal/mol in average
and up to 4\,kcal/mol, but one would improve the estimations in 67\,\%
of the cases by 0.6\,kcal/mol in average. The same comparison between
CAM-B3LYP and PBE0 (Fig.\,\ref{fig:pernot1-1}(c)) shows that there
is no strong basis to favor one of these method. 

\paragraph{Ranking.}

\begin{figure}[!tb]
\noindent \begin{centering}
\includegraphics[width=0.32\columnwidth]{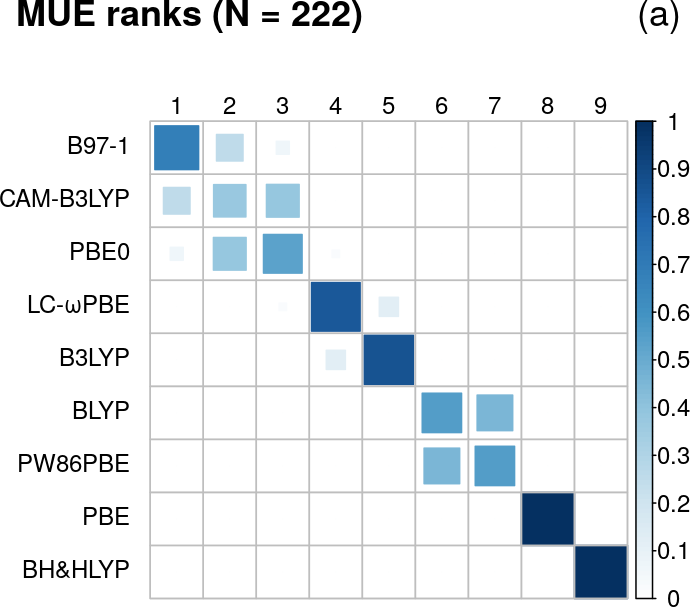}~\includegraphics[width=0.32\columnwidth]{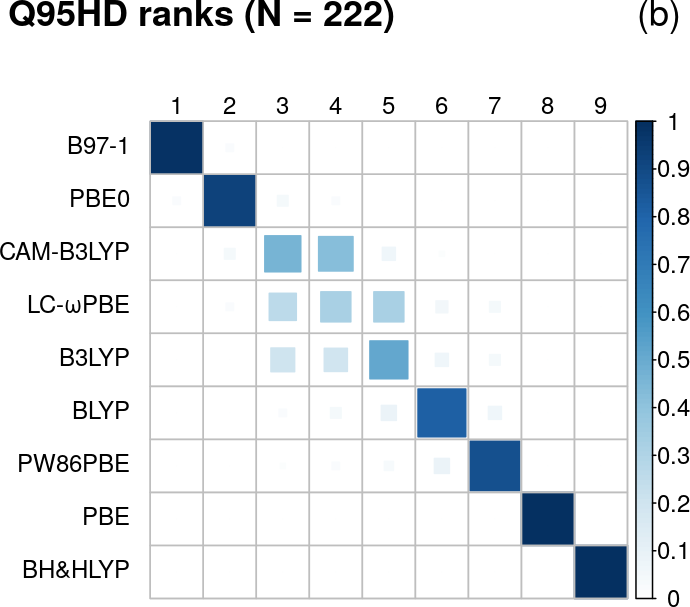}~\includegraphics[width=0.32\columnwidth]{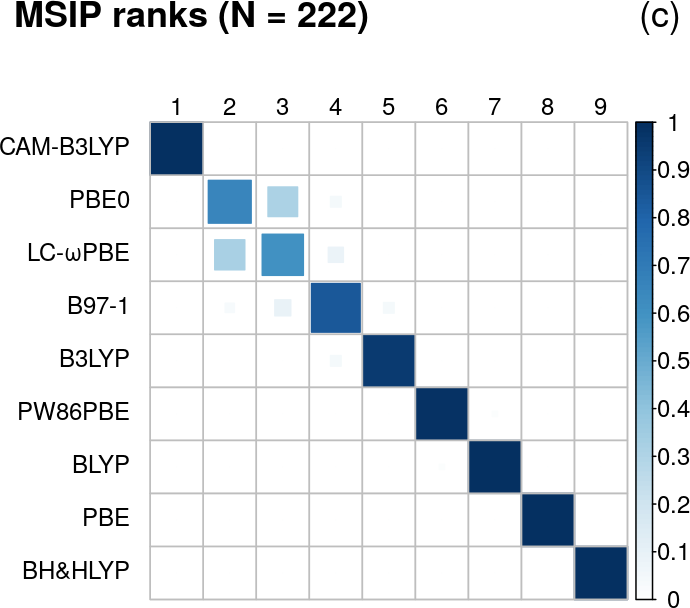}
\par\end{centering}
\noindent \centering{}\caption{\label{fig:pernot2}Case PER2018: ranking probability matrix for (a)
MUE, (b) $Q_{95}$ and (c) MSIP.}
\end{figure}

The ranking probability matrices (Fig.\,\ref{fig:pernot2}) confirm
the above analysis. The group of three methods (B97-1, CAM-B3LYP and
PBE0) at the top of the MUE ranking presents a blurred image (no clear
diagonal), whereas the first $Q_{95}$ rank of B97-1 is not ambiguous.
As expected, the MSIP ranking favors solidly CAM-B3LYP. Globally,
B97-1 should be preferred to minimize the risk of large errors, where
CAM-B3LYP would provide, overall, smaller absolute errors. 

\subsection{BOR2019\label{subsec:Borlido2019}}

Band gap estimations for a set of 471 systems\footnote{The original dataset contains 472 systems, but several values are
missing for \ce{NaYbP2S6}, which was excluded.} by 15 DFAs were extracted from the Supplementary Information of a
recent article by Borlido \emph{et al.} \citep{Borlido2019}. For
a full description of the dataset, we refer the reader to the original
article. 

The reference authors reported and analyzed relative errors, but as
there is a large range of band gaps in this set this causes a dispersion
of relative errors over six orders of magnitude, and an unsuitable
distortion of the errors distributions, with large relative errors
for small band gaps, and small relative errors for large band gaps.
It is true that for some methods (\emph{e.g.}, LDA) the errors increase
with the value of the band gap, but this is due mostly to a systematic
deviation (trend), not to an increase in the dispersion of the errors.
In consequence, we chose to treat here the raw errors. 

Borlido \emph{et al.} \citep{Borlido2019} discuss the uncertainties
on the reference band gaps in their dataset and estimate it to a few
tenths of eV. Without more detailed information, we assume that this
represents a uniform uncertainty for the dataset.

\paragraph{Correlations.}

\begin{figure}[!t]
\noindent \begin{centering}
\includegraphics[clip,width=0.5\textwidth]{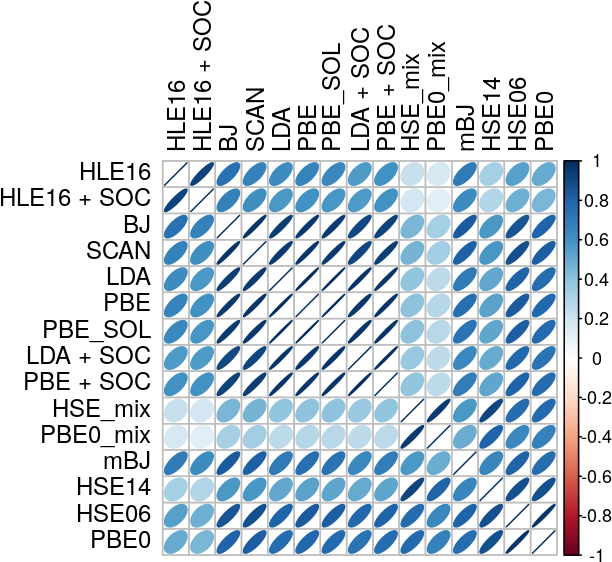}
\par\end{centering}
\noindent \centering{}\caption{\label{fig:borlido1}Case BOR2019 - rank correlation between errors
sets. The methods are ordered by a clustering algorithm using the
complete linkage method \citep{Defays1977} implemented in the \texttt{R}
function\texttt{ hclust} \citep{CiteR}. }
 
\end{figure}

One sees in Fig.\,\ref{fig:borlido1} that across the spectrum of
methods, all error sets correlation coefficients are positive, and
can reach very large values, up to 0.998. Only about 30\,\% of the
dataset pairs have correlation coefficients below 0.6, involving notably
PBE0\_mix and HSE\_mix. If the error sets are dominated by method
errors (\emph{i.e.}, there are no large reference data errors, nor
outliers), the correlation matrix can be used to infer a clustering
of methods, describing the relationships of the methods for the current
property/dataset. Error sets with large correlation coefficients are
related by a linear or monotonous transformation and the corresponding
methods are clustered together. The presence of well delimited clusters
indicates that the error sets are not dominated by reference data
errors. From the correlation matrix, the clusters would be (HLE16,
HLE16+SOC), (BJ, SCAN, LDA, PBE, PBE\_SOL, LDA+SOC, PBE+SOC), (HSE\_mix,
PBE0\_mix) and (HSE06,PBE0). mBJ and HSE14 stay alone. This clustering
seems to produce blocks that correspond to physical intuition: LDA,
PBE, SCAN... have all an electron-gas background. This is relaxed
for HLE16 that differs fro HLE16+SOC only by taking into account spin-orbit
coupling. These methods are further decoupled from hybrid methods
(PBE0, HSE06). 

\paragraph{Statistics.}

The values are reported in Table\,\ref{tab:borlido}. Although mBJ
presents the smallest MUE (0.50(2)\,eV), the value for HSE06 is very
close (0.53(5)\,eV), and one cannot exclude that the difference is
due to a mere sampling effect ($p_{g}\simeq2P_{inv}=0.16$). Besides,
HSE06 is the only method with a notably non-zero $P_{inv}$ value
with mBJ for the MUE. mBJ is also the method with the smallest $Q_{95}$,
and no other method is able to challenge this rank. mBJ has the largest
MSIP, but its value is moderate (0.7), indicating that mBJ does not
provide a full systematic improvement over (some of) the other methods.
 
\begin{table}[!t]
\begin{centering}
\begin{tabular}{llr@{\extracolsep{0pt}.}lr@{\extracolsep{0pt}.}lr@{\extracolsep{0pt}.}lr@{\extracolsep{0pt}.}lr@{\extracolsep{0pt}.}lr@{\extracolsep{0pt}.}lr@{\extracolsep{0pt}.}lr@{\extracolsep{0pt}.}lr@{\extracolsep{0pt}.}lr@{\extracolsep{0pt}.}l}
\hline 
Methods  &  & \multicolumn{2}{c}{MUE } & \multicolumn{2}{c}{$P_{inv}$} & \multicolumn{2}{c}{} & \multicolumn{2}{c}{$Q_{95}$} & \multicolumn{2}{c}{$P_{inv}$} & \multicolumn{2}{c}{} & \multicolumn{2}{c}{MSIP } & \multicolumn{2}{c}{SIP } & \multicolumn{2}{c}{MG } & \multicolumn{2}{c}{ML }\tabularnewline
 &  & \multicolumn{2}{c}{eV} & \multicolumn{2}{c}{} & \multicolumn{2}{c}{} & \multicolumn{2}{c}{eV} & \multicolumn{2}{c}{} & \multicolumn{2}{c}{} & \multicolumn{2}{c}{} & \multicolumn{2}{c}{} & \multicolumn{2}{c}{eV} & \multicolumn{2}{c}{eV}\tabularnewline
\cline{1-1} \cline{3-6} \cline{9-12} \cline{15-22} 
LDA  &  & 1&17(5)  & 0&00  & \multicolumn{2}{c}{} & 3&2(2)  & 0&00  & \multicolumn{2}{c}{} & 0&25(2)  & 0&84(2)  & -0&87(4)  & 0&41(4) \tabularnewline
LDA + SOC  &  & 1&24(5)  & 0&00  & \multicolumn{2}{c}{} & 3&3(2)  & 0&00  & \multicolumn{2}{c}{} & 0&16(2)  & 0&86(2)  & -0&92(4)  & 0&38(4) \tabularnewline
PBE  &  & 1&05(5)  & 0&00  & \multicolumn{2}{c}{} & 3&0(2)  & 0&00  & \multicolumn{2}{c}{} & 0&41(2)  & 0&82(2)  & -0&76(4)  & 0&40(3) \tabularnewline
PBE + SOC  &  & 1&12(5)  & 0&00  & \multicolumn{2}{c}{} & 3&0(2)  & 0&00  & \multicolumn{2}{c}{} & 0&30(2)  & 0&83(2)  & -0&82(4)  & 0&37(4) \tabularnewline
PBE\_SOL  &  & 1&12(5)  & 0&00  & \multicolumn{2}{c}{} & 3&1(2)  & 0&00  & \multicolumn{2}{c}{} & 0&30(2)  & 0&83(2)  & -0&82(4)  & 0&42(4) \tabularnewline
HLE16  &  & 0&60(4)  & 0&00  & \multicolumn{2}{c}{} & 1&9(2)  & 0&00  & \multicolumn{2}{c}{} & 0&66(2)  & 0&49(2)  & -0&44(4)  & 0&23(2) \tabularnewline
HLE16 + SOC  &  & 0&61(4)  & 0&00  & \multicolumn{2}{c}{} & 2&0(2)  & 0&00  & \multicolumn{2}{c}{} & 0&65(2)  & 0&49(2)  & -0&48(4)  & 0&25(2) \tabularnewline
BJ  &  & 0&79(4)  & 0&00  & \multicolumn{2}{c}{} & 2&3(2)  & 0&00  & \multicolumn{2}{c}{} & 0&55(2)  & 0&75(2)  & -0&49(3)  & 0&31(2) \tabularnewline
mBJ  &  & \textbf{0}&\textbf{50(2) } & \multicolumn{2}{c}{-} & \multicolumn{2}{c}{} & \textbf{1}&\textbf{41(7)}  & \multicolumn{2}{c}{-} & \multicolumn{2}{c}{} & \textbf{0}&\textbf{69(2)}  & \multicolumn{2}{c}{-} & \multicolumn{2}{c}{-} & \multicolumn{2}{c}{-}\tabularnewline
SCAN  &  & 0&81(4)  & 0&00  & \multicolumn{2}{c}{} & 2&4(2)  & 0&00  & \multicolumn{2}{c}{} & 0&55(2)  & 0&74(2)  & -0&53(3)  & 0&30(2) \tabularnewline
HSE06  &  & \textbf{0}&\textbf{53(3) } & 0&09  & \multicolumn{2}{c}{} & 1&7(2)  & 0&00  & \multicolumn{2}{c}{} & 0&68(2)  & 0&52(2)  & -0&28(3)  & 0&25(2) \tabularnewline
HSE14  &  & 0&57(3)  & 0&00  & \multicolumn{2}{c}{} & 1&8(1)  & 0&00  & \multicolumn{2}{c}{} & 0&63(2)  & 0&56(2)  & -0&38(2)  & 0&33(2) \tabularnewline
HSE06\_mix  &  & 0&64(3)  & 0&00  & \multicolumn{2}{c}{} & 2&0(1)  & 0&00  & \multicolumn{2}{c}{} & 0&60(2)  & 0&58(2)  & -0&51(3)  & 0&36(3) \tabularnewline
PBE0  &  & 0&78(3)  & 0&00  & \multicolumn{2}{c}{} & 1&8(1)  & 0&00  & \multicolumn{2}{c}{} & 0&44(2)  & 0&72(2)  & -0&57(2)  & 0&46(4) \tabularnewline
PBE0\_mix  &  & 0&82(4)  & 0&00  & \multicolumn{2}{c}{} & 2&4(2)  & 0&00  & \multicolumn{2}{c}{} & 0&47(2)  & 0&66(2)  & -0&67(4)  & 0&37(3) \tabularnewline
\hline 
\end{tabular}
\par\end{centering}
\noindent \centering{}\caption{\label{tab:borlido}Case BOR2019 - absolute error statistics: inversion
probabilities and SIP statistics for comparison with the DFA of smallest
MUE (mBJ). The best scores and the values for which $(p_{g}=2P_{inv})>0.05$
are in boldface.}
\end{table}

\paragraph{SIP analysis.}

The SIP values for mBJ lie between 0.49 and 0.86. The latter value
is against LDA+SOC, which means that for 14\,\% of the systems, LDA+SOC
achieves smaller absolute errors than mBJ, despite its poor scores.
Interestingly, small values, close to 0.5, are also observed against
HLS16, HLSE16+SOC and HSE06, indicating a notable risk of performance
loss when switching from one of these methods to mBJ.

\begin{figure}[!tb]
\noindent \centering{}\includegraphics[width=0.32\textwidth]{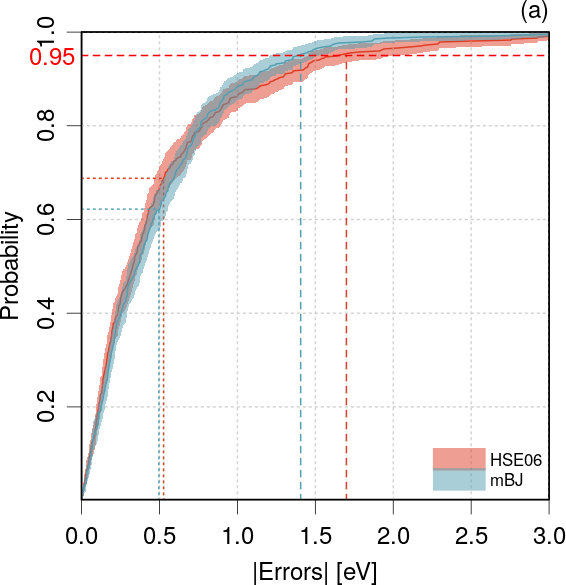}~\includegraphics[width=0.32\textwidth]{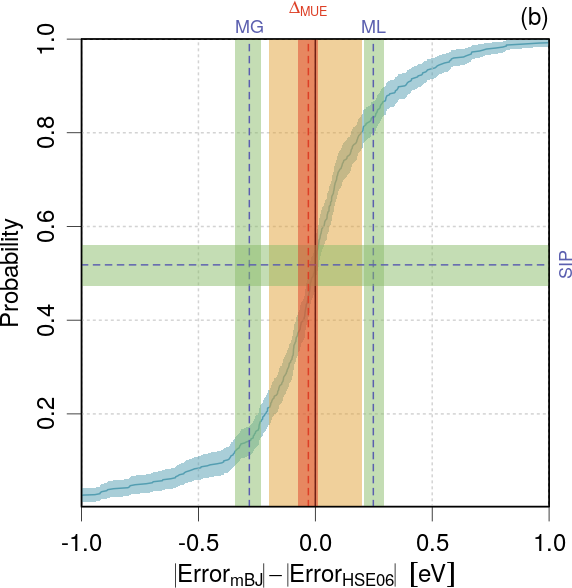}\caption{\label{fig:borlido2}Case BOR2019 - absolute errors statistics: (a)
ECDF of the absolute errors; (b) ECDF of the difference of absolute
errors for mBJ and HSE06. See Fig.\,\ref{fig:pernot1-1} for details.
The orange band depicts a reasonable level of uncertainty in the dataset
(0.2\,eV).}
\end{figure}

As seen in Paper\,I \citep{Pernot2020} (Fig.\,3), when going from
LDA to mBJ , one has about 15\,\% chance to perform better using
LDA, and the mean gain more than doubles the mean loss. By contrast,
the comparison of mBJ to HSE06 (Fig.\,\ref{fig:borlido2}(b)) is
an example of undecidability: the $\Delta_{MUE}$ is not significantly
different from zero, and one has as much to loose as to gain by switching
between both methods.

\begin{figure}[!tb]
\noindent \centering{}\includegraphics[width=0.45\textwidth]{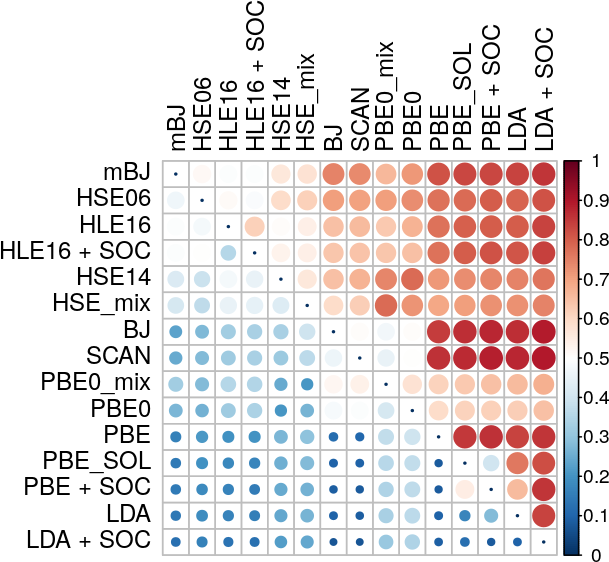}\caption{\label{fig:borlidoSIP}Case BOR2019 - SIP matrix. }
\end{figure}

The SIP matrix (Fig.\,\ref{fig:borlidoSIP}) provides a convenient
summary of these observations. The mBJ line is mostly reddish with
white spots indicating neutral comparisons. In contrast, the LDA+SOC
line is fully blueish, indicating that it is dominated by all other
methods. 
 
\paragraph{Ranking.}

\begin{figure}[!tb]
\noindent \begin{centering}
\begin{tabular}{ccc}
\includegraphics[width=0.32\columnwidth]{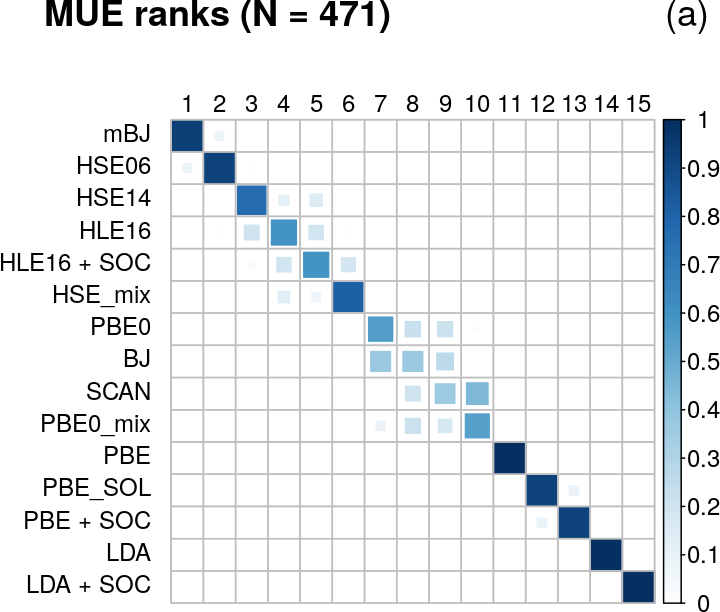} & ~\includegraphics[width=0.32\columnwidth]{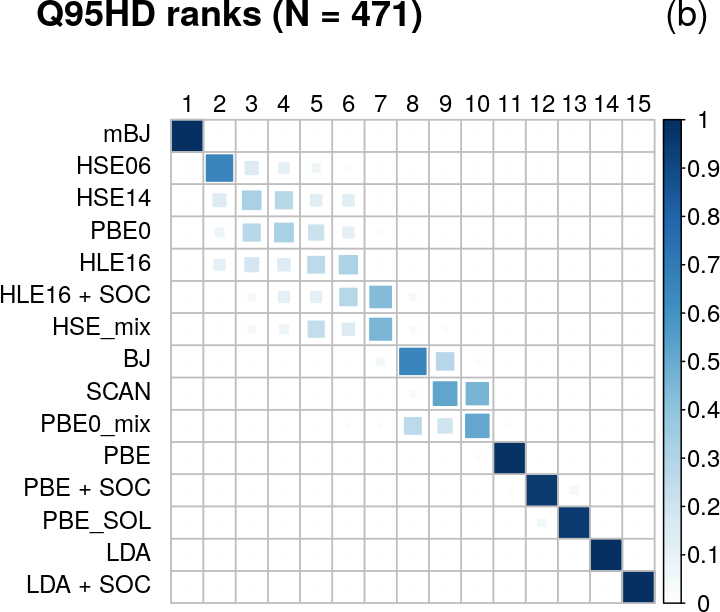} & ~\includegraphics[width=0.32\columnwidth]{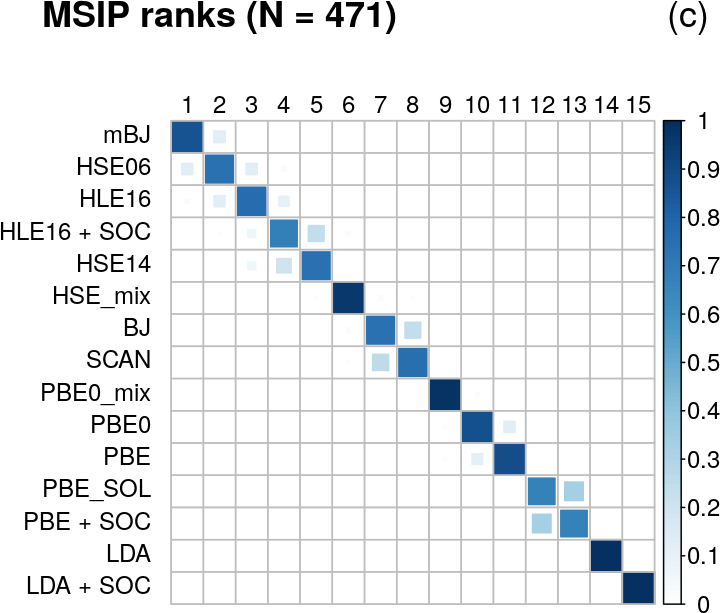}\tabularnewline
\includegraphics[width=0.32\columnwidth]{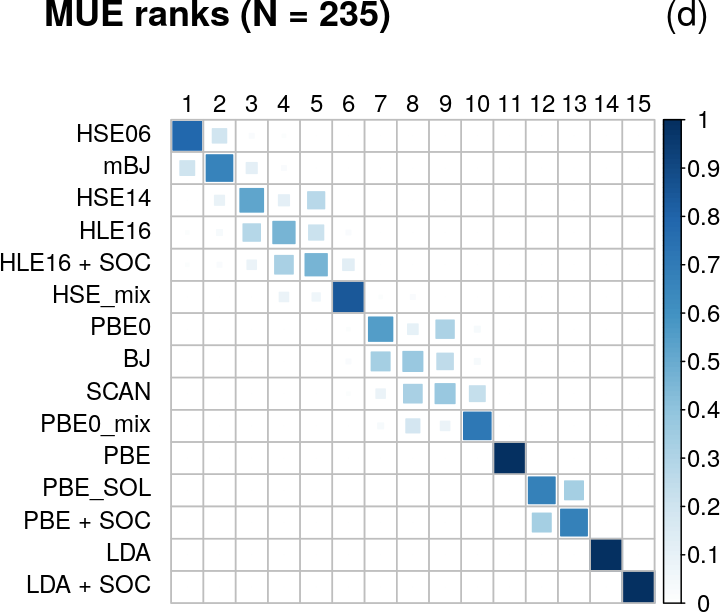} & ~\includegraphics[width=0.32\columnwidth]{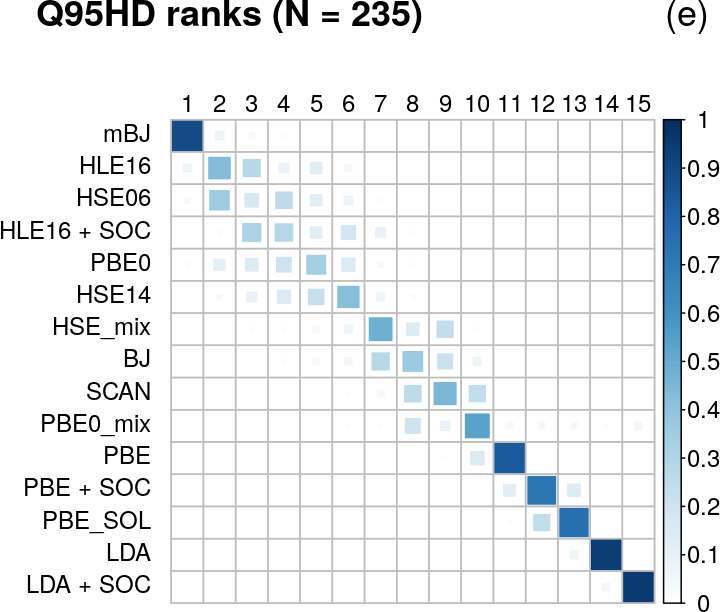} & ~\includegraphics[width=0.32\columnwidth]{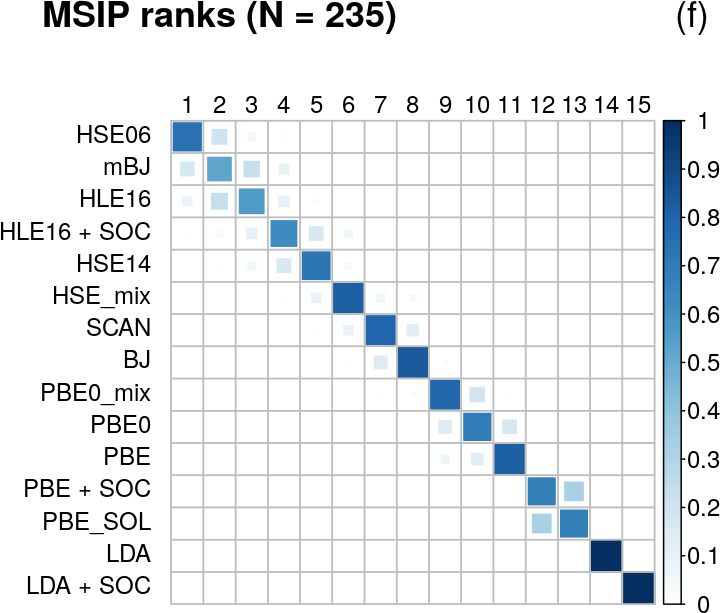}\tabularnewline
\includegraphics[width=0.32\columnwidth]{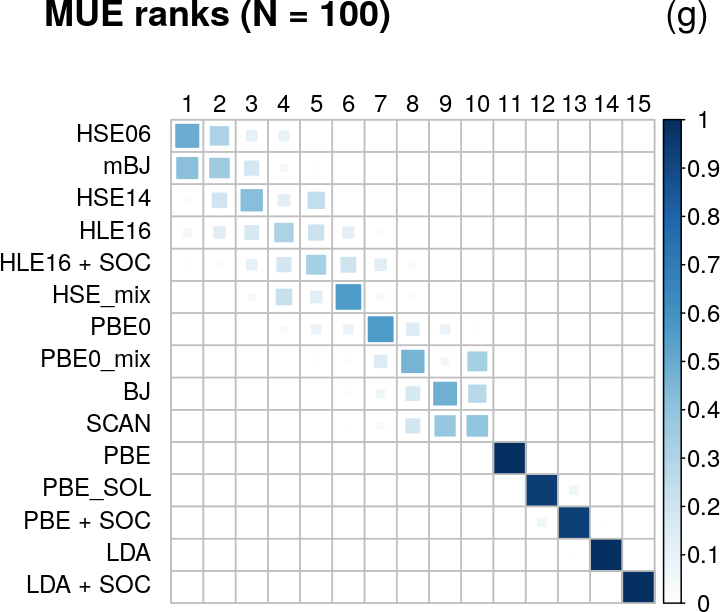} & ~\includegraphics[width=0.32\columnwidth]{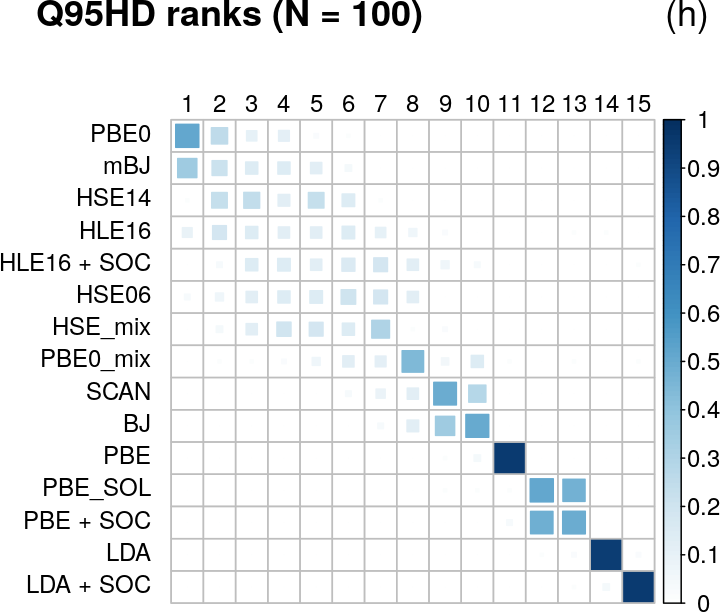} & ~\includegraphics[width=0.32\columnwidth]{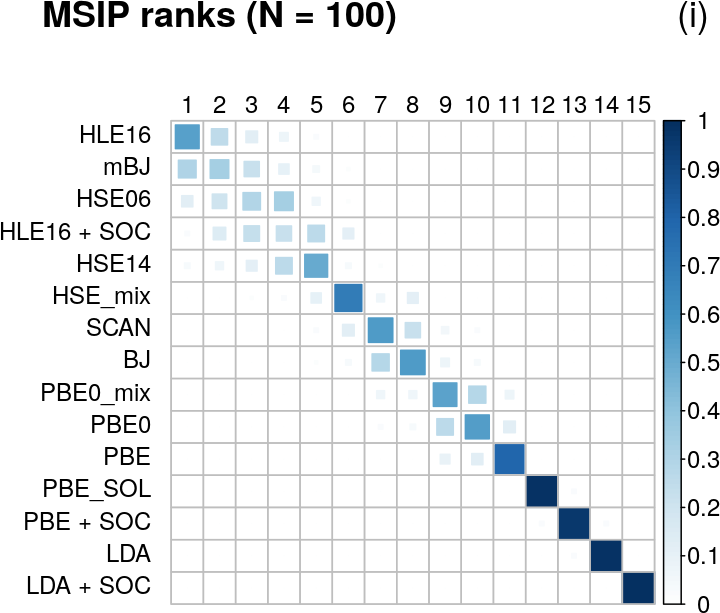}\tabularnewline
\end{tabular}
\par\end{centering}
\noindent \centering{}\caption{\label{fig:borlidoRPmat}Case BOR2019: ranking probability matrices
for the full dataset, $N=471$ (a-c), and for random reduced sets
$N=235$ (d-f) and $N=100$ (g-i). }
\end{figure}

Ranking probability matrices for the MUE, $Q_{95}$ and MSIP are presented
in Fig.\,\ref{fig:borlidoRPmat}(a-c). They illustrate the previous
results and show that ranking by MUE beyond the second place becomes
uncertain. This is even more notable for $Q_{95.}$. The MSIP ranking
selects the same group of five methods as the MUE ranking, with some
inversions. At the opposite, an end-group of five methods is rather
well ascertained for all three statistics. 

These matrices are a convenient tool to visualize the impact of dataset
size on the ranking quality. We estimated them for reduced error sets
($N=235$ and $N=100$), sampled randomly from the original one. The
impact is clearly visible in Fig.\,\ref{fig:borlidoRPmat}(d-i),
as the diagonal contributions get weaker when $N$ decreases. For
the MUE, the block of ranks 1 and 2 is quite robust, but the situation
deteriorates for the upper ranks. For $Q_{95}$, the first place of
mBJ is very stable, but the upper ranks become very uncertain, up
to the last ranks for $N=100$. As for the MUE, the MSIP ranking suffers
from the reduced datasets, but a head group of five methods is well
preserved. 

\subsection{NAR2019\label{subsec:Narayanan2019}}

The dataset contains the calculated enthalpies of formation by G4MP2
for 469 molecules having experimental values with small uncertainty
(Pedley test set) \citep{Narayanan2019}. The G4MP2 values are compared
with those of B3LYP, M06-2X and $\omega$B97X-D. 

\paragraph{Correlations.}

\begin{figure}[!tb]
\noindent \begin{centering}
\begin{tabular}{ccc}
\includegraphics[width=0.32\textwidth]{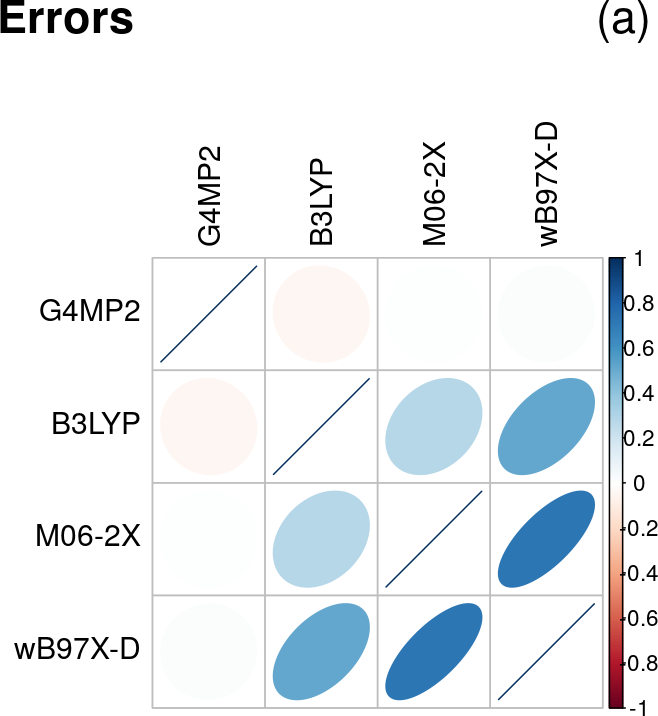} & ~\includegraphics[width=0.32\textwidth]{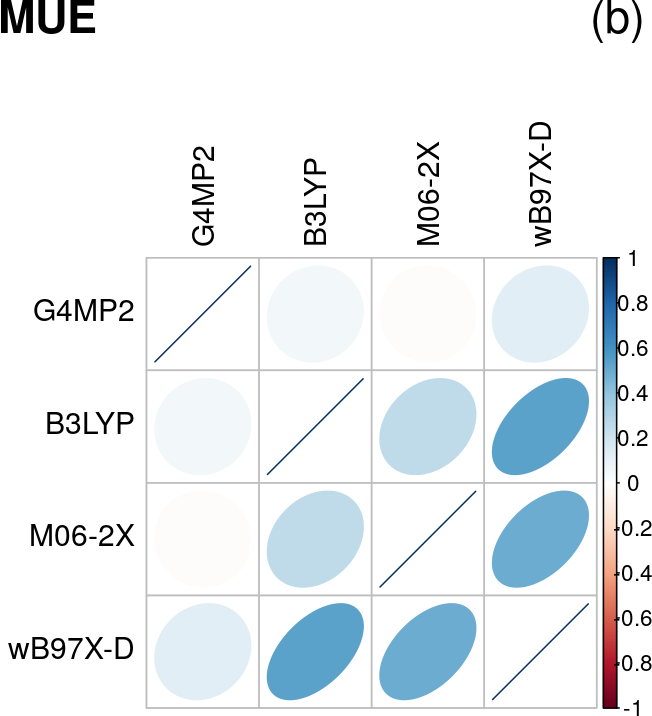} & ~\includegraphics[width=0.32\textwidth]{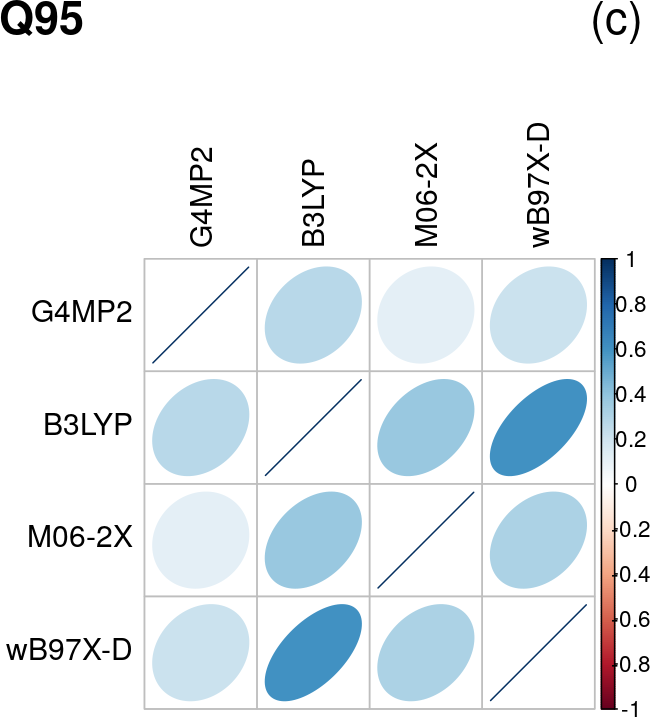}\tabularnewline
\end{tabular}
\par\end{centering}
\noindent \centering{}\caption{\label{fig:narayanan1}Case NAR2019 - rank correlation matrices: (a)
Errors; (b) MUE; (c) $Q_{95}$.}
\end{figure}

The most remarkable feature of the correlation matrices in Fig.\,\ref{fig:narayanan1}
is the decorrelation of G4MP2 errors from the other error sets. For
the MUE and $Q_{95}$, weak positive correlations appear, more notably
for $Q_{95}$. 

\paragraph{Statistics.}

\begin{figure}[!tb]
\noindent \begin{centering}
\includegraphics[width=0.32\textwidth]{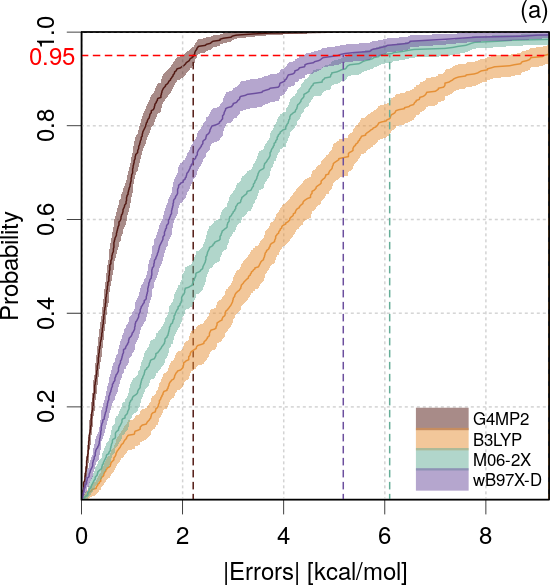}~\includegraphics[width=0.32\textwidth]{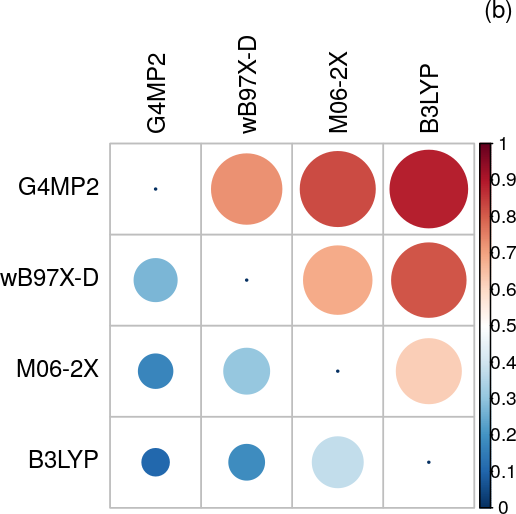}
\par\end{centering}
\noindent \begin{centering}
\includegraphics[width=0.32\textwidth]{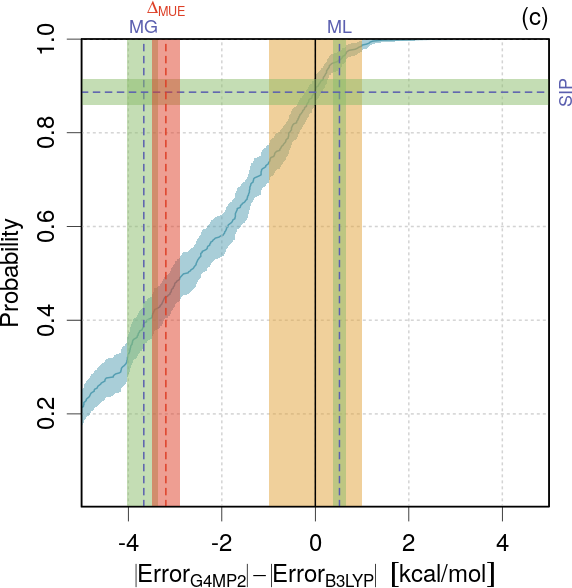}~\includegraphics[width=0.32\textwidth]{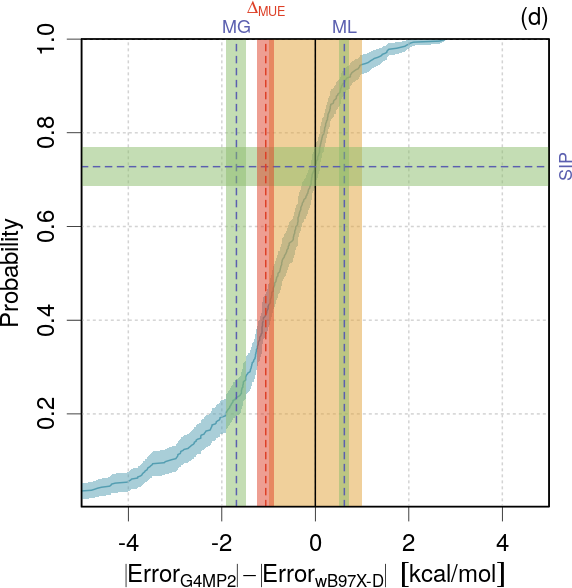}
\par\end{centering}
\noindent \centering{}\caption{\label{fig:narayanan2}Case NAR2019: (a) ECDF of the absolute errors;
(b) SIP matrix; (c,d) ECDF of the difference of absolute errors of
B3LYP and $\omega$B97X-D with respect to G4MP2 (see Fig.\,\ref{fig:pernot1-1}
for details).}
\end{figure}

The statistics reported in Table\,\ref{tab:narayanan} show the supremacy
of G4MP2 over the three DFAs for all statistics. Narayanan \emph{et
al.} \citep{Narayanan2019} claim an ``accuracy''\footnote{The MUE is sometimes abusively used to characterize the \emph{accuracy}
of a method, which cannot be the case when error distributions are
not zero-centered normal \citep{Pernot2015,Pernot2018}.} (MUE) of 0.79\,kcal/mol with G4MP2. However, a look at the absolute
errors CDFs (Fig.\,\ref{fig:narayanan2}(a)) shows that for G4MP2,
there is still a probability of about 20\,\% that the absolute errors
exceed 1 kcal/mol, and 5\,\% to exceed 2.2 kcal/mol. 
\begin{table}[!tb]
\begin{centering}
{\small{}}%
\begin{tabular}{llr@{\extracolsep{0pt}.}lr@{\extracolsep{0pt}.}lr@{\extracolsep{0pt}.}lr@{\extracolsep{0pt}.}lr@{\extracolsep{0pt}.}lr@{\extracolsep{0pt}.}lr@{\extracolsep{0pt}.}lr@{\extracolsep{0pt}.}lr@{\extracolsep{0pt}.}lr@{\extracolsep{0pt}.}l}
\hline 
Methods &  & \multicolumn{2}{c}{MUE } & \multicolumn{2}{c}{$P_{inv}$} & \multicolumn{2}{c}{} & \multicolumn{2}{c}{$Q_{95}$} & \multicolumn{2}{c}{$P_{inv}$} & \multicolumn{2}{c}{} & \multicolumn{2}{c}{MSIP} & \multicolumn{2}{c}{SIP } & \multicolumn{2}{c}{MG} & \multicolumn{2}{c}{ML}\tabularnewline
 &  & \multicolumn{2}{c}{kcal/mol} & \multicolumn{2}{c}{} & \multicolumn{2}{c}{} & \multicolumn{2}{c}{kcal/mol} & \multicolumn{2}{c}{} & \multicolumn{2}{c}{} & \multicolumn{2}{c}{} & \multicolumn{2}{c}{} & \multicolumn{2}{c}{kcal/mol} & \multicolumn{2}{c}{kcal/mol}\tabularnewline
\cline{1-1} \cline{3-6} \cline{9-12} \cline{15-22} 
G4MP2  &  & \textbf{0}&\textbf{79(3) } & \multicolumn{2}{c}{-} & \multicolumn{2}{c}{} & \textbf{2}&\textbf{21(9) } & \multicolumn{2}{c}{-} & \multicolumn{2}{c}{} & \textbf{0}&\textbf{81(2) } & \multicolumn{2}{c}{- } & \multicolumn{2}{c}{- } & \multicolumn{2}{c}{- }\tabularnewline
B3LYP  &  & 4&0(2)  & 0&0 & \multicolumn{2}{c}{} & 9&3(6)  & 0&0 & \multicolumn{2}{c}{} & 0&22(2)  & 0&89(1)  & -3&7(2)  & 0&52(7) \tabularnewline
M06-2X  &  & 2&71(10)  & 0&0 & \multicolumn{2}{c}{} & 6&1(5)  & 0&0 & \multicolumn{2}{c}{} & 0&37(2)  & 0&83(2)  & -2&5(1)  & 0&82(7) \tabularnewline
$\omega$B97X-D  &  & 1&85(9)  & 0&0 & \multicolumn{2}{c}{} & 5&2(4)  & 0&0 & \multicolumn{2}{c}{} & 0&59(2)  & 0&73(2)  & -1&7(1)  & 0&62(5) \tabularnewline
\hline 
\end{tabular}{\small\par}
\par\end{centering}
\noindent \centering{}\caption{\label{tab:narayanan}Case NAR2019 - absolute error statistics: inversion
probabilities and SIP statistics for comparison with the DFA of smallest
MUE (G4MP2). The best scores are in boldface.}
\end{table}

\paragraph{SIP analysis.}

G4MP2 presents a high degree of systematic improvement over the three
DFAs (MSIP = 0.81). Nonetheless, there is about 27\,\% probability
(1-0.73) that $\omega$B97X-D performs better, but with a rather small
value of ML (0.62\,kcal/mol), when compared to the chemical accuracy
(Fig.\,\ref{fig:narayanan2}(d)). In contrast, the mean gain when
using G4MP2 instead of $\omega$B97X-D is about -1.7\,kcal/mol for
73\,\% of the systems. The advantage of G4MP2 over B3LYP is more
spectacular (Fig.\,\ref{fig:narayanan2}(c)). 

\subsection{CAL2019\label{subsec:Caldeweyher2019}}

The impact of an atomic-charge dependent London dispersion correction
(D4 model) has been evaluated by Caldeweyher \emph{at al.} \citep{Caldeweyher2019}
on a large series of datasets. From those, we selected one of the
largest ones, \emph{i.e.}, the reference energies for the MOR41 transition
metal reaction benchmark set \citep{Dohm2018}, available as Tables\,14-18
in the Supplementary Information of the reference article.\footnote{Reproducibility note: these data are inconsistent with the results
reported in Fig.\,9 of the reference article and the subsequent discussion.
We contacted the corresponding author (S. Grimme) who kindly sent
us a corrected version of the Supplementary Information. } The reference data are calculated values, with a priori no significant
numerical uncertainty. The London dispersion corrections have been
tested on a series of 10 DFAs. Note that the nomenclature used here
for the corrections is the one provided in the SI table, which differs
somewhat from the one used in the reference article. 

\paragraph{Statistics.}

The results are reported in Tables\,\ref{tab:caldeweyher}-\ref{tab:caldeweyher-1}
, where DFT-D3 has been taken as reference throughout for $P_{inv}$
estimation. The aim here is to check if DFT-D4 brings significant
differences. It is notable that with a set of size 41, the sampling
uncertainty is rather large for both statistics (typically on the
second or first digit). Nevertheless, significant MUE improvements
are observed when passing from DFT-D3 do DFT-D4, except for revPBE
and PW6B95. In the latter case, the better MUE of the D3 calculations,
noted by the reference authors, might be due to a random effect of
dataset selection. Based on $Q_{95}$ the improvements due to D4 are
not significant, except for DOD-PBE, DSD-PBE and RPBE. Globally, DFT-D4
improves the MUE, but does not reduce the risk of large errors. Note
that comparisons of $Q_{95}$ values have to be taken with care, considering
the small size of the dataset. 
\begin{table}[t]
\begin{centering}
\begin{tabular}{llr@{\extracolsep{0pt}.}lr@{\extracolsep{0pt}.}lr@{\extracolsep{0pt}.}lr@{\extracolsep{0pt}.}lr@{\extracolsep{0pt}.}lr@{\extracolsep{0pt}.}lr@{\extracolsep{0pt}.}lr@{\extracolsep{0pt}.}lr@{\extracolsep{0pt}.}lr@{\extracolsep{0pt}.}l}
\hline 
{\footnotesize{}Methods} &  & \multicolumn{2}{c}{{\footnotesize{}MUE }} & \multicolumn{2}{c}{{\footnotesize{}$P_{inv}$}} & \multicolumn{2}{c}{} & \multicolumn{2}{c}{{\footnotesize{}$Q_{95}$}} & \multicolumn{2}{c}{{\footnotesize{}$P_{inv}$}} & \multicolumn{2}{c}{} & \multicolumn{2}{c}{{\footnotesize{}MSIP}} & \multicolumn{2}{c}{{\footnotesize{}SIP }} & \multicolumn{2}{c}{{\footnotesize{}MG}} & \multicolumn{2}{c}{{\footnotesize{}ML}}\tabularnewline
 &  & \multicolumn{2}{c}{{\footnotesize{}kcal/mol}} & \multicolumn{2}{c}{} & \multicolumn{2}{c}{} & \multicolumn{2}{c}{{\footnotesize{}kcal/mol}} & \multicolumn{2}{c}{} & \multicolumn{2}{c}{} & \multicolumn{2}{c}{} & \multicolumn{2}{c}{} & \multicolumn{2}{c}{{\footnotesize{}kcal/mol}} & \multicolumn{2}{c}{{\footnotesize{}kcal/mol}}\tabularnewline
\cline{1-1} \cline{3-6} \cline{9-12} \cline{15-22} 
{\footnotesize{}DOD-PBE-D4-ATM } &  & \textbf{\footnotesize{}2}&\textbf{\footnotesize{}1(4) } & {\footnotesize{}0}&{\footnotesize{}00 } & \multicolumn{2}{c}{} & \multicolumn{2}{c}{\textbf{\footnotesize{}7(2) }} & {\footnotesize{}0}&{\footnotesize{}00 } & \multicolumn{2}{c}{} & {\footnotesize{}0}&{\footnotesize{}63(6) } & \multicolumn{2}{c}{{\footnotesize{}-}} & \multicolumn{2}{c}{{\footnotesize{}-}} & \multicolumn{2}{c}{{\footnotesize{}-}}\tabularnewline
{\footnotesize{}DOD-PBE-D4-MBD } &  & \textbf{\footnotesize{}2}&\textbf{\footnotesize{}1(4) } & {\footnotesize{}0}&{\footnotesize{}00 } & \multicolumn{2}{c}{} & \multicolumn{2}{c}{\textbf{\footnotesize{}8(2) }} & {\footnotesize{}0}&{\footnotesize{}00 } & \multicolumn{2}{c}{} & {\footnotesize{}0}&{\footnotesize{}65(6) } & {\footnotesize{}0}&{\footnotesize{}44(8) } & {\footnotesize{}-0}&{\footnotesize{}28(4) } & {\footnotesize{}0}&{\footnotesize{}24(5) }\tabularnewline
{\footnotesize{}DOD-PBE-D3 } &  & {\footnotesize{}3}&{\footnotesize{}5(4) } & \multicolumn{2}{c}{{\footnotesize{}- }} & \multicolumn{2}{c}{} & \multicolumn{2}{c}{{\footnotesize{}10(2) }} & \multicolumn{2}{c}{{\footnotesize{}- }} & \multicolumn{2}{c}{} & {\footnotesize{}0}&{\footnotesize{}13(4) } & {\footnotesize{}0}&{\footnotesize{}83(6) } & {\footnotesize{}-1}&{\footnotesize{}8(3) } & {\footnotesize{}0}&{\footnotesize{}8(2) }\tabularnewline
\noalign{\vskip\doublerulesep}
{\footnotesize{}DSD-PBE-D4-ATM } &  & \textbf{\footnotesize{}2}&\textbf{\footnotesize{}9(5) } & {\footnotesize{}0}&{\footnotesize{}00 } & \multicolumn{2}{c}{} & \multicolumn{2}{c}{\textbf{\footnotesize{}11(3) }} & {\footnotesize{}0}&{\footnotesize{}00 } & \multicolumn{2}{c}{} & {\footnotesize{}0}&{\footnotesize{}35(4) } & \multicolumn{2}{c}{{\footnotesize{}-}} & \multicolumn{2}{c}{{\footnotesize{}-}} & \multicolumn{2}{c}{{\footnotesize{}-}}\tabularnewline
{\footnotesize{}DSD-PBE-D4-MBD } &  & \textbf{\footnotesize{}2}&\textbf{\footnotesize{}9(5) } & {\footnotesize{}0}&{\footnotesize{}00 } & \multicolumn{2}{c}{} & \multicolumn{2}{c}{\textbf{\footnotesize{}11(3) }} & {\footnotesize{}0}&{\footnotesize{}00 } & \multicolumn{2}{c}{} & {\footnotesize{}0}&{\footnotesize{}35(4) } & \multicolumn{2}{c}{{\footnotesize{}0 }} & \multicolumn{2}{c}{{\footnotesize{}0 }} & \multicolumn{2}{c}{{\footnotesize{}0 }}\tabularnewline
{\footnotesize{}DSD-PBE-D3 } &  & {\footnotesize{}3}&{\footnotesize{}7(5) } & \multicolumn{2}{c}{{\footnotesize{}- }} & \multicolumn{2}{c}{} & \multicolumn{2}{c}{{\footnotesize{}12(2) }} & \multicolumn{2}{c}{{\footnotesize{}- }} & \multicolumn{2}{c}{} & {\footnotesize{}0}&{\footnotesize{}29(6) } & {\footnotesize{}0}&{\footnotesize{}71(7) } & {\footnotesize{}-1}&{\footnotesize{}5(2) } & {\footnotesize{}0}&{\footnotesize{}7(1) }\tabularnewline
\noalign{\vskip\doublerulesep}
{\footnotesize{}B3LYP-D4-ATM } &  & \textbf{\footnotesize{}4}&\textbf{\footnotesize{}2(5) } & {\footnotesize{}0}&{\footnotesize{}00 } & \multicolumn{2}{c}{} & \multicolumn{2}{c}{\textbf{\footnotesize{}11(3) }} & {\footnotesize{}0}&{\footnotesize{}07 } & \multicolumn{2}{c}{} & {\footnotesize{}0}&{\footnotesize{}56(6) } & {\footnotesize{}0}&{\footnotesize{}41(8) } & {\footnotesize{}-0}&{\footnotesize{}22(4) } & {\footnotesize{}0}&{\footnotesize{}21(3) }\tabularnewline
{\footnotesize{}B3LYP-D4-MBD } &  & \textbf{\footnotesize{}4}&\textbf{\footnotesize{}2(5) } & {\footnotesize{}0}&{\footnotesize{}00 } & \multicolumn{2}{c}{} & \multicolumn{2}{c}{\textbf{\footnotesize{}11(3) }} & {\footnotesize{}0}&{\footnotesize{}11 } & \multicolumn{2}{c}{} & {\footnotesize{}0}&{\footnotesize{}56(6) } & \multicolumn{2}{c}{{\footnotesize{}-}} & \multicolumn{2}{c}{{\footnotesize{}-}} & \multicolumn{2}{c}{{\footnotesize{}-}}\tabularnewline
{\footnotesize{}B3LYP-D3 } &  & {\footnotesize{}4}&{\footnotesize{}8(6) } & \multicolumn{2}{c}{{\footnotesize{}- }} & \multicolumn{2}{c}{} & \multicolumn{2}{c}{\textbf{\footnotesize{}13(3) }} & \multicolumn{2}{c}{{\footnotesize{}- }} & \multicolumn{2}{c}{} & {\footnotesize{}0}&{\footnotesize{}26(6) } & {\footnotesize{}0}&{\footnotesize{}71(7) } & {\footnotesize{}-1}&{\footnotesize{}1(2) } & {\footnotesize{}0}&{\footnotesize{}8(2) }\tabularnewline
\noalign{\vskip\doublerulesep}
{\footnotesize{}PBE0-D4-ATM } &  & \textbf{\footnotesize{}2}&\textbf{\footnotesize{}3(3) } & {\footnotesize{}0}&{\footnotesize{}01 } & \multicolumn{2}{c}{} & \multicolumn{2}{c}{\textbf{\footnotesize{}8(1) }} & {\footnotesize{}0}&{\footnotesize{}08 } & \multicolumn{2}{c}{} & {\footnotesize{}0}&{\footnotesize{}30(4) } & \multicolumn{2}{c}{{\footnotesize{}-}} & \multicolumn{2}{c}{{\footnotesize{}-}} & \multicolumn{2}{c}{{\footnotesize{}-}}\tabularnewline
{\footnotesize{}PBE0-D4-MBD } &  & \textbf{\footnotesize{}2}&\textbf{\footnotesize{}3(3) } & {\footnotesize{}0}&{\footnotesize{}01 } & \multicolumn{2}{c}{} & \multicolumn{2}{c}{\textbf{\footnotesize{}8(1) }} & {\footnotesize{}0}&{\footnotesize{}08 } & \multicolumn{2}{c}{} & {\footnotesize{}0}&{\footnotesize{}30(5) } & \multicolumn{2}{c}{{\footnotesize{}0 }} & \multicolumn{2}{c}{{\footnotesize{}0 }} & \multicolumn{2}{c}{{\footnotesize{}0 }}\tabularnewline
{\footnotesize{}PBE0-D3 } &  & {\footnotesize{}2}&{\footnotesize{}6(4) } & \multicolumn{2}{c}{{\footnotesize{}- }} & \multicolumn{2}{c}{} & \multicolumn{2}{c}{\textbf{\footnotesize{}8(1) }} & \multicolumn{2}{c}{{\footnotesize{}- }} & \multicolumn{2}{c}{} & {\footnotesize{}0}&{\footnotesize{}29(6) } & {\footnotesize{}0}&{\footnotesize{}61(8) } & {\footnotesize{}-0}&{\footnotesize{}7(1) } & {\footnotesize{}0}&{\footnotesize{}4(1) }\tabularnewline
\noalign{\vskip\doublerulesep}
{\footnotesize{}PW6B95-D4-ATM } &  & \textbf{\footnotesize{}3}&\textbf{\footnotesize{}2(4) } & {\footnotesize{}0}&{\footnotesize{}02 } & \multicolumn{2}{c}{} & \textbf{\footnotesize{}7}&\textbf{\footnotesize{}9(9) } & {\footnotesize{}0}&{\footnotesize{}30 } & \multicolumn{2}{c}{} & {\footnotesize{}0}&{\footnotesize{}35(6) } & {\footnotesize{}0}&{\footnotesize{}56(8) } & {\footnotesize{}-1}&{\footnotesize{}6(2) } & {\footnotesize{}1}&{\footnotesize{}0(2) }\tabularnewline
{\footnotesize{}PW6B95-D4-MBD } &  & \textbf{\footnotesize{}3}&\textbf{\footnotesize{}0(4) } & {\footnotesize{}0}&{\footnotesize{}08 } & \multicolumn{2}{c}{} & \textbf{\footnotesize{}7}&\textbf{\footnotesize{}8(8) } & {\footnotesize{}0}&{\footnotesize{}31 } & \multicolumn{2}{c}{} & {\footnotesize{}0}&{\footnotesize{}48(6) } & {\footnotesize{}0}&{\footnotesize{}54(8) } & {\footnotesize{}-1}&{\footnotesize{}3(2) } & {\footnotesize{}1}&{\footnotesize{}0(2) }\tabularnewline
{\footnotesize{}PW6B95-D3 } &  & \textbf{\footnotesize{}2}&\textbf{\footnotesize{}7(4) } & \multicolumn{2}{c}{{\footnotesize{}- }} & \multicolumn{2}{c}{} & \textbf{\footnotesize{}7}&\textbf{\footnotesize{}4(9) } & \multicolumn{2}{c}{{\footnotesize{}- }} & \multicolumn{2}{c}{} & {\footnotesize{}0}&{\footnotesize{}55(6) } & \multicolumn{2}{c}{{\footnotesize{}-}} & \multicolumn{2}{c}{{\footnotesize{}-}} & \multicolumn{2}{c}{{\footnotesize{}-}}\tabularnewline
\noalign{\vskip\doublerulesep}
{\footnotesize{}CAM-B3LYP-D4-ATM } &  & \textbf{\footnotesize{}3}&\textbf{\footnotesize{}7(4) } & {\footnotesize{}0}&{\footnotesize{}00 } & \multicolumn{2}{c}{} & \multicolumn{2}{c}{\textbf{\footnotesize{}9(1) }} & {\footnotesize{}0}&{\footnotesize{}04 } & \multicolumn{2}{c}{} & {\footnotesize{}0}&{\footnotesize{}38(4) } & \multicolumn{2}{c}{{\footnotesize{}-}} & \multicolumn{2}{c}{{\footnotesize{}-}} & \multicolumn{2}{c}{{\footnotesize{}-}}\tabularnewline
{\footnotesize{}CAM-B3LYP-D4-MBD } &  & \textbf{\footnotesize{}3}&\textbf{\footnotesize{}7(4) } & {\footnotesize{}0}&{\footnotesize{}00 } & \multicolumn{2}{c}{} & \multicolumn{2}{c}{\textbf{\footnotesize{}9(1) }} & {\footnotesize{}0}&{\footnotesize{}04 } & \multicolumn{2}{c}{} & {\footnotesize{}0}&{\footnotesize{}38(4) } & \multicolumn{2}{c}{{\footnotesize{}0 }} & \multicolumn{2}{c}{{\footnotesize{}0 }} & \multicolumn{2}{c}{{\footnotesize{}0 }}\tabularnewline
{\footnotesize{}CAM-B3LYP-D3 } &  & {\footnotesize{}4}&{\footnotesize{}3(4) } & \multicolumn{2}{c}{{\footnotesize{}- }} & \multicolumn{2}{c}{} & \multicolumn{2}{c}{\textbf{\footnotesize{}10(1) }} & \multicolumn{2}{c}{{\footnotesize{}- }} & \multicolumn{2}{c}{} & {\footnotesize{}0}&{\footnotesize{}20(5) } & {\footnotesize{}0}&{\footnotesize{}76(7) } & {\footnotesize{}-0}&{\footnotesize{}8(1) } & {\footnotesize{}0}&{\footnotesize{}5(1) }\tabularnewline
\noalign{\vskip\doublerulesep}
{\footnotesize{}revPBE-D4-ATM } &  & \textbf{\footnotesize{}3}&\textbf{\footnotesize{}3(5) } & {\footnotesize{}0}&{\footnotesize{}11 } & \multicolumn{2}{c}{} & \multicolumn{2}{c}{\textbf{\footnotesize{}12(2) }} & {\footnotesize{}0}&{\footnotesize{}33 } & \multicolumn{2}{c}{} & {\footnotesize{}0}&{\footnotesize{}43(6) } & {\footnotesize{}0}&{\footnotesize{}54(8) } & {\footnotesize{}-0}&{\footnotesize{}27(6) } & {\footnotesize{}0}&{\footnotesize{}28(6) }\tabularnewline
{\footnotesize{}revPBE-D4-MBD } &  & \textbf{\footnotesize{}3}&\textbf{\footnotesize{}3(6) } & {\footnotesize{}0}&{\footnotesize{}08 } & \multicolumn{2}{c}{} & \multicolumn{2}{c}{\textbf{\footnotesize{}12(2) }} & {\footnotesize{}0}&{\footnotesize{}39 } & \multicolumn{2}{c}{} & {\footnotesize{}0}&{\footnotesize{}54(7) } & \multicolumn{2}{c}{{\footnotesize{}-}} & \multicolumn{2}{c}{{\footnotesize{}-}} & \multicolumn{2}{c}{{\footnotesize{}-}}\tabularnewline
{\footnotesize{}revPBE-D3 } &  & \textbf{\footnotesize{}3}&\textbf{\footnotesize{}8(6) } & \multicolumn{2}{c}{{\footnotesize{}- }} & \multicolumn{2}{c}{} & \multicolumn{2}{c}{\textbf{\footnotesize{}12(1) }} & \multicolumn{2}{c}{{\footnotesize{}- }} & \multicolumn{2}{c}{} & {\footnotesize{}0}&{\footnotesize{}46(7) } & {\footnotesize{}0}&{\footnotesize{}54(8) } & {\footnotesize{}-2}&{\footnotesize{}0(4) } & {\footnotesize{}1}&{\footnotesize{}3(3) }\tabularnewline
\hline 
\end{tabular}
\par\end{centering}
\noindent \centering{}\caption{\label{tab:caldeweyher}Case CAL2019 - absolute error statistics:
inversion probabilities are calculated for comparison with DFT-D3,
for each DFT. The SIP statistics are calculated for comparison with
the smallest MUE within each DFT. The best scores and the values for
which $p_{g}>0.05$ are in boldface.}
\end{table}
 
\begin{table}[t]
\begin{centering}
\begin{tabular}{llr@{\extracolsep{0pt}.}lr@{\extracolsep{0pt}.}lr@{\extracolsep{0pt}.}lr@{\extracolsep{0pt}.}lr@{\extracolsep{0pt}.}lr@{\extracolsep{0pt}.}lr@{\extracolsep{0pt}.}lr@{\extracolsep{0pt}.}lr@{\extracolsep{0pt}.}lr@{\extracolsep{0pt}.}l}
\hline 
{\footnotesize{}Methods} &  & \multicolumn{2}{c}{{\footnotesize{}MUE }} & \multicolumn{2}{c}{{\footnotesize{}$P_{inv}$}} & \multicolumn{2}{c}{} & \multicolumn{2}{c}{{\footnotesize{}$Q_{95}$}} & \multicolumn{2}{c}{{\footnotesize{}$P_{inv}$}} & \multicolumn{2}{c}{} & \multicolumn{2}{c}{{\footnotesize{}MSIP}} & \multicolumn{2}{c}{{\footnotesize{}SIP }} & \multicolumn{2}{c}{{\footnotesize{}MG}} & \multicolumn{2}{c}{{\footnotesize{}ML}}\tabularnewline
 &  & \multicolumn{2}{c}{{\footnotesize{}kcal/mol}} & \multicolumn{2}{c}{} & \multicolumn{2}{c}{} & \multicolumn{2}{c}{{\footnotesize{}kcal/mol}} & \multicolumn{2}{c}{} & \multicolumn{2}{c}{} & \multicolumn{2}{c}{} & \multicolumn{2}{c}{} & \multicolumn{2}{c}{{\footnotesize{}kcal/mol}} & \multicolumn{2}{c}{{\footnotesize{}kcal/mol}}\tabularnewline
\cline{1-1} \cline{3-6} \cline{9-12} \cline{15-22} 
\noalign{\vskip\doublerulesep}
{\footnotesize{}M06L-D4-ATM } &  & \textbf{\footnotesize{}5}&\textbf{\footnotesize{}1(6) } & {\footnotesize{}0}&{\footnotesize{}00 } & \multicolumn{2}{c}{} & \multicolumn{2}{c}{\textbf{\footnotesize{}13(1) }} & {\footnotesize{}0}&{\footnotesize{}08 } & \multicolumn{2}{c}{} & {\footnotesize{}0}&{\footnotesize{}35(4) } & \multicolumn{2}{c}{{\footnotesize{}-}} & \multicolumn{2}{c}{{\footnotesize{}-}} & \multicolumn{2}{c}{{\footnotesize{}-}}\tabularnewline
{\footnotesize{}M06L-D4-MBD } &  & \textbf{\footnotesize{}5}&\textbf{\footnotesize{}1(6) } & {\footnotesize{}0}&{\footnotesize{}00 } & \multicolumn{2}{c}{} & \multicolumn{2}{c}{\textbf{\footnotesize{}13(1) }} & {\footnotesize{}0}&{\footnotesize{}08 } & \multicolumn{2}{c}{} & {\footnotesize{}0}&{\footnotesize{}35(4) } & \multicolumn{2}{c}{{\footnotesize{}0 }} & \multicolumn{2}{c}{{\footnotesize{}0 }} & \multicolumn{2}{c}{{\footnotesize{}0 }}\tabularnewline
{\footnotesize{}M06L-D3 } &  & {\footnotesize{}5}&{\footnotesize{}5(6) } & \multicolumn{2}{c}{{\footnotesize{}- }} & \multicolumn{2}{c}{} & \multicolumn{2}{c}{\textbf{\footnotesize{}14(1) }} & \multicolumn{2}{c}{{\footnotesize{}- }} & \multicolumn{2}{c}{} & {\footnotesize{}0}&{\footnotesize{}22(5) } & {\footnotesize{}0}&{\footnotesize{}71(7) } & {\footnotesize{}-0}&{\footnotesize{}7(1) } & {\footnotesize{}0}&{\footnotesize{}5(2) }\tabularnewline
\noalign{\vskip\doublerulesep}
{\footnotesize{}PBE-D4-ATM } &  & \textbf{\footnotesize{}3}&\textbf{\footnotesize{}5(5) } & {\footnotesize{}0}&{\footnotesize{}00 } & \multicolumn{2}{c}{} & \multicolumn{2}{c}{\textbf{\footnotesize{}12(2) }} & {\footnotesize{}0}&{\footnotesize{}34 } & \multicolumn{2}{c}{} & {\footnotesize{}0}&{\footnotesize{}45(6) } & {\footnotesize{}0}&{\footnotesize{}51(8) } & {\footnotesize{}-0}&{\footnotesize{}20(5) } & {\footnotesize{}0}&{\footnotesize{}16(2) }\tabularnewline
{\footnotesize{}PBE-D4-MBD } &  & \textbf{\footnotesize{}3}&\textbf{\footnotesize{}4(5) } & {\footnotesize{}0}&{\footnotesize{}00 } & \multicolumn{2}{c}{} & \multicolumn{2}{c}{\textbf{\footnotesize{}12(2) }} & {\footnotesize{}0}&{\footnotesize{}48 } & \multicolumn{2}{c}{} & {\footnotesize{}0}&{\footnotesize{}60(6) } & \multicolumn{2}{c}{{\footnotesize{}-}} & \multicolumn{2}{c}{{\footnotesize{}-}} & \multicolumn{2}{c}{{\footnotesize{}-}}\tabularnewline
{\footnotesize{}PBE-D3 } &  & {\footnotesize{}3}&{\footnotesize{}9(5) } & \multicolumn{2}{c}{{\footnotesize{}- }} & \multicolumn{2}{c}{} & \multicolumn{2}{c}{\textbf{\footnotesize{}12(2) }} & \multicolumn{2}{c}{{\footnotesize{}- }} & \multicolumn{2}{c}{} & {\footnotesize{}0}&{\footnotesize{}30(6) } & {\footnotesize{}0}&{\footnotesize{}68(7) } & {\footnotesize{}-1}&{\footnotesize{}0(1) } & {\footnotesize{}0}&{\footnotesize{}5(2) }\tabularnewline
\noalign{\vskip\doublerulesep}
{\footnotesize{}RPBE-D4-ATM } &  & \textbf{\footnotesize{}3}&\textbf{\footnotesize{}4(6) } & {\footnotesize{}0}&{\footnotesize{}00 } & \multicolumn{2}{c}{} & \multicolumn{2}{c}{\textbf{\footnotesize{}12(2) }} & {\footnotesize{}0}&{\footnotesize{}00 } & \multicolumn{2}{c}{} & {\footnotesize{}0}&{\footnotesize{}48(2) } & \multicolumn{2}{c}{{\footnotesize{}-}} & \multicolumn{2}{c}{{\footnotesize{}-}} & \multicolumn{2}{c}{{\footnotesize{}-}}\tabularnewline
{\footnotesize{}RPBE-D4-MBD } &  & \textbf{\footnotesize{}3}&\textbf{\footnotesize{}4(6) } & {\footnotesize{}0}&{\footnotesize{}00 } & \multicolumn{2}{c}{} & \multicolumn{2}{c}{\textbf{\footnotesize{}12(2) }} & {\footnotesize{}0}&{\footnotesize{}00 } & \multicolumn{2}{c}{} & {\footnotesize{}0}&{\footnotesize{}48(2) } & \multicolumn{2}{c}{{\footnotesize{}0 }} & \multicolumn{2}{c}{{\footnotesize{}0 }} & \multicolumn{2}{c}{{\footnotesize{}0 }}\tabularnewline
{\footnotesize{}RPBE-D3 } &  & {\footnotesize{}8}&{\footnotesize{}3(9) } & \multicolumn{2}{c}{{\footnotesize{}- }} & \multicolumn{2}{c}{} & \multicolumn{2}{c}{{\footnotesize{}20(5) }} & \multicolumn{2}{c}{{\footnotesize{}- }} & \multicolumn{2}{c}{} & {\footnotesize{}0}&{\footnotesize{}05(3) } & {\footnotesize{}0}&{\footnotesize{}95(3) } & {\footnotesize{}-5}&{\footnotesize{}3(7) } & \multicolumn{2}{c}{{\footnotesize{}2(1) }}\tabularnewline
\hline 
\end{tabular}
\par\end{centering}
\noindent \centering{}\caption{\label{tab:caldeweyher-1}Case CAL2019 - Table\,\ref{tab:caldeweyher},
continued.}
\end{table}

\paragraph{SIP analysis.}

\begin{figure}[!tb]
\noindent \centering{}\includegraphics[width=0.32\textwidth]{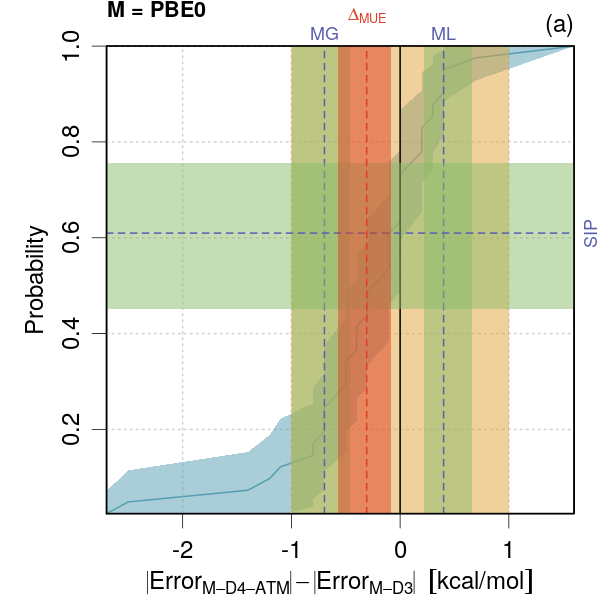}~\includegraphics[width=0.32\textwidth]{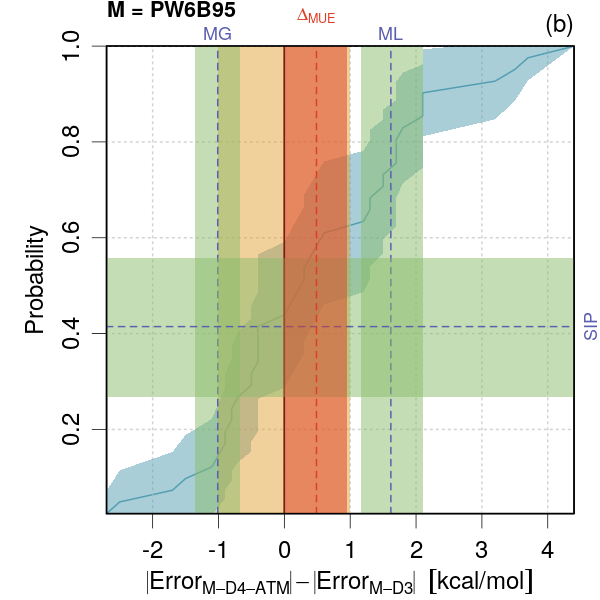}~\includegraphics[width=0.32\textwidth]{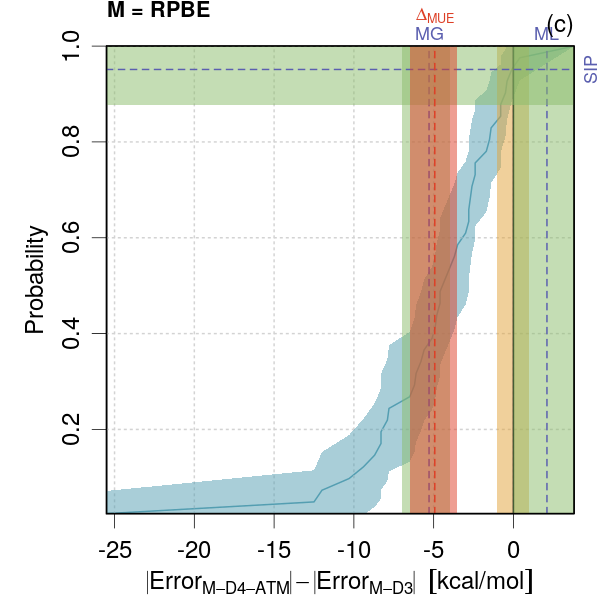}\caption{\label{fig:caldeweyher1}Case CAL2019 - selected SIP plots. The orange
band depicts the chemical accuracy (1\,kcal/mol).}
\end{figure}

Let us consider several examples with the SIP approach:
\begin{itemize}
\item \textbf{PBE0-Dn}. Inspection of Fig.\,\ref{fig:caldeweyher1}(a)
shows that the 95\,\% confidence interval (CI) for the SIP value
of 0.61 for PBE0-D4-ATM over PBE0-D3 does not exclude the neutral
value (0.5), with a tiny advantage of the mean gain over the mean
loss. One can note also that, despite the large uncertainty on the
MUE values 2.3(3) and 2.6(4), the small difference $\Delta_{MUE}=0.3$
between these two methods is significantly different from 0 (its 95\,\%
confidence interval excludes 0), an effect of the strong positive
correlation between the error sets (0.98) as discussed in Paper\,I
\citep{Pernot2020}.
\item \textbf{PW6B95-Dn}. This case is an inversion of the previous one,
where the confidence interval on the SIP value of nearly 0.4 (disadvantaging
D4) does not exclude the neutral value, and the CI on the MUE difference
$\Delta_{MUE}$ does not exclude 0. One cannot firmly conclude that
the D3 version performs better than the D4 ones for this DFA.
\item \textbf{RPBE-Dn}. For this case, one has a rare instance where D4
improves almost systematically over D3, with a SIP of 0.95(3), and
a mean gain overwhelming the mean loss.
\end{itemize}
Except for RPBE-Dn, where the SIP value of D4 over D3 is about 0.95,
and DOD-PBE ($\mathrm{SIP}=0.83$), all the estimated SIP values lie
near or below 0.75, down to 0.45, meaning that there is no systematic
improvement when passing from D3 to D4. In several cases, the uncertainty
due to the limited set size does not allow to conclude clearly. 

\paragraph{Ranking.}

\begin{figure}[!tb]
\noindent \begin{centering}
\includegraphics[width=0.32\columnwidth]{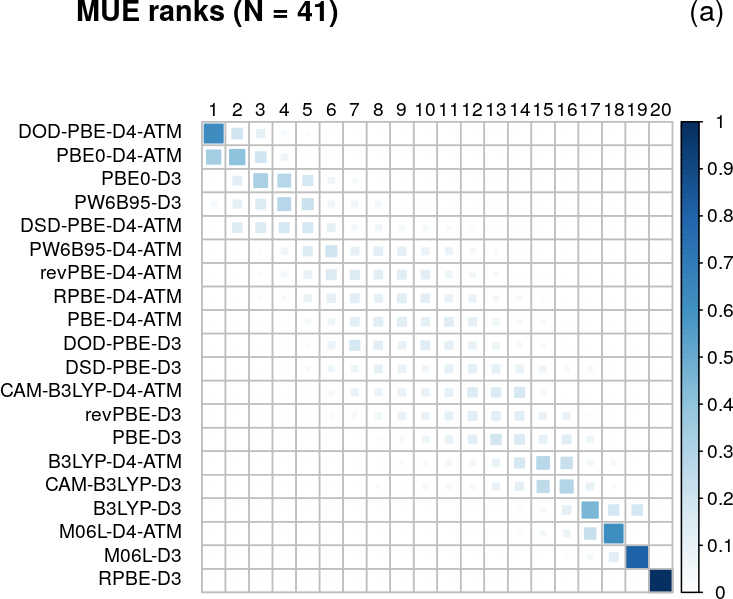}~\includegraphics[width=0.32\columnwidth]{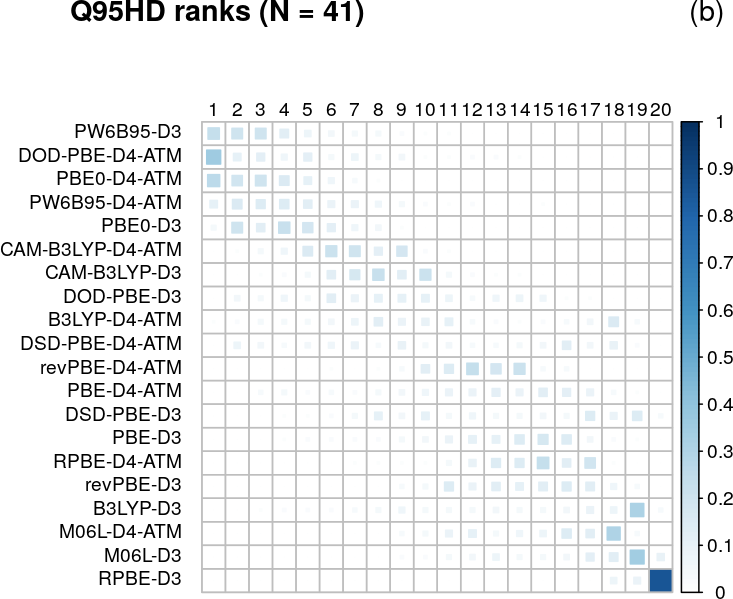}~\includegraphics[width=0.32\columnwidth]{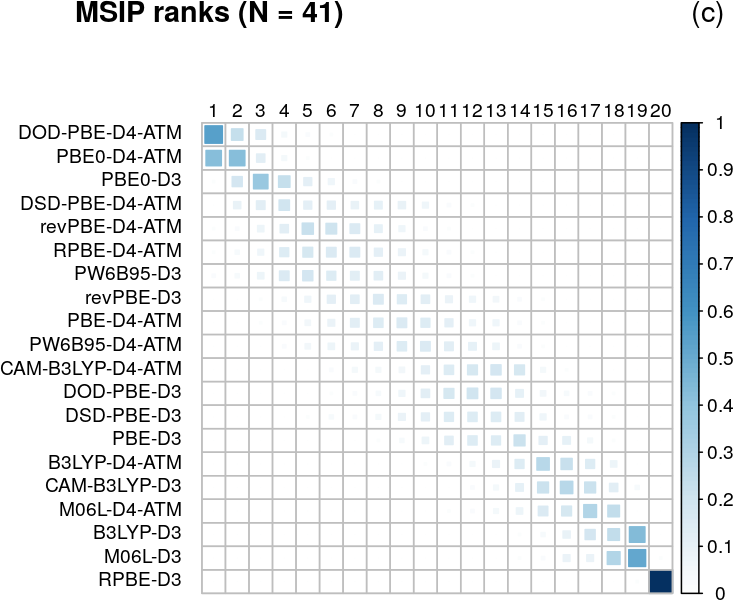}
\par\end{centering}
\noindent \begin{centering}
\includegraphics[width=0.32\columnwidth]{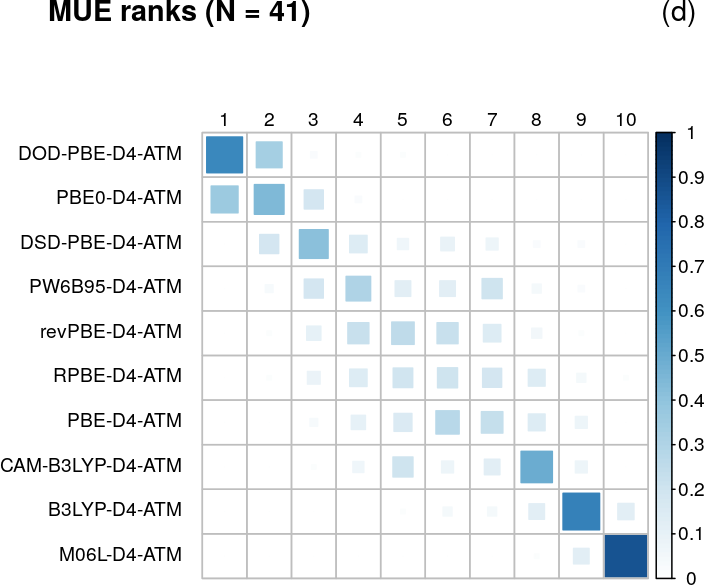}~\includegraphics[width=0.32\columnwidth]{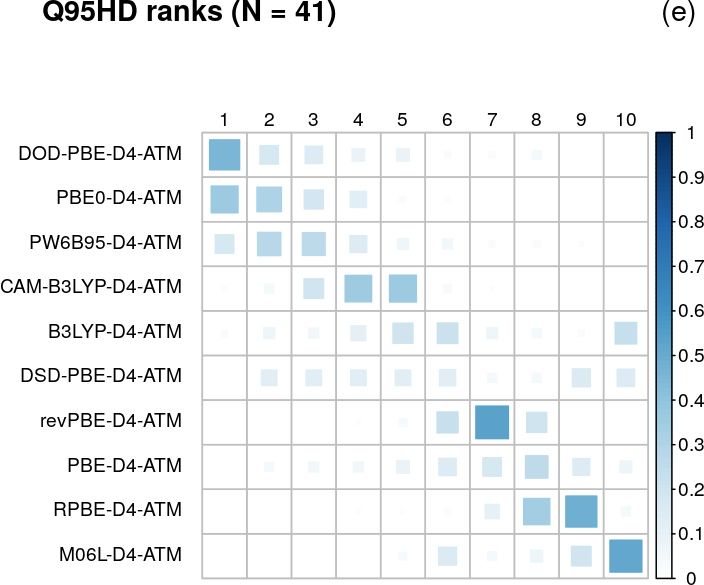}~\includegraphics[width=0.32\columnwidth]{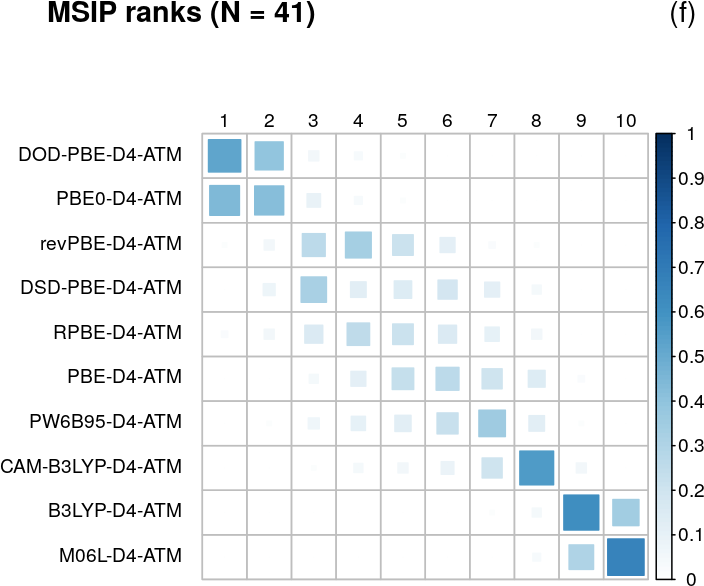}
\par\end{centering}
\noindent \centering{}\caption{\label{fig:caldewheyerRPmat}Case CAL2019: ranking probability matrices
for (a-c) DFT-D3 and DFT-D4-ATM methods, and (d-f) DFT-D4-ATM methods
only. }
\end{figure}

Considering that both DFT-D4 options are mostly indiscernible, we
built global ranking probability matrices for the DFT-D3 and DFT-D4-ATM
data. The results are reported in Fig.\,\ref{fig:caldewheyerRPmat}(top).
Although the rankings of the Dn options for each DFA are mostly unambiguous,
a global ranking is clearly very uncertain. Based on the MUE, DOD-PBE-D4-ATM
and PBE0-D4-ATM would share the leading places. Beyond that, the situation
is utterly scrambled, the only clear point being the last ranks for
M06-L-D3 and RPBE-D3. The picture based on $Q_{95}$ is even less
well defined, with no clear leading method within a group of five.
The MSIP ranking is akin to the MUE ranking.

If one restricts the methods to DFT-D4-ATM (Fig.\,\ref{fig:caldewheyerRPmat}
bottom), the situation is slightly better defined for the leading
and tailing places for the three scores, but remains very undecidable
in intermediate ranks. This illustrates how, for a given dataset,
the uncertainty in ranking is also affected by the number of methods
to be ranked. 

\subsection{JEN2018\label{subsec:Jensen2018}}

This dataset contains non-covalent interaction energies estimated
by M06-L with six different basis sets for 66 systems in the S66 dataset
\citep{Rezac2011,Rezac2011_Erratum}. This is a part of the results
reported in Table 8 of a recent article by Jensen \citep{Jensen2018},
and available as Supplementary Information to this article. This dataset
was used by Jensen to study the impact of error cancellations when
using standard or optimized medium-sized basis sets. Six basis sets
are considered (pop2 = 6-31G(d,p), pop3 = 6-311G(2df,2pd), pcseg-1,
pcseg-4, pop2-opt and pcseg1-opt), where the \textquotedbl -opt\textquotedbl{}
ones have optimized contraction coefficients with respect to the reference
data.

\paragraph{Correlations.}

\begin{figure}[tb]
\noindent \begin{centering}
\includegraphics[width=0.32\textwidth]{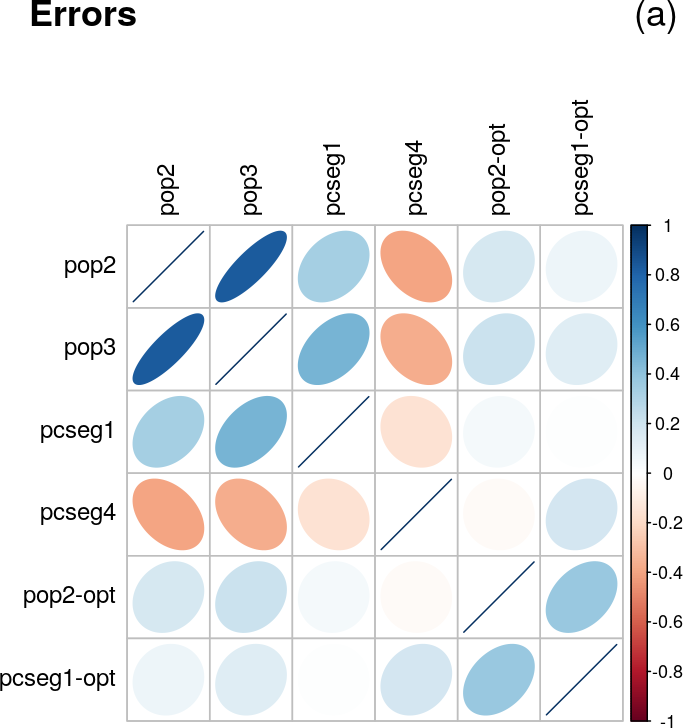}~\includegraphics[width=0.32\textwidth]{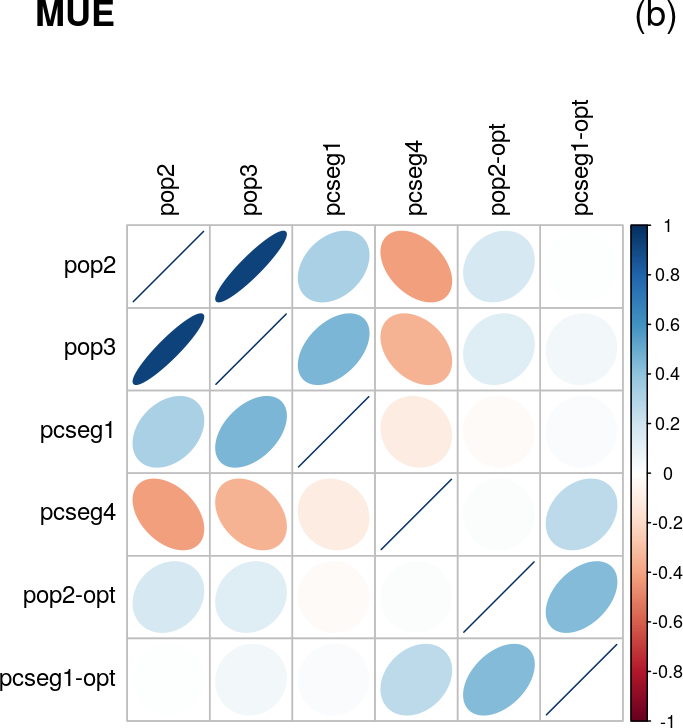}~\includegraphics[width=0.32\textwidth]{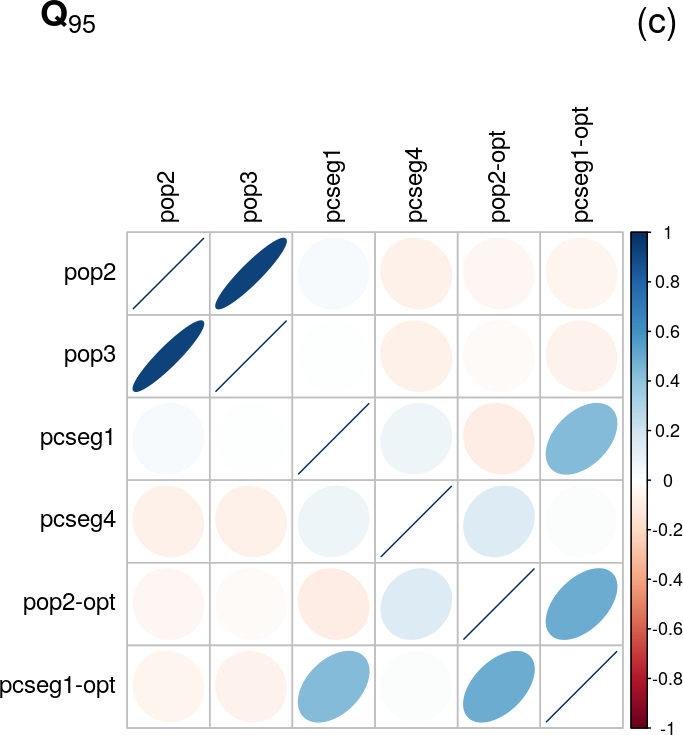}
\par\end{centering}
\noindent \centering{}\caption{\label{fig:jensenCorrmat}Case JEN2018 - rank correlation matrices:
(a) Errors; (b) MUE; (c) $Q_{95}$.}
\end{figure}

The error sets of the \textquotedbl -opt\textquotedbl{} methods are
practically uncorrelated to the other sets (Fig.\,\ref{fig:jensenCorrmat}(a)),
and in the remaining methods, pcseg4 errors are anti-correlated with
the other ones. A striking feature of this dataset is that this negative
correlation persists for the MUE, contradicting the trends observed
in Appendix\,B of Paper\,I \citep{Pernot2020}. Otherwise, the
correlations globally weaken for $Q_{95}$, except for the pop2/pop3
and pcseg1/pcseg1-opt cases, for which the correlation is stronger
as the one between the error sets. 

\paragraph{Statistics.}

The statistics in Table\,\ref{tab:jensen} show the strong impact
of basis-set optimization, both optimized basis sets provide comparable
results for the MUE and $Q_{95}$. All statistics show that the ranking
between both \textquotedbl -opt\textquotedbl{} methods is not strict.
\begin{table}[!tb]
\begin{centering}
{\small{}}%
\begin{tabular}{llr@{\extracolsep{0pt}.}lr@{\extracolsep{0pt}.}lr@{\extracolsep{0pt}.}lr@{\extracolsep{0pt}.}lr@{\extracolsep{0pt}.}lr@{\extracolsep{0pt}.}lr@{\extracolsep{0pt}.}lr@{\extracolsep{0pt}.}lr@{\extracolsep{0pt}.}lr@{\extracolsep{0pt}.}l}
\hline 
Methods  &  & \multicolumn{2}{c}{MUE } & \multicolumn{2}{c}{$P_{inv}$} & \multicolumn{2}{c}{} & \multicolumn{2}{c}{$Q_{95}$} & \multicolumn{2}{c}{$P_{inv}$} & \multicolumn{2}{c}{} & \multicolumn{2}{c}{MSIP} & \multicolumn{2}{c}{SIP } & \multicolumn{2}{c}{MG} & \multicolumn{2}{c}{ML}\tabularnewline
 &  & \multicolumn{2}{c}{kJ/mol} & \multicolumn{2}{c}{} & \multicolumn{2}{c}{} & \multicolumn{2}{c}{kJ/mol} & \multicolumn{2}{c}{} & \multicolumn{2}{c}{} & \multicolumn{2}{c}{} & \multicolumn{2}{c}{} & \multicolumn{2}{c}{kJ/mol} & \multicolumn{2}{c}{kJ/mol}\tabularnewline
\cline{1-1} \cline{3-6} \cline{9-12} \cline{15-22} 
pop2  &  & 2&9(3)  & 0&00  & \multicolumn{2}{c}{} & 7&2(7)  & 0&00  & \multicolumn{2}{c}{} & 0&35(5)  & 0&77(5)  & -2&9(3)  & 0&8(2) \tabularnewline
pop3  &  & 2&4(3)  & 0&00  & \multicolumn{2}{c}{} & 6&4(7)  & 0&00  & \multicolumn{2}{c}{} & 0&47(5)  & 0&74(5)  & -2&3(3)  & 0&8(1) \tabularnewline
pcseg1  &  & 2&5(2)  & 0&00  & \multicolumn{2}{c}{} & 5&6(4)  & 0&00  & \multicolumn{2}{c}{} & 0&42(5)  & 0&76(5)  & -2&3(2)  & 0&9(2) \tabularnewline
pcseg4  &  & 2&5(1)  & 0&00  & \multicolumn{2}{c}{} & 4&8(4)  & 0&00  & \multicolumn{2}{c}{} & 0&33(5)  & 0&89(4)  & -1&8(1)  & 0&6(2) \tabularnewline
pop2-opt  &  & \textbf{1}&\textbf{06(10)}  & 0&05  & \multicolumn{2}{c}{} & \textbf{2}&\textbf{6(2)}  & 0&24  & \multicolumn{2}{c}{} & 0&67(5)  & 0&62(6)  & -0&66(8)  & 0&65(9) \tabularnewline
pcseg1-opt  &  & \textbf{0}&\textbf{90(9) } & \multicolumn{2}{c}{-} & \multicolumn{2}{c}{} & \textbf{2}&\textbf{5(3) } & \multicolumn{2}{c}{-} & \multicolumn{2}{c}{} & \textbf{0}&\textbf{76(5) } & \multicolumn{2}{c}{- } & \multicolumn{2}{c}{- } & \multicolumn{2}{c}{- }\tabularnewline
\hline 
\end{tabular}{\small\par}
\par\end{centering}
\noindent \centering{}\caption{\label{tab:jensen}Case JEN2018 - absolute error statistics: inversion
probabilities and SIP statistics for comparison with the method of
smallest MUE (pcseg1-opt). The best scores and the values for which
$(p_{g}=2P_{inv})>0.05$ are in boldface.}
\end{table}

\paragraph{SIP analysis.}

\begin{figure}[!tb]
\noindent \begin{centering}
\includegraphics[width=0.32\textwidth]{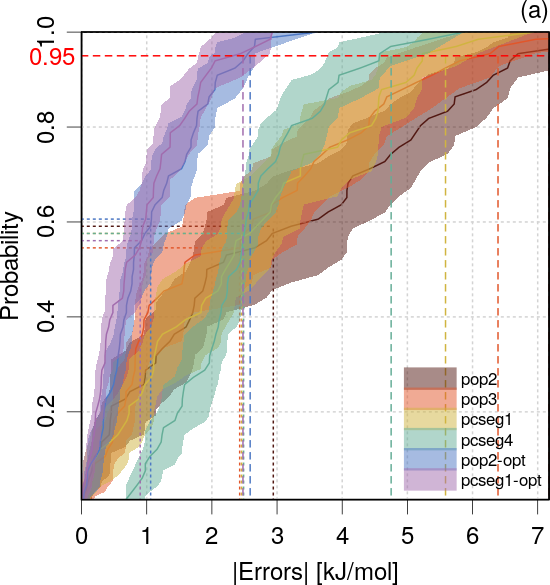}~\includegraphics[width=0.32\textwidth]{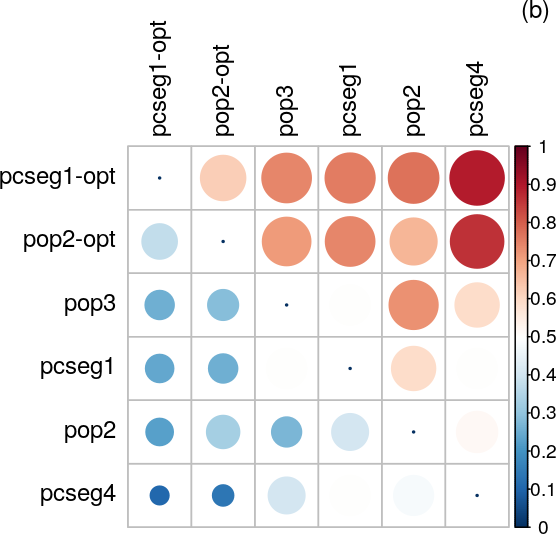}
\par\end{centering}
\noindent \begin{centering}
\includegraphics[width=0.32\textwidth]{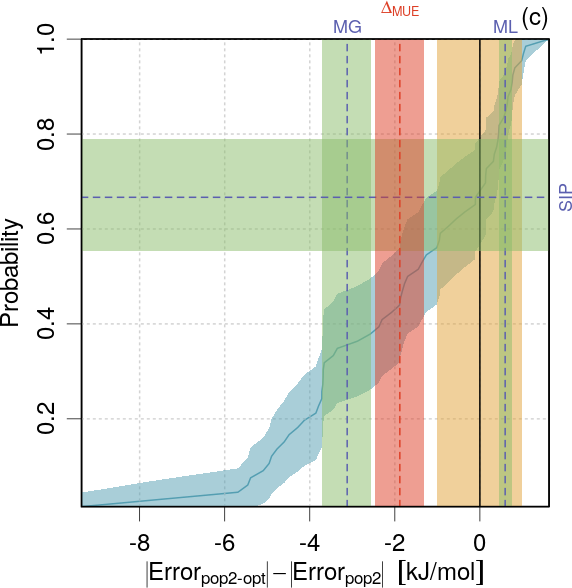}~\includegraphics[width=0.32\textwidth]{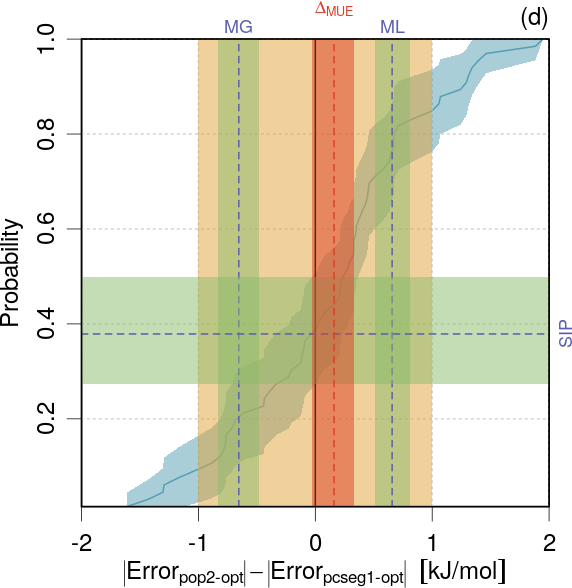}
\par\end{centering}
\noindent \centering{}\caption{\label{fig:jensen1}Case JEN2018: (a) ECDF of the absolute errors;
(b) SIP matrix; (c,d) ECDF of the difference of absolute errors of
pop2 and pcseg1-opt with respect to pop2-opt (see Fig.\,\ref{fig:pernot1-1}
for details). The orange bar represents a chemical accuracy of 1\,kJ/mol.}
\end{figure}

They both also stand out by their MSIP, with a slight advantage for
pcseg1-opt. Once again, the importance of error cancellations stands
out through the medium values of the SIP of pcseg1-opt over the other
cases. The strongest improvement is 0.9 over pcseg4, the smallest
0.6 over pop2-opt. The plots in Fig.\,\ref{fig:jensen1} illustrate
these features. The SIP matrix shows clearly that the optimized basis
sets provide a partial improvement, and a slight advantage of pcseg1-opt
over pop2-opt. The major gain when going to pop2 to pop2-opt is visible
in Fig.\,\ref{fig:jensen1}(c) where the medium SIP (\textasciitilde 0.7)
is compensated by the very small mean loss (0.6\,kJ/mol). In contrast,
Fig.\,\ref{fig:jensen1}(d) shows that the improvement of pcseg1-opt
over pop2-opt is marginal, with SIP values close to the neutral value
(0.5) and symmetrical MG and ML values. 

\paragraph{Ranking.}

\begin{figure}[!tb]
\noindent \begin{centering}
\includegraphics[width=0.32\columnwidth]{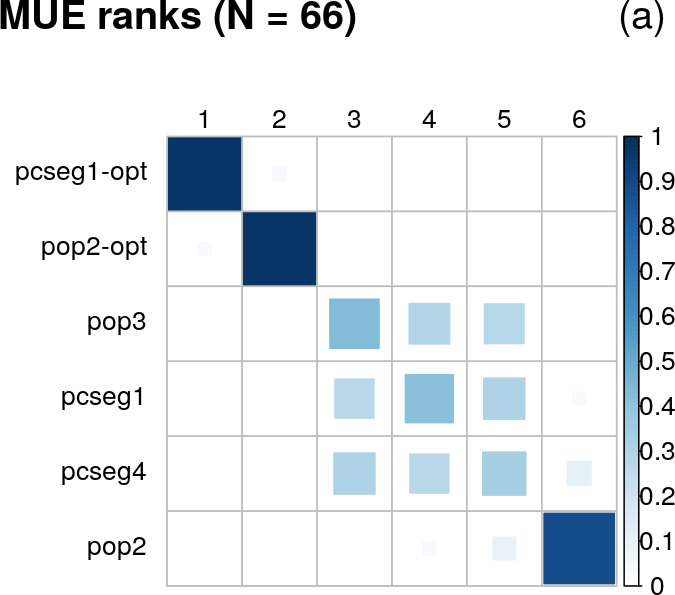}~\includegraphics[width=0.32\columnwidth]{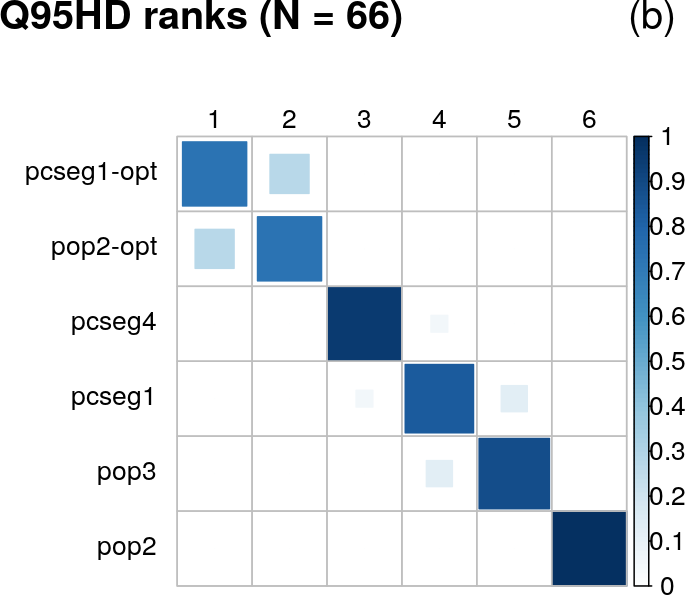}~\includegraphics[width=0.32\columnwidth]{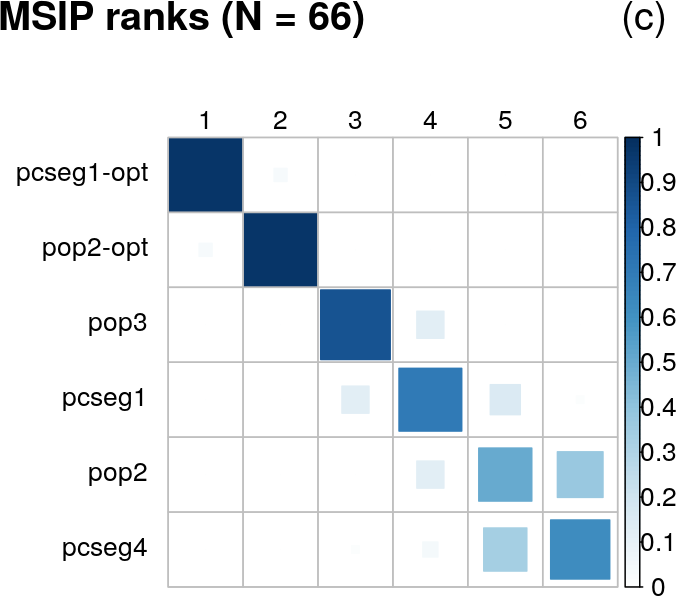}
\par\end{centering}
\noindent \centering{}\caption{\label{fig:jensenRanking}Case JEN2018: ranking probability matrices. }
\end{figure}

The leading position of the \textquotedbl -opt\textquotedbl{} methods
is solid and confirmed by our three scores (Fig.\,\ref{fig:jensenRanking}).

\subsection{DAS2019\label{subsec:Das2019}}

A set of 24 dielectric constants for 3D metal oxides has been reported
by Das \emph{et al.} \citep{Das2019} in their Table 3. One of the
experimental values being unknown, this limits the dataset to 23 values.
Experimental uncertainties are not specified. The predictions by six
DFAs are reported, three global hybrids (PBE0, B3LYP and DD-B3LYP)
and three range-separated hybrids (SC-BLYP, DD-SCBLYP and DD-CAM-B3LYP).
This is a small dataset, below the standards required for low type
I errors (false positive) in the comparison of MUE ($N>30$) and $Q_{95}$
($N>60$) (Paper\,I \citep{Pernot2020}\,-\,Appendix\,C).

\paragraph{Correlations.}

\begin{figure}[!tb]
\noindent \centering{}%
\begin{tabular}{ccc}
\includegraphics[width=0.32\textwidth]{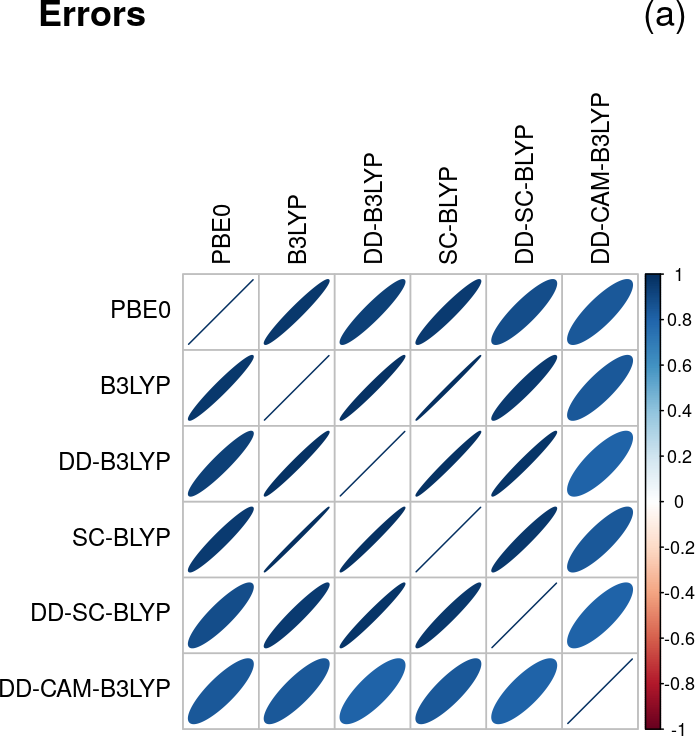} & ~\includegraphics[width=0.32\textwidth]{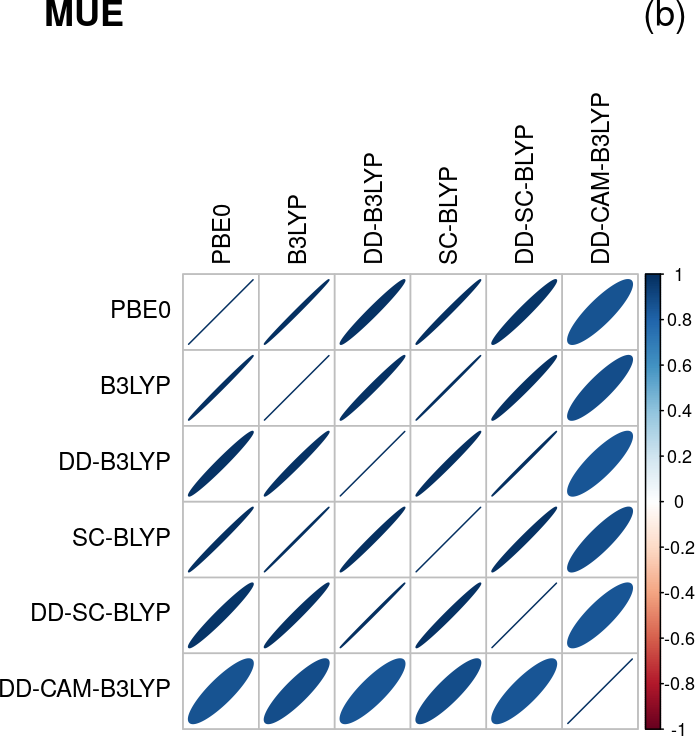} & ~\includegraphics[width=0.32\textwidth]{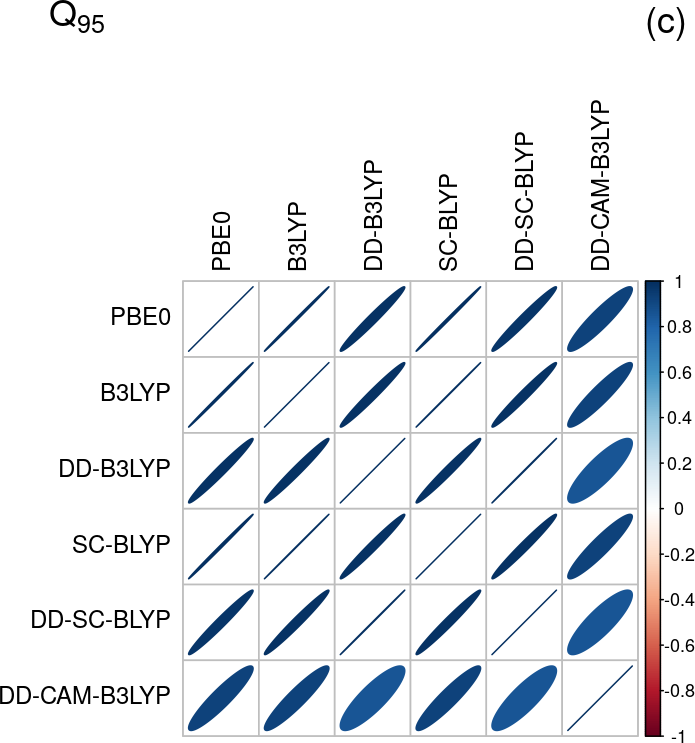}\tabularnewline
\includegraphics[width=0.32\textwidth]{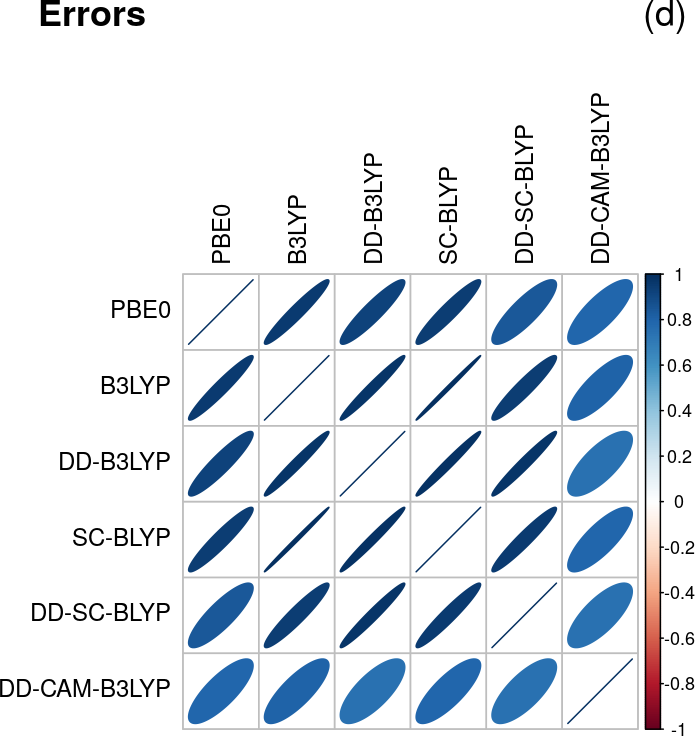} & ~\includegraphics[width=0.32\textwidth]{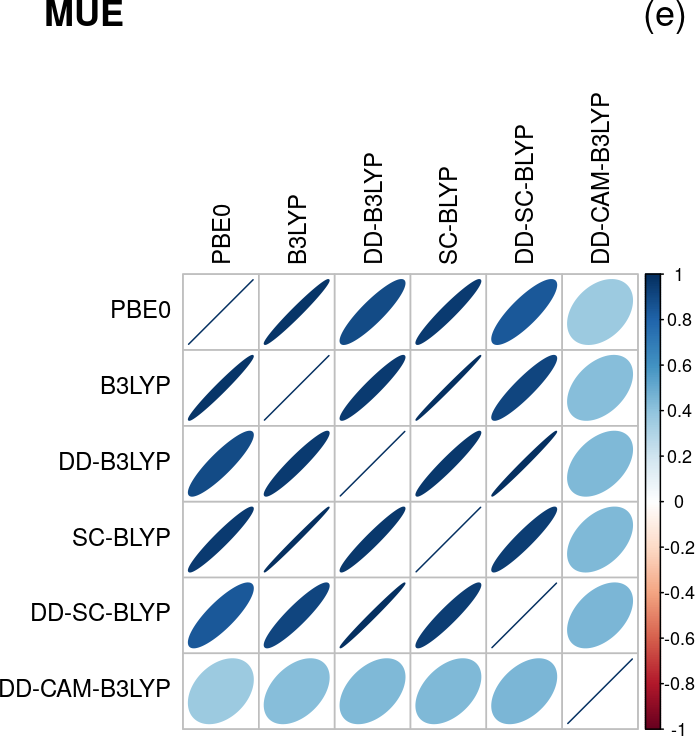} & ~\includegraphics[width=0.32\textwidth]{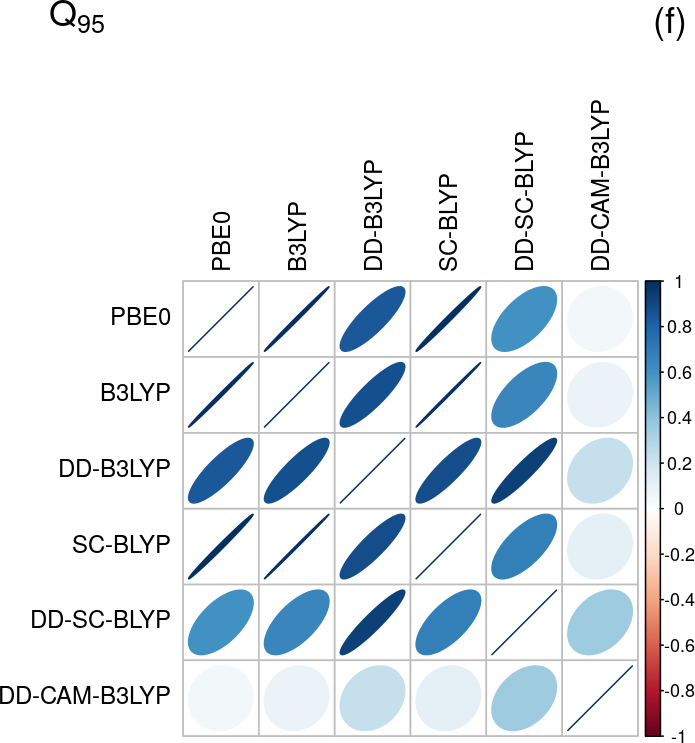}\tabularnewline
\end{tabular}\caption{\label{fig:dasCorrmat}Case DAS2019 - rank correlation matrices: (a-c)
original data set ($N=23$); (d-f) after removal of two outliers ($N=21$). }
\end{figure}

\begin{figure}[!tb]
\noindent \begin{centering}
\includegraphics[width=0.4\textwidth]{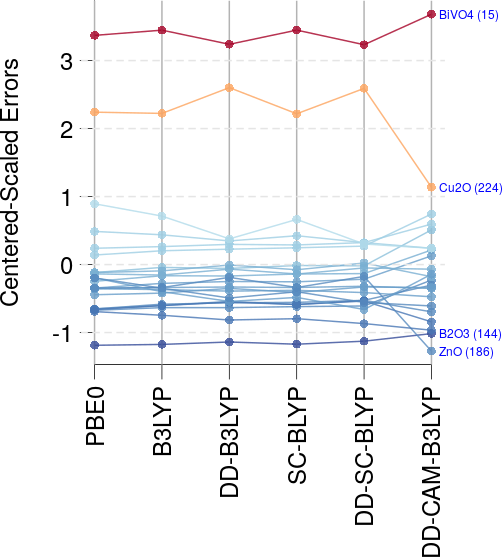}
\par\end{centering}
\noindent \centering{}\caption{\label{fig:dasCorrmat-1}Case DAS2019: parallel plot of scaled and
centered error sets, used to identify global outliers.}
\end{figure}

The correlation matrices of the errors, MUE and $Q_{95}$ have uniformly
strongly positive elements (Fig.\,\ref{fig:dasCorrmat}-top). 
 This is an unusual situation when compared to the previous cases.
Knowing that correlation coefficients are sensitive to outliers (even
if rank correlation is a little more robust), we explored the dataset
for outliers. A parallel plot (Fig.\,\ref{fig:dasCorrmat-1}) of
the scaled and centered error sets enables to identify systems which
deviate significantly from the core distribution for all methods (global
outliers). Two such systems exist for all methods: \ce{BiVO4} and
\ce{Cu2O}. After removal of these two points, the correlation matrix
for the errors is slightly relaxed (the smallest correlation coefficient
decreases from 0.81 to 0.74), but those for MUE and $Q_{95}$ are
visibly more affected ((Fig.\,\ref{fig:dasCorrmat}-bottom)). In
fact, the parallel plot reflects the strong correlations between all
errors sets (many quasi-parallel horizontal lines), except for DD-CAM-B3LYP.
The pruned dataset ($N=21$) is used in the following. 

\paragraph{Statistics.}

Considering the small size of the sample, few clear-cut conclusions
are possible. Only DD-CAM-B3LYP stands out significantly, either by
its MUE, Q95 or MSIP values (Table\,\ref{tab:das}). At the opposite,
although its MUE and $Q_{95}$ values are not distinguishable from
those of PBE0, B3LYP, SC-BLYP and DD-SC-BLYP, DD-B3LYP is the worst
performer of the group based on the SIP statistics. 
\begin{table}[!tb]
\begin{centering}
{\small{}}%
\begin{tabular}{llcr@{\extracolsep{0pt}.}lr@{\extracolsep{0pt}.}lr@{\extracolsep{0pt}.}lr@{\extracolsep{0pt}.}lr@{\extracolsep{0pt}.}lr@{\extracolsep{0pt}.}lr@{\extracolsep{0pt}.}lr@{\extracolsep{0pt}.}lr@{\extracolsep{0pt}.}l}
\hline 
Methods  &  & MUE  & \multicolumn{2}{c}{$P_{inv}$} & \multicolumn{2}{c}{} & \multicolumn{2}{c}{$Q_{95}$} & \multicolumn{2}{c}{$P_{inv}$} & \multicolumn{2}{c}{} & \multicolumn{2}{c}{MSIP} & \multicolumn{2}{c}{SIP } & \multicolumn{2}{c}{MG} & \multicolumn{2}{c}{ML}\tabularnewline
 &  & a.u. & \multicolumn{2}{c}{} & \multicolumn{2}{c}{} & a&u. & \multicolumn{2}{c}{} & \multicolumn{2}{c}{} & \multicolumn{2}{c}{} & \multicolumn{2}{c}{} & a&u. & a&u.\tabularnewline
\cline{1-1} \cline{3-21} 
PBE0  &  & 0.66(9)  & 0&00  & \multicolumn{2}{c}{} & 1&6(2)  & 0&00  & \multicolumn{2}{c}{} & 0&47(9)  & 0&76(9)  & -0&44(8)  & 0&19(4) \tabularnewline
B3LYP  &  & 0.61(8)  & 0&00  & \multicolumn{2}{c}{} & 1&4(2)  & 0&00  & \multicolumn{2}{c}{} & 0&49(8)  & 0&76(10)  & -0&38(6)  & 0&21(6) \tabularnewline
DD-B3LYP  &  & 0.70(7)  & 0&00  & \multicolumn{2}{c}{} & 1&30(7)  & 0&00  & \multicolumn{2}{c}{} & 0&19(8)  & 0&90(6)  & -0&41(6)  & 0&4(1) \tabularnewline
SC-BLYP  &  & 0.58(8)  & 0&00  & \multicolumn{2}{c}{} & 1&3(1)  & 0&00  & \multicolumn{2}{c}{} & 0&62(8)  & 0&76(9)  & -0&36(6)  & 0&22(7) \tabularnewline
DD-SC-BLYP  &  & 0.68(7)  & 0&00  & \multicolumn{2}{c}{} & 1&23(5)  & 0&00  & \multicolumn{2}{c}{} & 0&29(8)  & 0&90(6)  & -0&39(6)  & 0&4(1) \tabularnewline
DD-CAM-B3LYP  &  & \textbf{0.36(6)}  & \multicolumn{2}{c}{-} & \multicolumn{2}{c}{} & \textbf{0}&\textbf{83(7)}  & \multicolumn{2}{c}{-} & \multicolumn{2}{c}{} & \textbf{0}&\textbf{82(8)}  & \multicolumn{2}{c}{-} & \multicolumn{2}{c}{-} & \multicolumn{2}{c}{-}\tabularnewline
\hline 
\end{tabular}{\small\par}
\par\end{centering}
\noindent \centering{}\caption{\label{tab:das}Case DAS2019 - absolute error statistics for the pruned
dataset ($N=21$): inversion probabilities and SIP statistics for
comparison with the DFA of smallest MUE (DD-CAM-B3LYP). The best scores
are in boldface.}
\end{table}

\paragraph{SIP analysis.}

\begin{figure}[!tb]
\noindent \centering{}\includegraphics[width=0.32\textwidth]{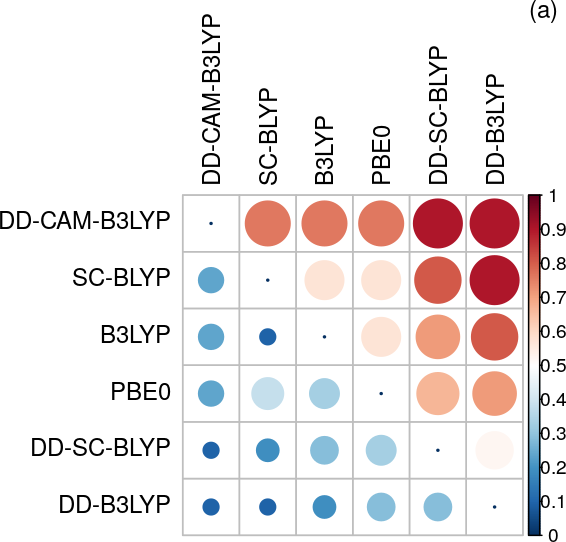}~\includegraphics[width=0.32\textwidth]{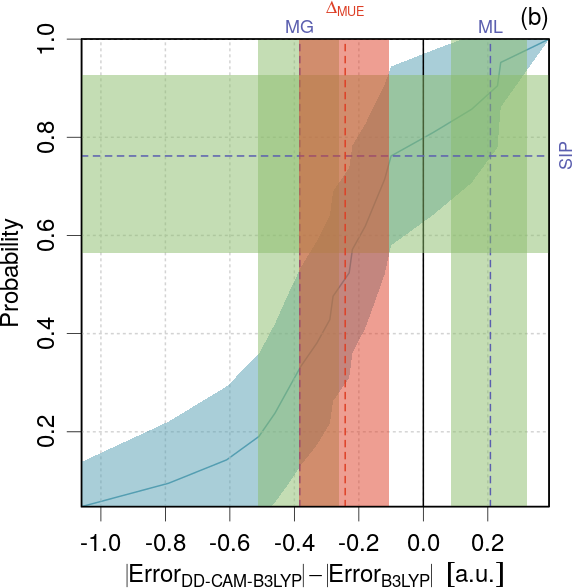}~\includegraphics[width=0.32\textwidth]{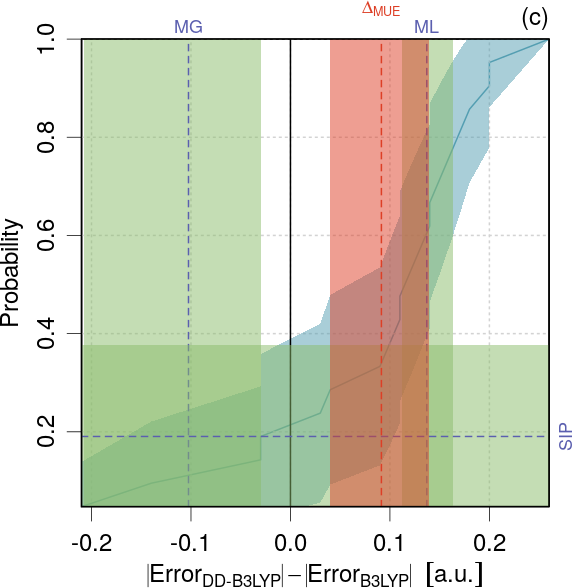}\caption{\label{fig:das}Case DAS2019: (a) SIP matrix; (b) ECDF of the difference
of absolute errors of methods DD-CAMB3LYP and B3LYP; (c) idem for
DD-B3LYP and B3LYP (see Fig.\,\ref{fig:pernot1-1} for details).}
\end{figure}

The best and worse methods are clearly identifiable in the SIP matrix
(Fig.\,\ref{fig:das}(a)), with a full reddish line for DD-CAMB3LYP,
and a full blueish line for DD-B3LYP. The impact of the small set
size on this conclusion is illustrated in Fig.\,\ref{fig:das}(b,c),
where the ECDFs of the differences of absolute errors are plotted
for DD-CAM-B3LYP \emph{vs}. B3LYP and DD-B3LYP \emph{vs}. B3LYP. Despite
being very large, the error bars on the statistics enable to validate
these conclusions. 

\paragraph{Ranking.}

\begin{figure}[!tb]
\noindent \begin{centering}
\includegraphics[width=0.32\columnwidth]{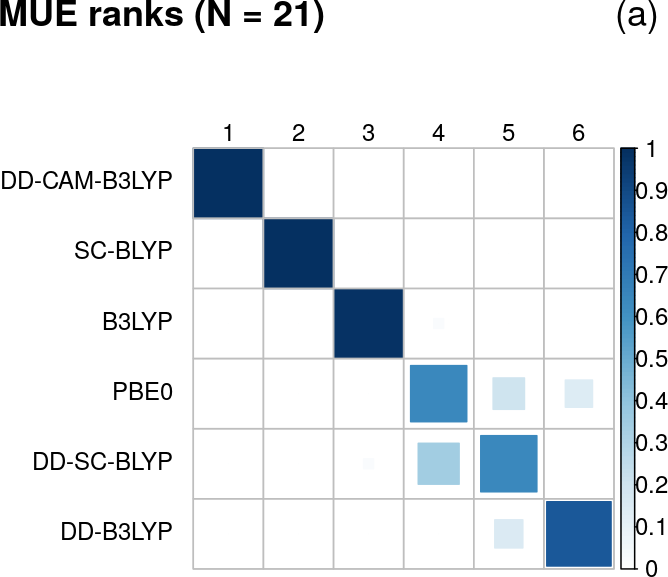}~\includegraphics[width=0.32\columnwidth]{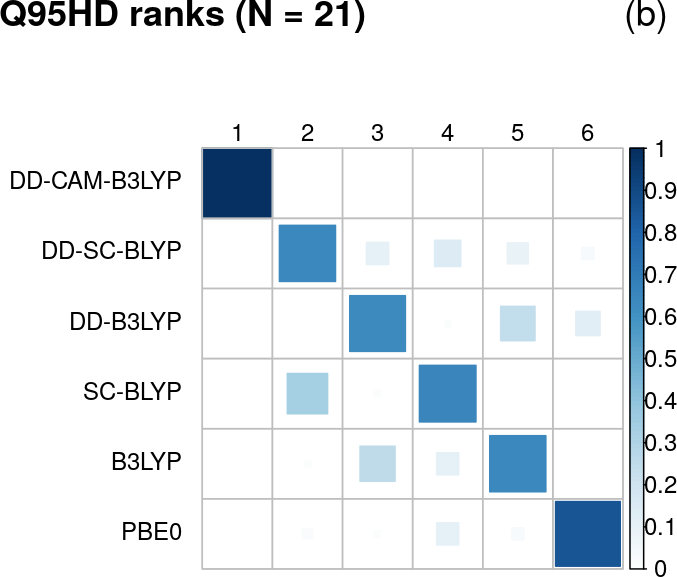}~\includegraphics[width=0.32\columnwidth]{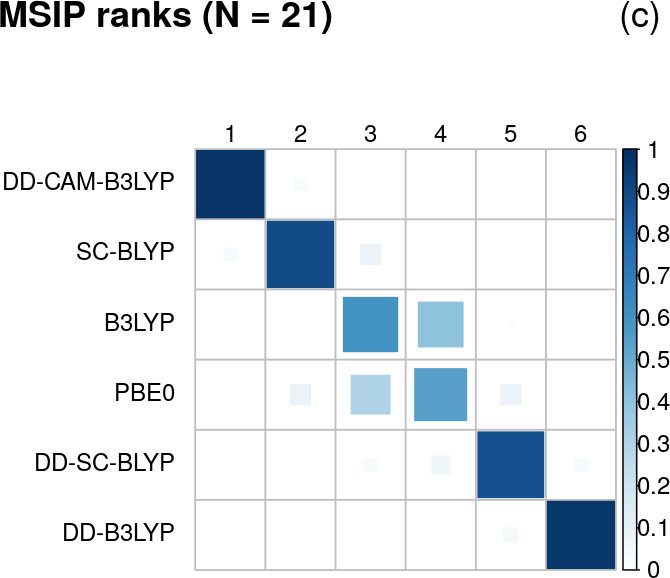}
\par\end{centering}
\noindent \begin{centering}
\includegraphics[width=0.32\columnwidth]{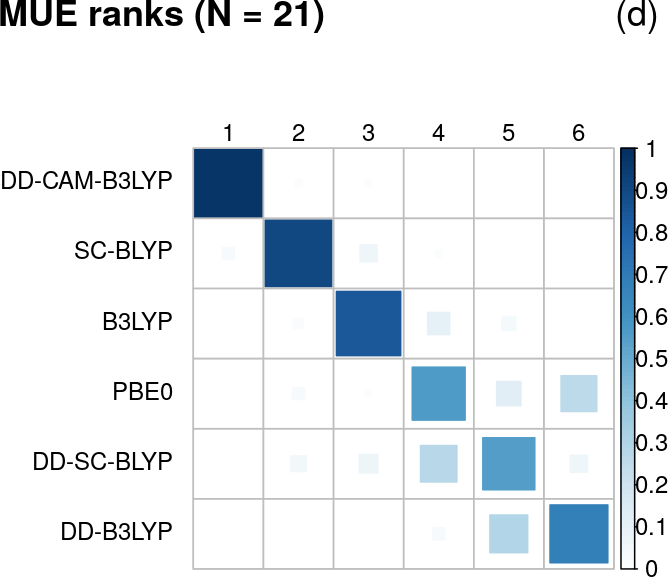}~\includegraphics[width=0.32\columnwidth]{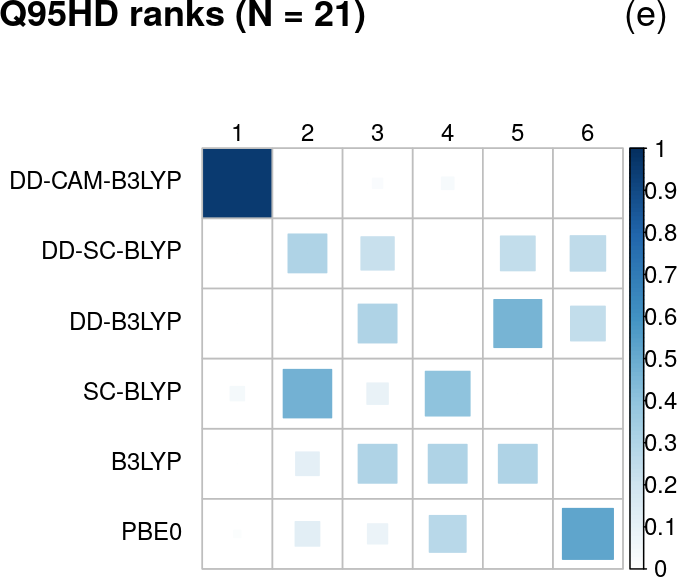}~\includegraphics[width=0.32\columnwidth]{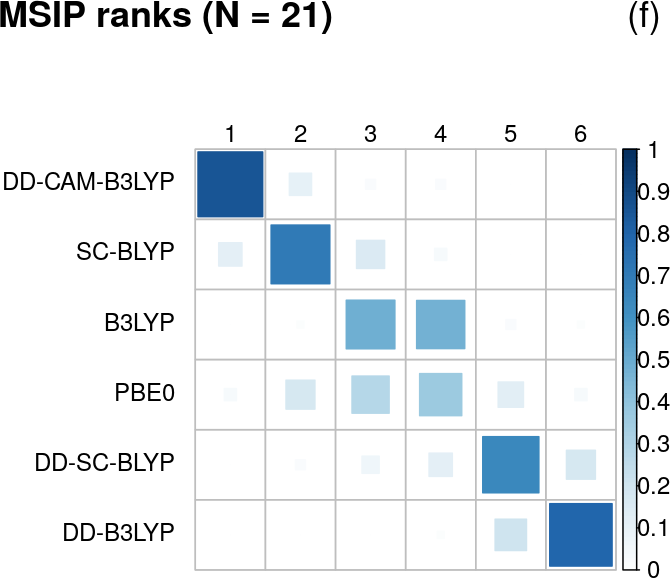}
\par\end{centering}
\noindent \centering{}\caption{\label{fig:dasRanking-1}Case DAS2019 - ranking probability matrices:
(a-c) $N$-out\,of-$N$ bootstrap; (d-f) $N/3$-out\,of-$N$ bootstrap.}
\end{figure}

All ranking matrices confirm a solid leading place for DD-CAM-B3LYP
(Fig.\,\ref{fig:dasRanking-1}). The MUE and MSIP rankings would
then favor SC-BLYP and B3LYP, in disagreement with the $Q_{95}$ ranking,
for which the three DD-X methods have leading ranks. An example of
a $N'$-out of-$N$ bootstrap ($N'=N/3$) is shown on the bottom row.
The uncertainty is slightly enhanced, notably for the $Q_{95}$ ranks
above the first, but the main features are mostly preserved. 

\subsection{THA2015 \& WU2015\label{subsec:Thakkar2015}}

Thakkar \emph{et al. }\citep{Thakkar2015} compiled a database of
polarizabilities for 135 molecules, from triatomics to 26-atoms systems.
The experimental data are given with their uncertainty, and computational
results are provided for 7 methods. Dataset THA2015 for our study
was extracted from Tables II-IV of the reference article. The raw
errors present a dispersion increasing with the polarizability, hence
relative errors are used in the reference article and this study. 

The relative uncertainties for the reference experimental data cover
a large range, from 0.09\,\% to 12.4\,\%, the median value is 1.7\,\%.
The authors identified 8 outliers, and a total of 32 systems in need
of further experimental study. The outliers do not contain the points
with the extreme uncertainties, so that, even after removal of the
32 problematic systems, the range of relative uncertainties stays
the same. The dispersion of uncertainties would certainly justify
the use of weighted statistics. This was not the choice of Thakkar
\emph{et al.}, and we proceed with unweighted statistics, keeping
in mind that the results might be influenced by reference data errors
instead of model errors.

In a complementary study, Wu \emph{et al.} \citep{Wu2015b} calculated
polarizabilities for a set of 145 molecules with HF, MP2, CCSD(T)
and 34 DFAs. In this study, CCSD(T) was used as reference to evaluate
the other methods. In the following, we select the subset of 7 methods
common to both datasets (WU2015). This enables us to study the impact
of the reference data (experimental \emph{vs.} calculated) on the
correlation and ranking matrices. 

\paragraph{Correlations.}

\begin{figure}[!tb]
\noindent \begin{centering}
\begin{tabular}{ccc}
\includegraphics[width=0.32\columnwidth]{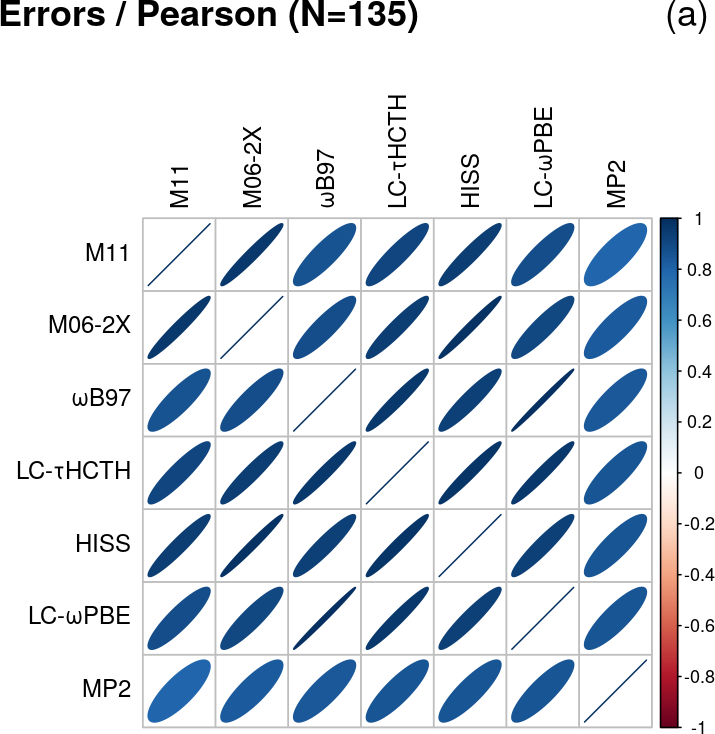} & ~\includegraphics[width=0.32\columnwidth]{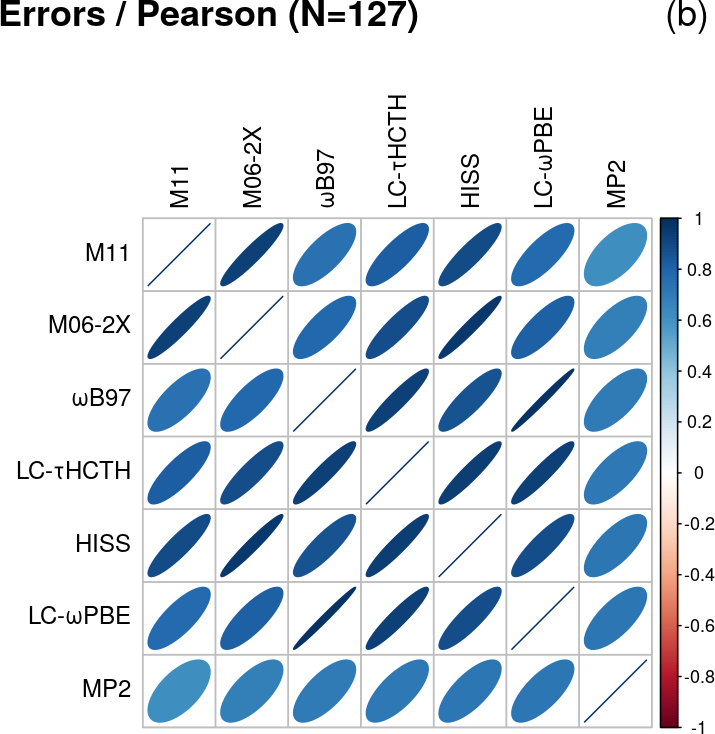} & ~\includegraphics[width=0.32\columnwidth]{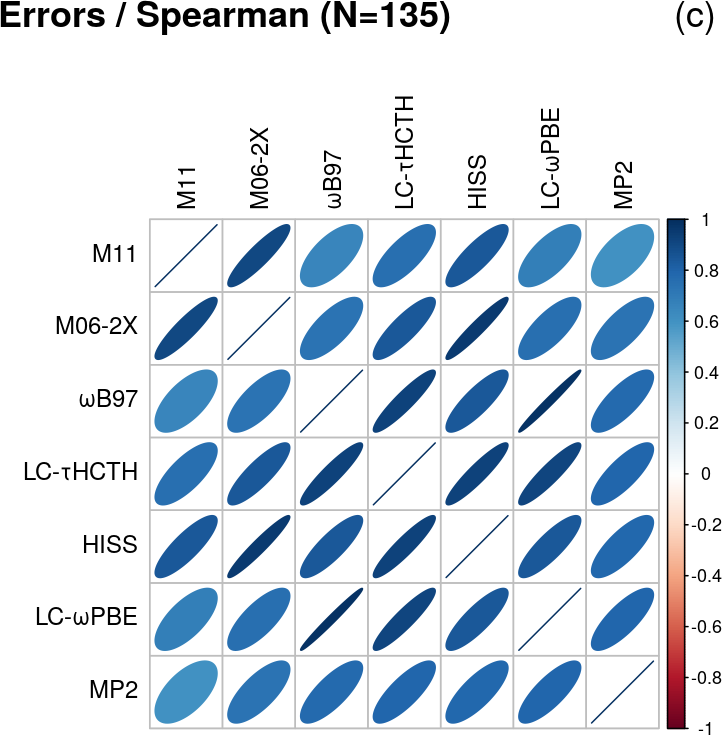}\tabularnewline
\includegraphics[width=0.32\columnwidth]{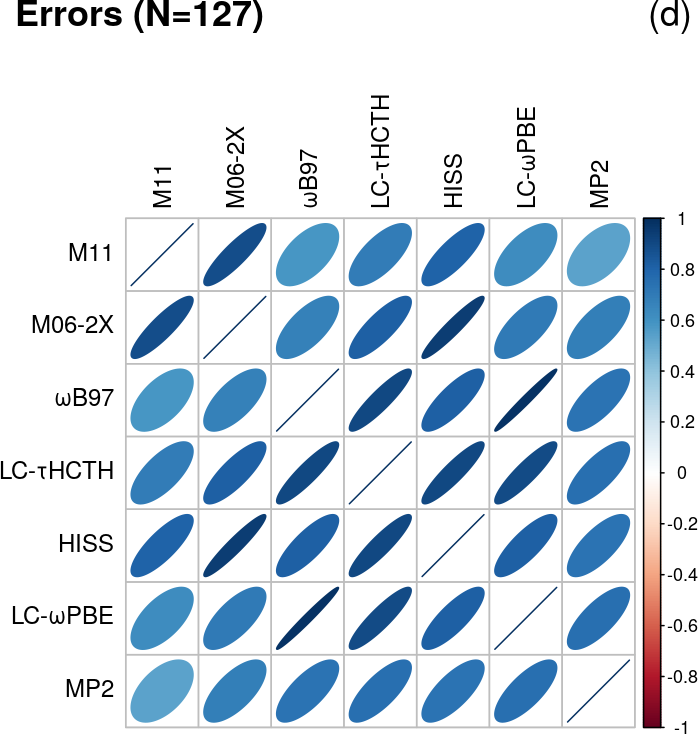} & ~\includegraphics[width=0.32\columnwidth]{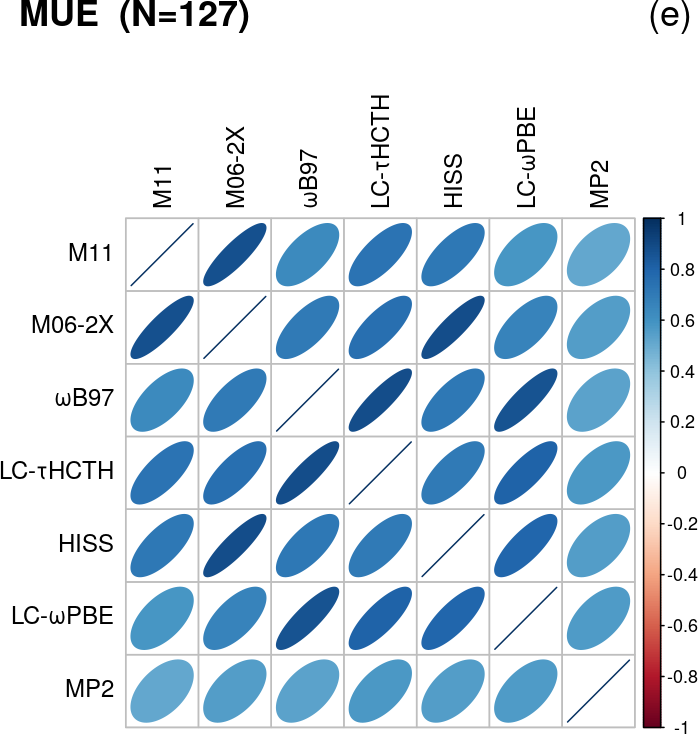} & ~\includegraphics[width=0.32\columnwidth]{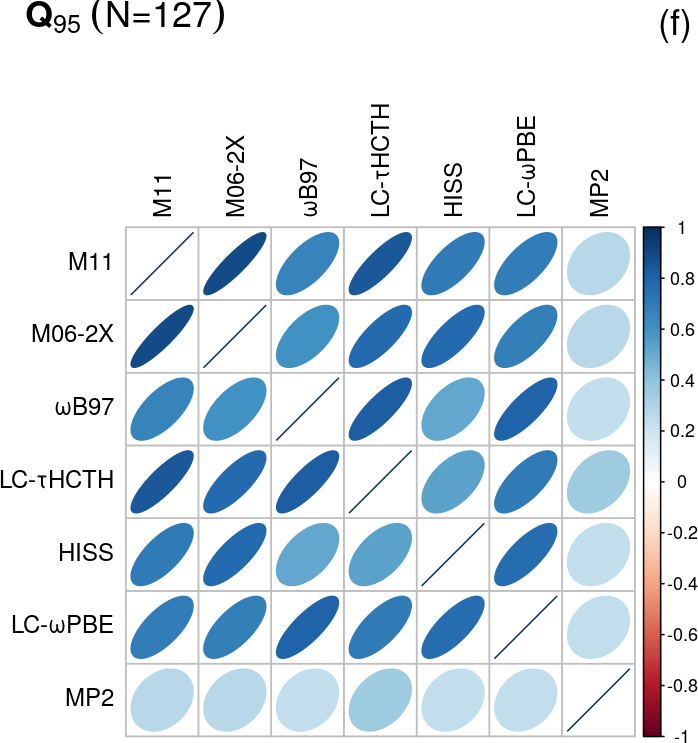}\tabularnewline
\end{tabular}
\par\end{centering}
\caption{\label{fig:thakkar2}Case THA2015 - correlation matrix: (a) Pearson
correlation of the full data set ($N=135$); (b) Pearson correlation
of the pruned dataset ($N=127$); (c) Spearman/rank correlation of
the full data set; (d): Errors rank correlation; (e): MUE rank correlation;
(f) $Q_{95}$ rank correlation.}
\end{figure}

The Pearson correlation matrix of the error sets (Fig.\,\ref{fig:thakkar2}(a))
is uniformly strongly positive. The smallest CC value is 0.8. To appreciate
the role of data points with large deviations (outliers) in these
strong correlations, we removed a set of 8 outliers identified by
Thakkar \emph{et al.}\citep{Thakkar2015} ((Fig.\,\ref{fig:thakkar2}(b)).
Most of the correlations weaken notably. For comparison, the rank
correlation matrix was calculated for the full dataset ((Fig.\,\ref{fig:thakkar2}(c)).
This matrix is very similar to the one with outliers removed, illustrating
the better resilience of rank correlations to outliers. Finally, the
errors, MUE and $Q_{95}$ rank correlation matrices were estimated
on the pruned ($N=127$) dataset (Fig.\,\ref{fig:thakkar2}(d-f)).
Globally, the structure of the errors correlation matrix seems to
be transferred to the statistics, with attenuated correlation intensities.

\begin{figure}[!tb]
\noindent \begin{centering}
\includegraphics[width=0.32\textwidth]{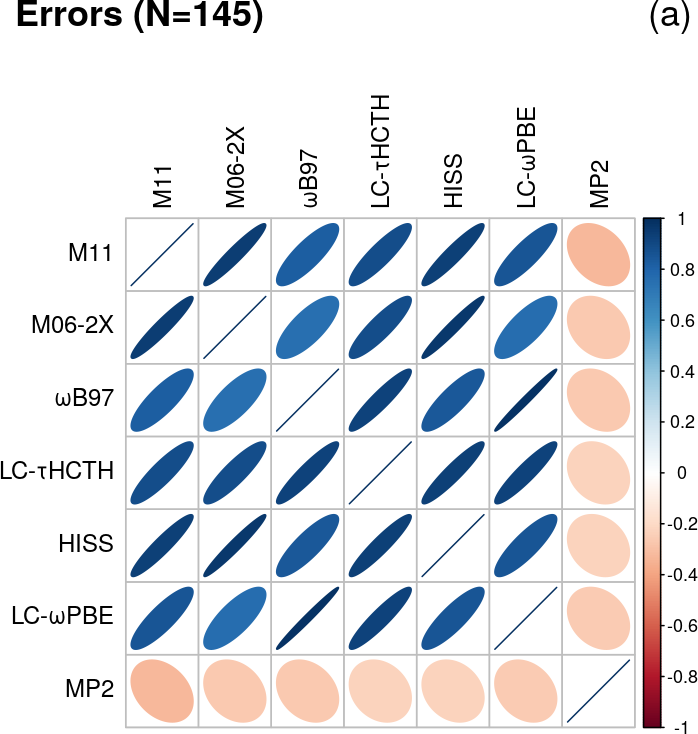}~\includegraphics[width=0.32\textwidth]{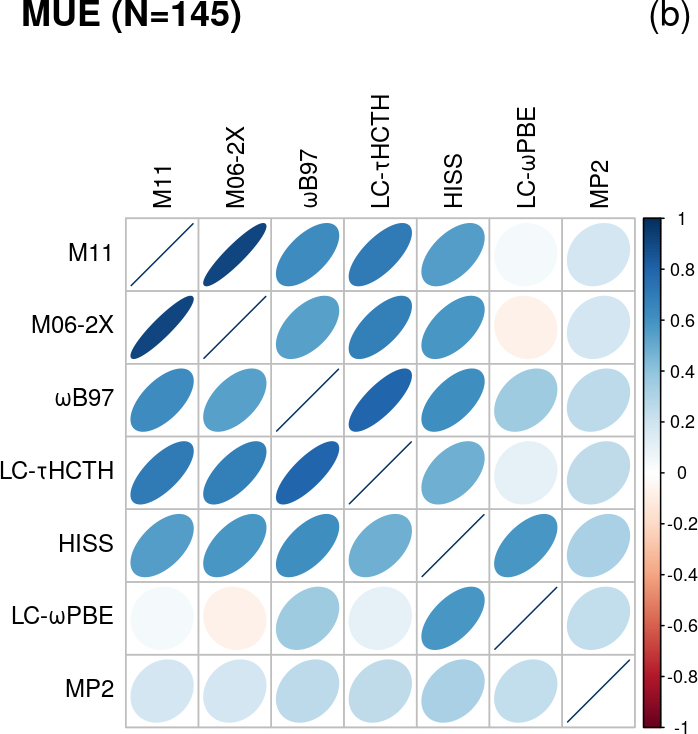}~\includegraphics[width=0.32\textwidth]{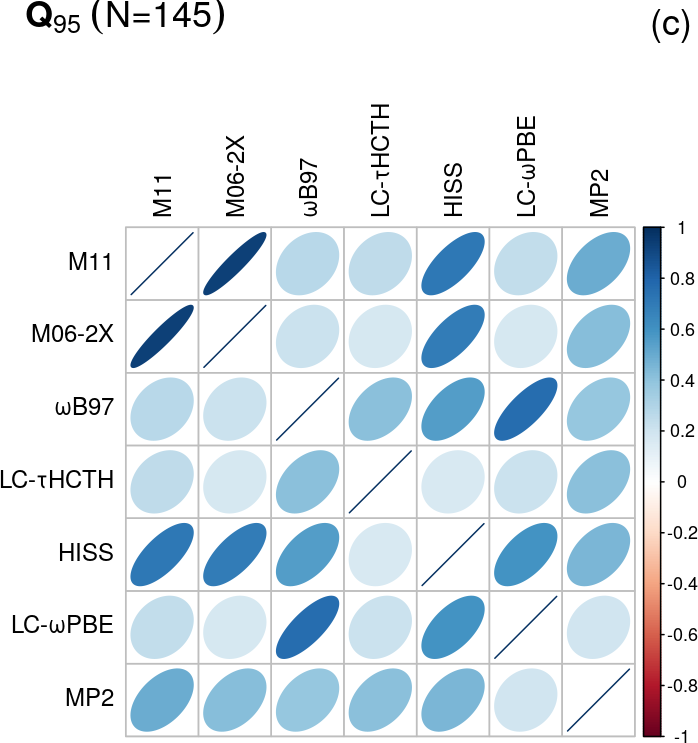}
\par\end{centering}
\caption{\label{fig:thakkar2-1}Case WU2015 - rank correlation matrix: (a)
Errors; (b) MUE; (c) $Q_{95}$.}
\end{figure}

The error, MUE and $Q_{95}$ rank correlation matrices were also calculated
for the WU2015 dataset (Fig.\,\ref{fig:thakkar2-1}). In the absence
of reference data uncertainties, MP2 errors are now weakly anticorrelated
to the other error sets, while all DFAs remain positively correlated.

The differences between both sets of correlation matrices, notably
when MP2 is concerned, might be due in a large part to the presence
of large experimental errors in the THA2015 dataset.

\paragraph{Statistics. }

\begin{figure}[!tb]
\noindent \begin{centering}
\includegraphics[clip,width=0.32\columnwidth]{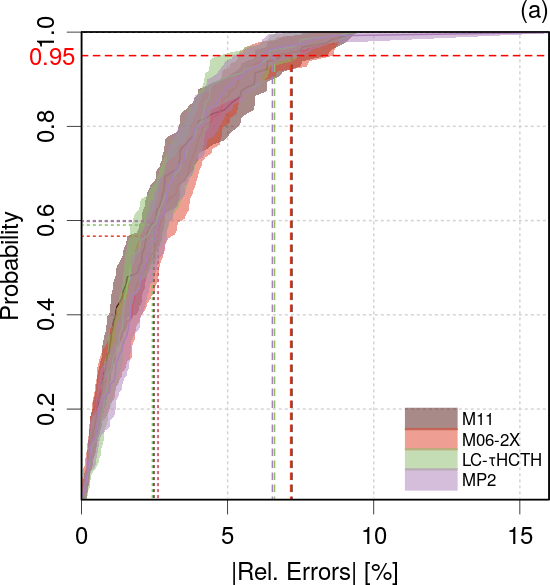}~\includegraphics[clip,width=0.32\columnwidth]{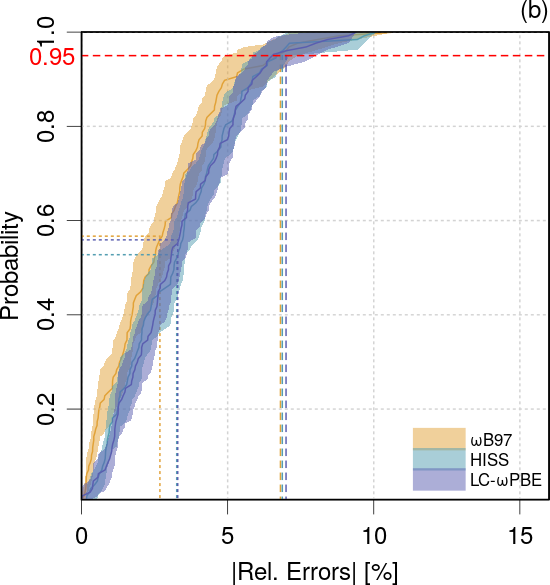}
\par\end{centering}
\caption{\label{fig:thakkar1}Case THA2015 - ECDFs of absolute relative errors:
(a) methods with smallest, indiscernible, MUE values; (b) other methods. }
\end{figure}

The values of MUE and $Q_{95}$ for the full THA2015 dataset are reported
in Table\,\ref{tab:thakkar}. The MUE values agree with those of
the reference article, but the uncertainty bears on the second digit,
showing that a third digit is essentially irrelevant. The analysis
of $P_{inv}$ for the MUE leads us to conclude that there is a group
of four methods (M11, M06-2X, LC-$\tau$HCTH and MP2) with similar
performances, which is confirmed by the comparison of their empirical
cumulated distribution functions \citep{Pernot2018} (Fig.\,\ref{fig:thakkar1}).
These ECDFs overlap over the whole error range. Besides, these methods
cannot be discriminated on the basis of their $Q_{95}$ values, as
it appears that all values are indiscernible. These conclusions are
unchanged when one removes the 8 outliers identified by Thakkar \emph{et
al. }(not shown). 
\begin{table}[!tb]
\begin{centering}
\begin{tabular}{llr@{\extracolsep{0pt}.}lr@{\extracolsep{0pt}.}lr@{\extracolsep{0pt}.}lr@{\extracolsep{0pt}.}lr@{\extracolsep{0pt}.}lr@{\extracolsep{0pt}.}lr@{\extracolsep{0pt}.}lr@{\extracolsep{0pt}.}lr@{\extracolsep{0pt}.}lr@{\extracolsep{0pt}.}l}
\hline 
Methods  &  & \multicolumn{2}{c}{MUE } & \multicolumn{2}{c}{$P_{inv}$} & \multicolumn{2}{c}{} & \multicolumn{2}{c}{$Q_{95}$} & \multicolumn{2}{c}{$P_{inv}$} & \multicolumn{2}{c}{} & \multicolumn{2}{c}{MSIP } & \multicolumn{2}{c}{SIP } & \multicolumn{2}{c}{MG } & \multicolumn{2}{c}{ML }\tabularnewline
 &  & \multicolumn{2}{c}{\%} & \multicolumn{2}{c}{} & \multicolumn{2}{c}{} & \multicolumn{2}{c}{\%} & \multicolumn{2}{c}{} & \multicolumn{2}{c}{} & \multicolumn{2}{c}{} & \multicolumn{2}{c}{} & \multicolumn{2}{c}{\%} & \multicolumn{2}{c}{\%}\tabularnewline
\cline{1-1} \cline{3-6} \cline{9-12} \cline{15-22} 
M11  &  & \textbf{3}&\textbf{1(3)}  & 0&34  & \multicolumn{2}{c}{} & \multicolumn{2}{c}{\textbf{10(1) }} & \multicolumn{2}{c}{-} & \multicolumn{2}{c}{} & 0&58(4)  & 0&47(4)  & -1&4(1)  & 1&16(10) \tabularnewline
M06-2X  &  & \textbf{3}&\textbf{2(3)}  & 0&09  & \multicolumn{2}{c}{} & \multicolumn{2}{c}{\textbf{10(2) }} & 0&50  & \multicolumn{2}{c}{} & 0&57(4)  & 0&53(4)  & -1&2(1)  & 1&0(1) \tabularnewline
$\omega$B97  &  & 3&3(3)  & 0&00  & \multicolumn{2}{c}{} & \multicolumn{2}{c}{\textbf{11(2) }} & 0&21  & \multicolumn{2}{c}{} & 0&53(4)  & 0&59(4)  & -0&94(7)  & 0&72(7) \tabularnewline
LC-$\tau$HCTH  &  & \textbf{3}&\textbf{0(3)}  & \multicolumn{2}{c}{-} & \multicolumn{2}{c}{} & \multicolumn{2}{c}{\textbf{10(2) }} & 0&30  & \multicolumn{2}{c}{} & \textbf{0}&\textbf{59(4) } & \multicolumn{2}{c}{-} & \multicolumn{2}{c}{-} & \multicolumn{2}{c}{-}\tabularnewline
HISS  &  & 3&8(3)  & 0&00  & \multicolumn{2}{c}{} & \multicolumn{2}{c}{\textbf{10(2) }} & 0&38  & \multicolumn{2}{c}{} & 0&34(4)  & 0&72(4)  & -1&62(10)  & 1&5(1) \tabularnewline
LC-$\omega$PBE  &  & 3&9(3)  & 0&00  & \multicolumn{2}{c}{} & \multicolumn{2}{c}{\textbf{11(1) }} & 0&25  & \multicolumn{2}{c}{} & 0&31(4)  & 0&78(3)  & -1&39(8)  & 1&2(1) \tabularnewline
MP2  &  & \textbf{3}&\textbf{2(3) } & 0&22  & \multicolumn{2}{c}{} & \multicolumn{2}{c}{\textbf{11(2) }} & 0&34  & \multicolumn{2}{c}{} & 0&56(4)  & 0&45(4)  & -1&3(3)  & 0&8(1) \tabularnewline
\hline 
\end{tabular}
\par\end{centering}
\caption{\label{tab:thakkar}Case THA2015 - absolute error statistics for the
full dataset ($N=145)$: inversion probabilities and SIP statistics
for comparison with the DFA of smallest MUE (LC-$\tau$HCTH), except
for $Q_{95}$ inversion probability, where the reference is the DFA
with smallest $Q_{95}$. The best scores and the values for which
$(p_{g}=2P_{inv})>0.05$ are in boldface.}
\end{table}
 
\paragraph{SIP analysis.}

\begin{figure}[!tb]
\noindent \centering{}\includegraphics[width=0.32\textwidth]{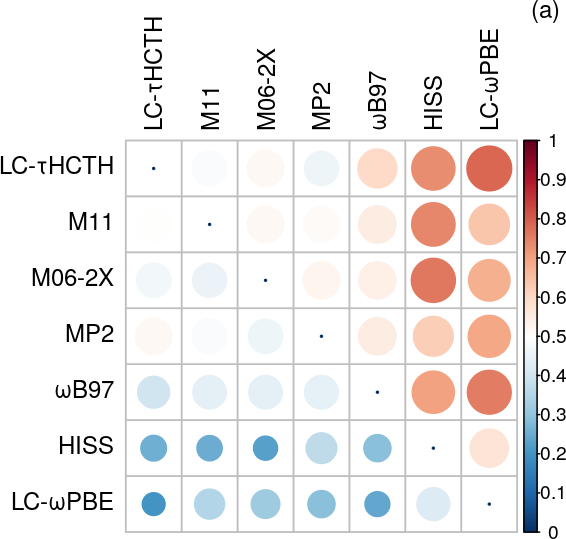}~\includegraphics[width=0.32\textwidth]{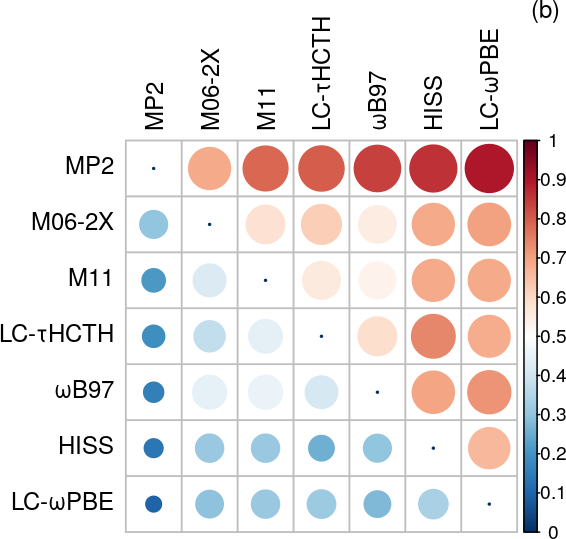}\caption{\label{fig:thakkarSIP}SIP matrix: (a) case THA2015 ($N=127$); (b)
case WU2015. The methods are sorted by decreasing MSIP value.}
\end{figure}

The SIP matrix (Fig.\,\ref{fig:thakkarSIP}(a)) for the THA2015 dataset
reveals a leading group of four methods identical to those identified
above. When passing to WU2015 (Fig.\,\ref{fig:thakkarSIP}(b)), there
is a better discrimination between methods, and MP2 presents SIP values
over all the other methods. 

\paragraph{Ranking.}

\begin{figure}[!tb]
\noindent \begin{centering}
\begin{tabular}{ccc}
\includegraphics[width=0.32\columnwidth]{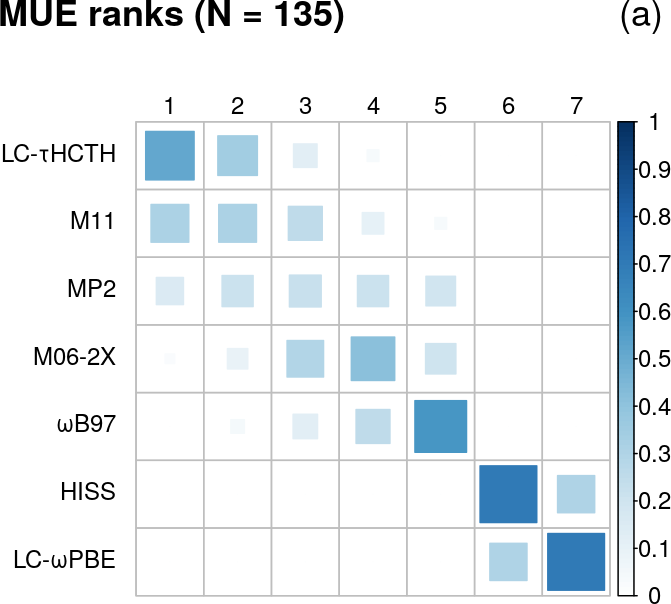} & ~\includegraphics[width=0.32\columnwidth]{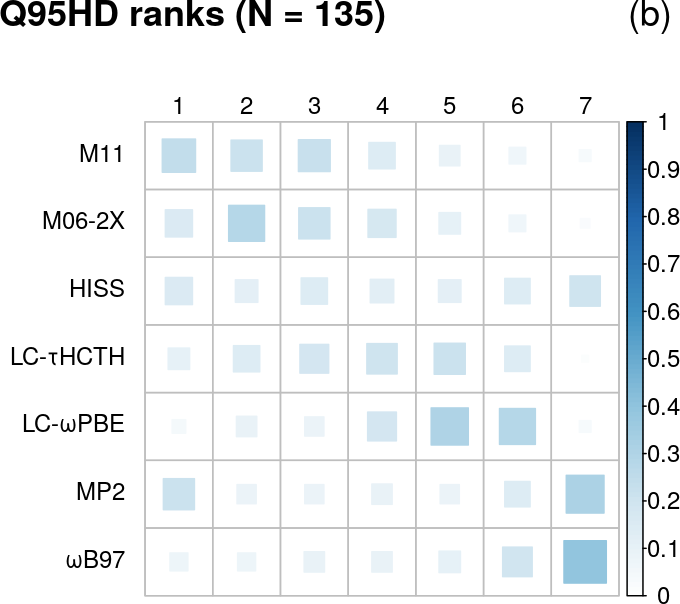} & ~\includegraphics[width=0.32\columnwidth]{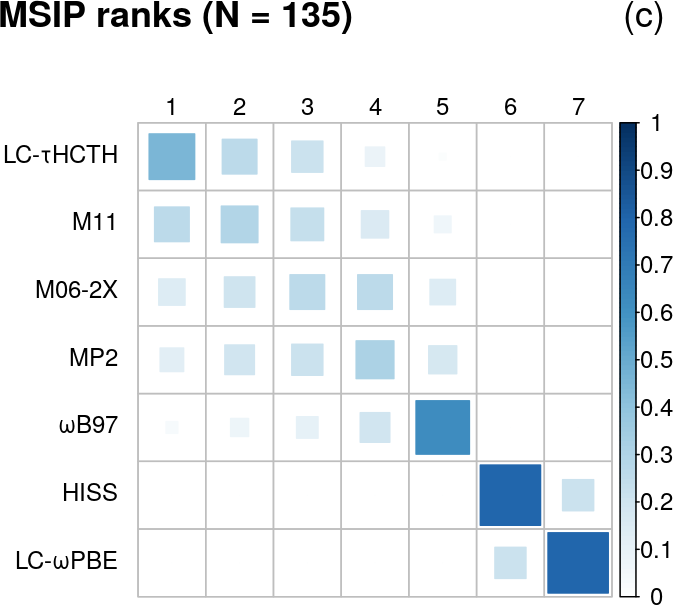}\tabularnewline
\includegraphics[width=0.32\columnwidth]{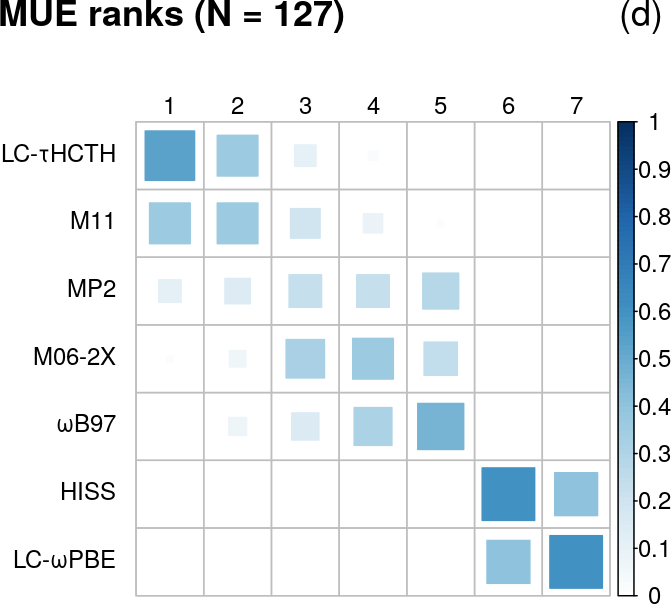} & ~\includegraphics[width=0.32\columnwidth]{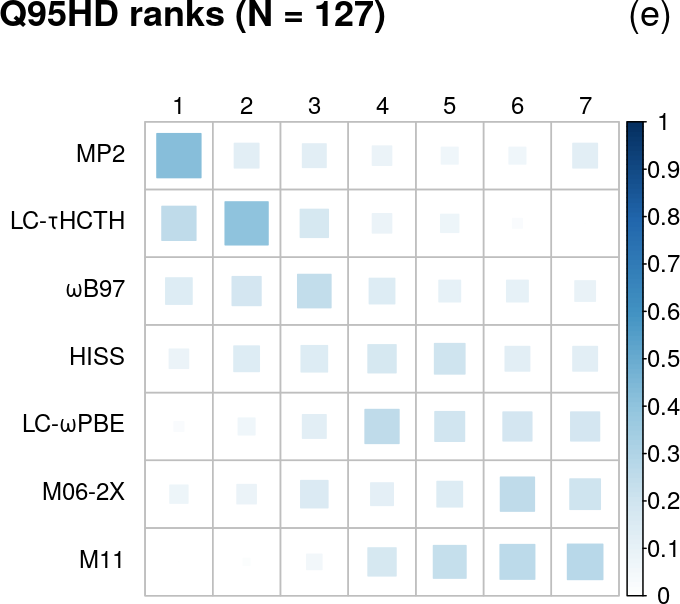} & ~\includegraphics[width=0.32\columnwidth]{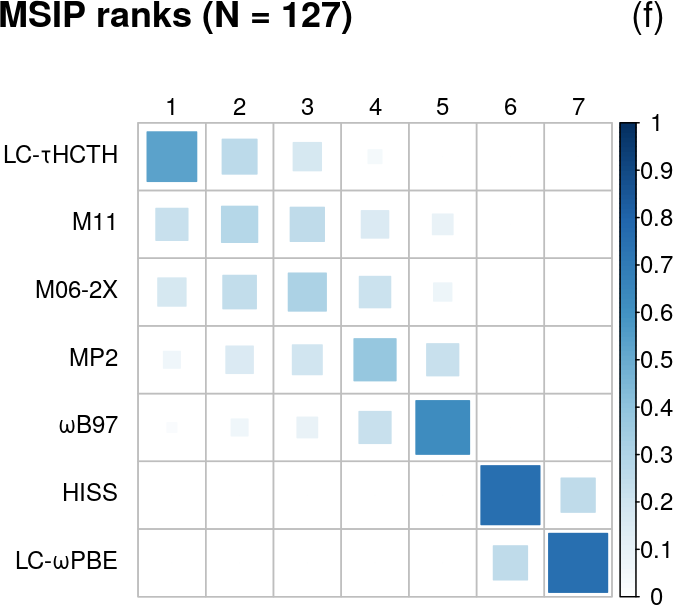}\tabularnewline
\includegraphics[width=0.32\columnwidth]{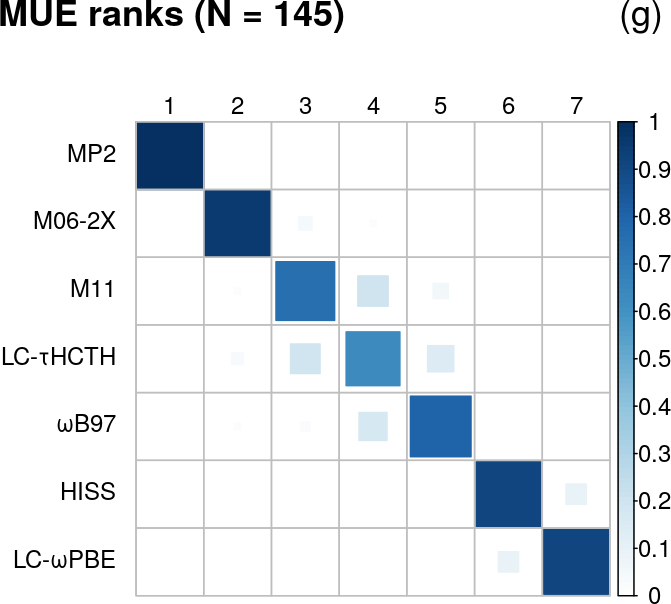} & ~\includegraphics[width=0.32\columnwidth]{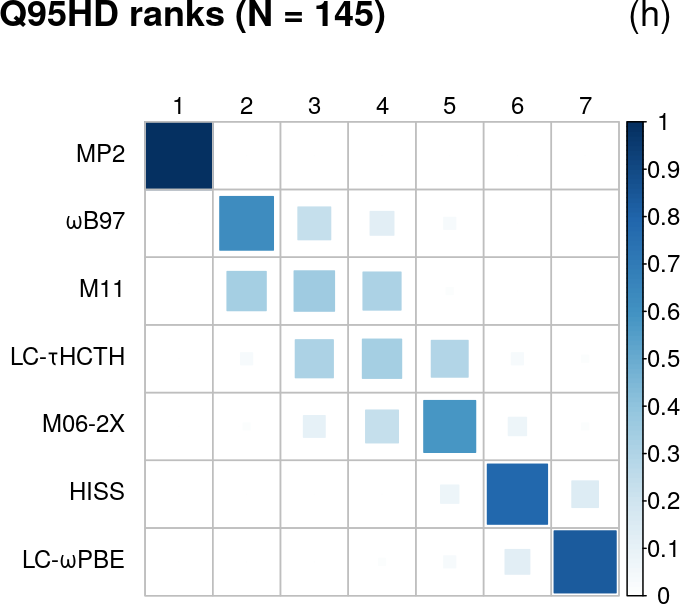} & ~\includegraphics[width=0.32\columnwidth]{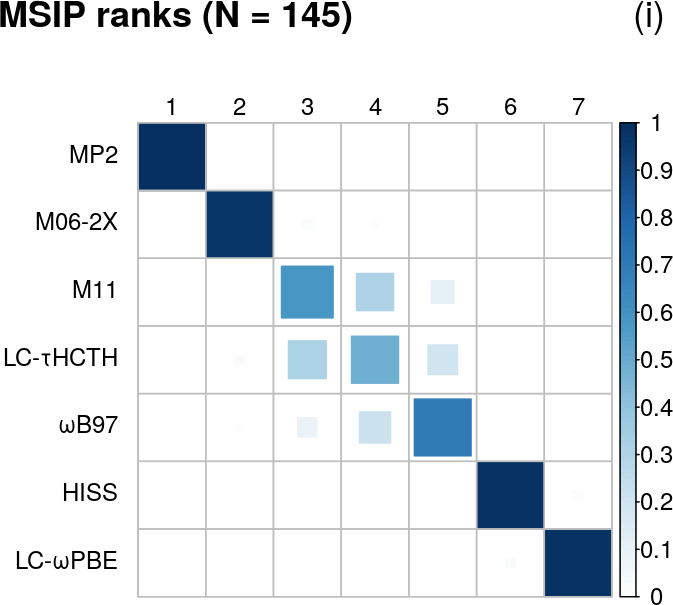}\tabularnewline
\end{tabular}
\par\end{centering}
\noindent \centering{}\caption{\label{fig:thakkarRPmat}Ranking probability matrices: (a-c) case
THA2015 full dataset ($N=135$); (d-f) case THA2015 dataset pruned
from 8 outliers ($N=127$); (g-i) case WU2015 ($N=145$).}
\end{figure}

The ranking matrices are plotted in Fig.\,\ref{fig:thakkarRPmat}.
The top row concerns dataset THA2015. The ranking probability matrices
for the MUE confirm the problem seen above for the four best methods.
It shows also that the rank of MP2 is quite ill-defined. For $Q_{95,}$,
as expected, any ranking seems illusory. The same matrices have been
estimated after the removal of 8 outliers defined above (Fig.\,\ref{fig:thakkarRPmat}-middle
row). This has a negligible impact on the MUE ranking, but fully scrambles
the $Q_{95}$ one, M11 passing from the first to the last place, MP2
from the 8th to the first, and so on. In fact, ill-defined ranking
matrices can be expected to be very sensitive to any alteration of
the dataset. 

When considering the WU2015 dataset, the ranking matrices (Fig.\,\ref{fig:thakkarRPmat}-bottom
row) show much less dispersion, underlining the deleterious role of
experimental errors on ranking. Note that there remains a notable
uncertainty to rank $\omega$B97, M11, M06-2X and LC-$\tau$HCTH using
$Q_{95}$. 

Depending on the reference dataset (experimental or CCSD(T)) one obtains
different rankings: LC-$\tau$HCTH seems a better option to predict
experimental values (possibly an artifact due to some large experimental
reference data errors), whereas MP2 is a better proxy for CCSD(T)
calculations. 

\subsection{ZAS2019\label{subsec:Zas2019}}

The effective atomization energies ($E^{*}$) for the QM7b dataset
\citep{Montavon2013}, for 7211 molecules up to 7 heavy atoms (C,
N, O, S or Cl) are available for several basis sets (STO-3g, 6-31g,
and cc-pvdz), three quantum chemistry methods (HF, MP2 and CCSD(T))
and four machine learning algorithms (CM-L1, CM-L2, SLATM-L1 and SLATM-L2).
The data have been provided on request by the authors of Zaspel \emph{et
al}.\,\citep{Zaspel2019}. The machine learning methods have been
trained over a random sample of 1000 CCSD(T) energies (learning set),
and the test set contains the prediction errors for the 6211 remaining
systems \citep{Zaspel2019}. We retain here only HF, MP2 and SLATM-L2
and compare their ability to predict CCSD(T) values.

\paragraph{Correlations.}

\begin{figure}[!tb]
\noindent \begin{centering}
\includegraphics[width=0.32\textwidth]{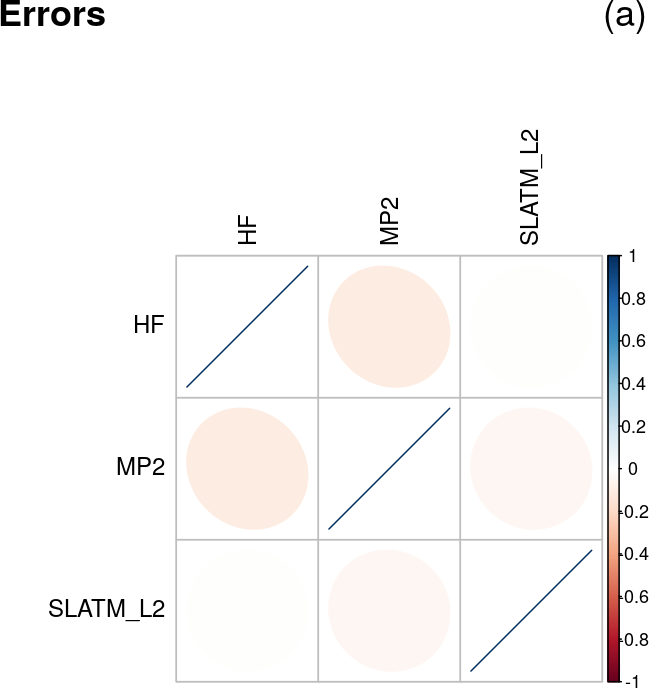}~\includegraphics[width=0.32\textwidth]{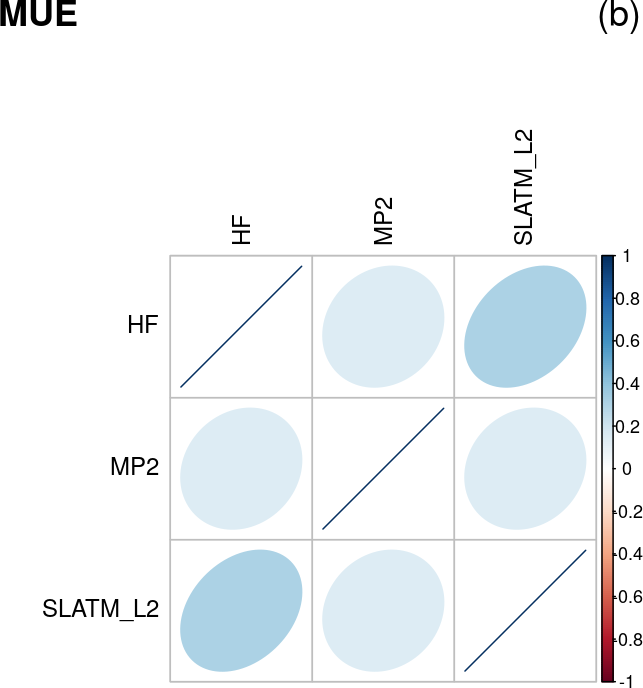}~\includegraphics[width=0.32\textwidth]{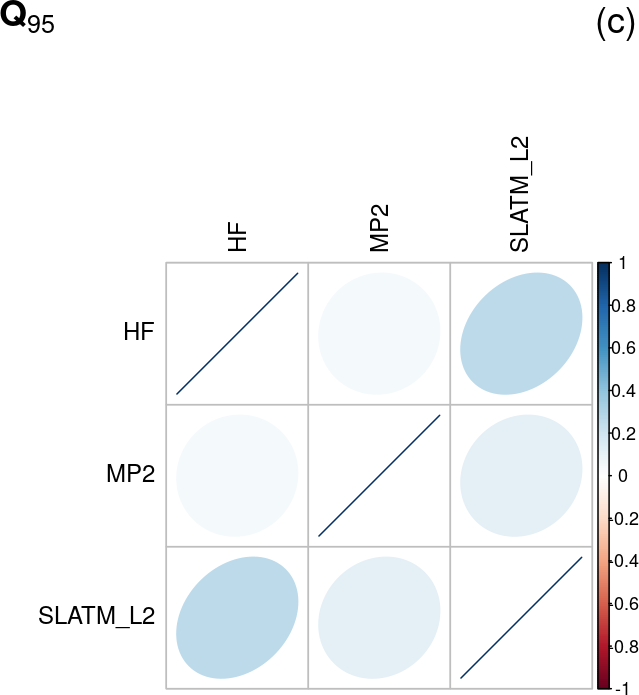}
\par\end{centering}
\noindent \centering{}\caption{\label{fig:zasCorrmat}Case ZAS2019 - rank correlation matrices: (a)
Errors; (b) MUE; (c) $Q_{95}$.}
\end{figure}

The error sets are essentially uncorrelated (Fig.\,\ref{fig:zasCorrmat}),
whereas small positive correlations can be noted for the MUE and $Q_{95}$.

\paragraph{Statistics.}

\begin{figure}[!tb]
\noindent \begin{centering}
\includegraphics[width=0.32\textwidth]{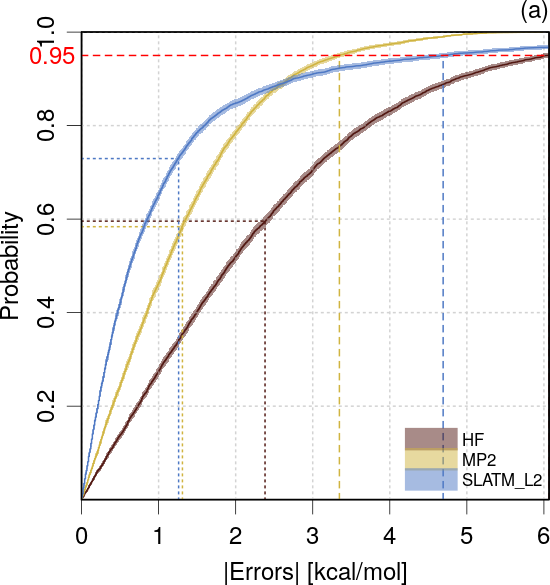}~\includegraphics[width=0.32\textwidth]{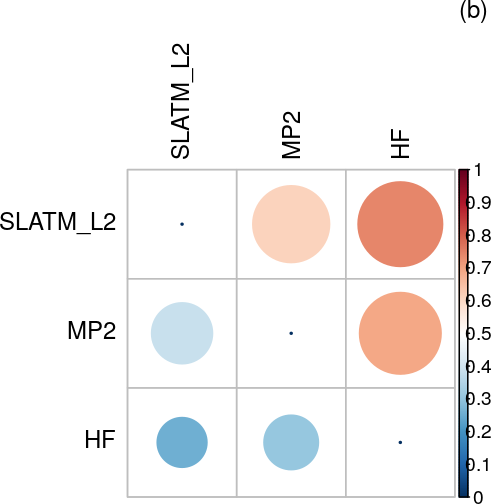}
\par\end{centering}
\noindent \centering{}\includegraphics[width=0.32\textwidth]{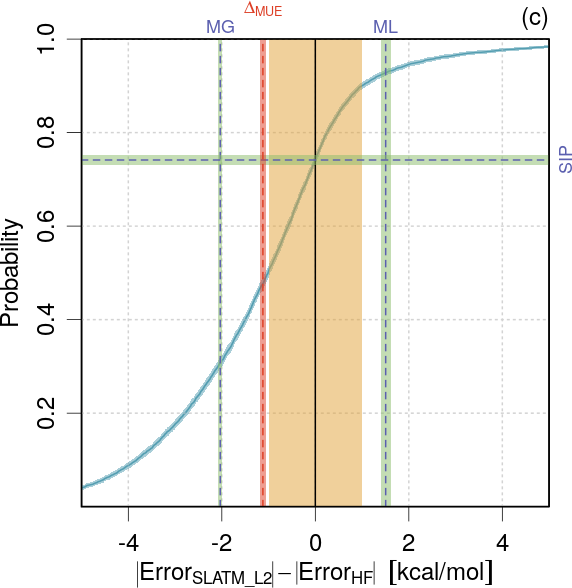}~\includegraphics[width=0.32\textwidth]{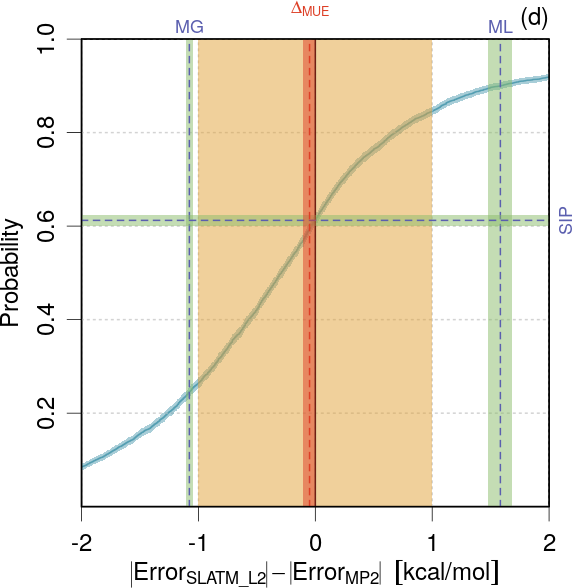}\caption{\label{fig:zas}Case ZAS2019: : (a) ECDF of the absolute errors; (b)
SIP matrix; (c,d) ECDF of the difference of absolute errors of HF
(c) and MP2 (d) with respect to SLATM-L2 (see Fig.\,\ref{fig:pernot1-1}
for details). The orange band represents the chemical accuracy (1\,kcal/mol).}
\end{figure}

The values are reported in Table\,\ref{tab:zas}. There is a contrast
between the MUE and $Q_{95}$. SLATM-L2 and MP2 have close MUE values,
with an above-threshold $p$-value ($p_{g}\simeq2P_{inv}=0.06$),
and a slight advantage for SLATM-L2. However, MP2 has a significantly
smaller $Q_{95}$. As seen on the absolute errors ECDFs (Fig.\ref{fig:zas}(a)),
SLATM-L2 has indeed a pronounced tail of large errors. 
\begin{table}[!tb]
\begin{centering}
{\small{}}%
\begin{tabular}{llr@{\extracolsep{0pt}.}lr@{\extracolsep{0pt}.}lr@{\extracolsep{0pt}.}lr@{\extracolsep{0pt}.}lr@{\extracolsep{0pt}.}lr@{\extracolsep{0pt}.}lr@{\extracolsep{0pt}.}lr@{\extracolsep{0pt}.}lr@{\extracolsep{0pt}.}lr@{\extracolsep{0pt}.}l}
\hline 
Methods &  & \multicolumn{2}{c}{MUE } & \multicolumn{2}{c}{$P_{inv}$} & \multicolumn{2}{c}{} & \multicolumn{2}{c}{$Q_{95}$} & \multicolumn{2}{c}{$P_{inv}$} & \multicolumn{2}{c}{} & \multicolumn{2}{c}{MSIP} & \multicolumn{2}{c}{SIP } & \multicolumn{2}{c}{MG} & \multicolumn{2}{c}{ML}\tabularnewline
 &  & \multicolumn{2}{c}{kcal/mol} & \multicolumn{2}{c}{} & \multicolumn{2}{c}{} & \multicolumn{2}{c}{kcal/mol} & \multicolumn{2}{c}{} & \multicolumn{2}{c}{} & \multicolumn{2}{c}{} & \multicolumn{2}{c}{} & \multicolumn{2}{c}{kcal/mol} & \multicolumn{2}{c}{kcal/mol}\tabularnewline
\cline{1-1} \cline{3-6} \cline{9-12} \cline{15-22} 
HF  &  & 2&38(3)  & 0&00  & \multicolumn{2}{c}{} & 6&1(1)  & 0&00  & \multicolumn{2}{c}{} & 0&283(5)  & 0&743(6)  & -2&03(2)  & 1&50(5) \tabularnewline
MP2  &  & \textbf{1}&\textbf{31(1) } & 0&03  & \multicolumn{2}{c}{} & \textbf{3}&\textbf{35(5)}  & \multicolumn{2}{c}{-} & \multicolumn{2}{c}{} & 0&538(5)  & 0&613(6)  & -1&08(2)  & 1&58(5) \tabularnewline
SLATM-L2  &  & \textbf{1}&\textbf{26(3) } & \multicolumn{2}{c}{-} & \multicolumn{2}{c}{} & 4&7(1)  & 0&00  & \multicolumn{2}{c}{} & \textbf{0}&\textbf{678(5)}  & \multicolumn{2}{c}{-} & \multicolumn{2}{c}{-} & \multicolumn{2}{c}{-}\tabularnewline
\hline 
\end{tabular}{\small\par}
\par\end{centering}
\noindent \centering{}\caption{\label{tab:zas} Case ZAS2019 - absolute error statistics: inversion
probabilities and SIP statistics for comparison with the DFA of smallest
MUE (SLATM-L2), except for $Q_{95}$ inversion probability, where
the reference is the DFA with smallest $Q_{95}$ (MP2). The best scores
and the values for which $(p_{g}=2P_{inv})>0.05$ are in boldface.}
\end{table}

This case emphasizes the fact that similar values of the MUE can result
by chance from very distinct error distributions, and that no conclusion
should be taken on the basis of MUE alone. 

\paragraph{SIP analysis.}

The SIP matrix (Fig.\,\ref{fig:zas}(b)) shows that SLATM-L2 presents
a notable improvement probability ($\sim0.75$) over HF and a moderate
one aver MP2 ($\sim0.61$). Even if SLATM-L2 has significantly better
statistics than HF (Fig.\,\ref{fig:zas}(c)), there remains a 25\,\%
chance that the latter provides smaller absolute errors. In most case
studies presented above, the mean gain was larger in absolute value
than the mean loss. In the comparison between SLATM-L2 and MP2, one
observes the opposite: by choosing SLATM-L2 over MP2 (Fig.\,\ref{fig:zas}(d)),
one has 61\,\% chance to get better results, with a mean gain $\mathrm{MG}\simeq-1.1$\,kcal/mol,
and 39\,\% chance to deteriorate the MP2 values with a mean loss
$\mathrm{ML}\simeq1.6$\,kcal/mol. In agreement with the $Q_{95}$
analysis, this is due to the notable tail of large errors of SLATM-L2.

\section{Discussion}

\subsection{Extracting data from articles and supplementary material}

The raw data of benchmark studies are important assets for the community,
and their accessibility and reusability are essential for intercomparison
studies or the development of alternative statistical analyses, as
performed in this study. When gathering the data, we found that many
benchmarking studies have practically inaccessible data, failing the
FAIR principle of Open Data \citep{Wilkinson2016}. Besides the trivial
case of non-available data, we have stumbled on data stored in complex
databases and requiring non-trivial coding for their extraction, or
data stored in inappropriate formats, such as PDF (a Page Description
Format), instead of recognized machine-readable data storage formats,
such as CSV tables. 

Note that for some of the cases we gathered here, we were able to
extract data from PDF articles or supplementary information files,
but not without some difficulty, involving several steps of manual
operations. Typical problems for the data extraction from tables in
PDF documents are: excessive numerical truncation, empty cells or
complex table mapping, typographical ($-$) instead of numerical (-)
minus sign, rotated tables, compact notations for uncertainty (either
123(4) or 123$\pm$4), bibliographical references attached to the
data (generally processed by extraction tools as spurious decimals)...
Most of these features preclude fully automated data extraction and
require error-prone human processing.

So, unless the structure of the data is complex, and this should not
be the case for most benchmark studies, it is warmly recommended to
use ``flat'' numerical tables stored in an open format, such as
CSV, and to avoid to put more than one information per table cell.
``Think Open, think FAIR !'' 

\subsection{Impact of dataset size}

The examples above have shown that dataset size impacts considerably
the ability to rank methods or to assert the impact of an improved
method. Size effect on the uncertainty of statistics is well known
for the mean value, and similar formulae can be derived for other
statistics under normality hypotheses. However, the non-normality
of error sets requires the use of numerical methods, typically bootstrap
sampling. This enables to show how the usual benchmark statistics
are affected by sample size. We have seen, for instance, that there
is a notable probability to conclude erroneously that two $Q_{95}$
values are different when they are not (type I errors or false positive)
if $N<60$ (Paper\,I\, \citep{Pernot2020}-\,Appendix\,C). For
the MUE, this limit is smaller ($N=30$). Moreover, for small datasets
(a few tens of points), even the first digit of the statistics is
often affected by the uncertainty. 

It is practically impossible to predict the dataset size required
for a stable and robust ranking. Many factors other than set size
are involved, notably the number and nature of methods to be ranked.
When a lot of DFAs are compared, a hierarchical ranking is often performed,
for instance by first choosing the best method at each rung of the
Jacob's ladder, and then comparing these methods together \citep{Wu2015b}.
This is one way to reduce the ranking uncertainty that is likely to
result from the direct comparison of a large number of methods, as
illustrated for instance in case CAL2019 (Section\,\ref{subsec:Caldeweyher2019},
Fig.\,\ref{fig:caldewheyerRPmat}).

\subsection{The correlation matrix as a sanity check}

When we started this study, the correlation matrices were mainly intended
to illustrate the importance to consider correlation when comparing
statistics. When cumulating the case studies, we realized that errors
correlation matrices may contain pertinent information on the quality
of the benchmark dataset. Considering that model errors in computational
chemistry are mostly systematic, one expects that error patterns over
a dataset are characteristic of each method or family of methods.
This seems to be a basic requirement for sound benchmarking studies.
One should thus expect that closely related methods produce similar
error patterns and have strongly correlated error sets, the correlation
level decreasing with a ``distance'' between methods. This is clearly
illustrated in case BOR2019, where the correlation matrix clusters
nicely into relevant DFA groups. There seems also to be a genuine
decorrelation between MP2 or MP2-based methods and DFAs (NAR2019,
WU2015). Similarly, one observes no correlation between HF, MP2 and
a machine-learning method calibrated on CCSD(T) in case ZAS2019.

As a consequence, when the methods set contains unrelated methods,
a uniform strongly positive correlation matrix should raise an alert.
We have seen in cases DAS2019 and THA2015/WU2015 that outliers and/or
large reference data errors could dominate the correlation matrix
and influence the benchmark statistics. Outliers common to all error
sets (global outliers) can be efficiently identified on a parallel
plot, as shown in case DAS2019 (Fig.\,\ref{fig:dasCorrmat-1}). If
the ranking study is to reflect the methods performances, the curation
and possible pruning of the dataset from such global outliers is a
necessary preliminary step. Otherwise, more complex statistical models
have to be used to alleviate the impact of those points (see Paper\,I
\citep{Pernot2020}-Appendix\,A and references \citep{Lejaeghere2014,Pernot2015,Proppe2017}).

Note that strongly correlated error sets do not imply similar performances.
For instance a set of linearly scaled harmonic vibrational frequencies
typically has better statistics than the unscaled set \citep{Scott1996},
whereas their correlation coefficient is 1 because of the linear transformation
between both error sets. One should also remember that the correlation
coefficient between calculated and reference values that is still
presented in some benchmarks is not a reliable performance statistic
\citep{Bland1986}. At most, it reveals a linear (Pearson) or monotonic
(Spearman, Kendall) association between datasets, but its their proximity
to the identity line. 

\subsection{Impact of error sets correlation on ranking}

The correlation between error sets is partially or totally transferred
to benchmark statistics. Except for linear transformations of the
errors, where the transfer is trivial, one has to use Monte Carlo
methods to estimate it. In many cases, such as for normal, Student's-$t$
or g-and-h error distributions \citep{Hoaglin1985}, one observes
that the correlation intensity mainly decreases when passing from
errors to MUE to $Q_{95}$. The case studies above show however that
there are exceptions to this ideal trend. We cannot presently rationalize
the observed exceptions. In a vast majority of the cases studied above,
the correlation matrices for MUE and $Q_{95}$ have positive coefficients.
These contribute to a reduction of the uncertainty on statistics differences,
with better discernibility between uncertain statistics. Globally,
positive correlations increase the robustness of rankings.

However, unlike for the error correlations, the visualization and
analysis of correlations between statistics might be of secondary
interest for benchmarks. In fact, the paired samples bootstrap algorithms
used in this study enable to account directly for these correlations,
without having to estimate intermediate correlation matrices. 

\subsection{Systematic improvement analysis}

We introduced a new criterion, the systematic improvement probability
(SIP), which has the major advantage to be independent of the usual
descriptive statistics. It is based on a sign statistic of the differences
of absolute error pairs. It is a useful complement to the MUE, as
it enables to analyze MUE differences. All the case studied above
show that a decrease of MUE results from a balance between gains and
losses. Only two methods pairs were found, in cases PER2018 (Section\,\ref{subsec:Pernot2018})
and CAL2019 (Section\,\ref{subsec:Caldeweyher2019}), with SIP values
reaching 0.95, close to the full systematic improvement. We did not
find a ``best method'' which fully improves the results of all lower
rank methods. Because of the well known error compensations in computational
chemistry methods \citep{Dunning2000}, even physics-based improvements
in DFAs do not lead to systematic improvements for all systems. Of
course, this balance is not a discovery, but the SIP enables to quantify
it, and provides a basis for the user to estimate the risk taken when
switching from an old,faithful, method, to a new one. We have seen
for instance that for band gaps, mBJ degrades LDA predictions for
16\,\% of the systems (BOR2019). In fact, there is often a non-negligible
percentage of systems for which a ``bad'' method is better than
a ``good'' one, all across Jacob's ladder. 

We have also introduced the mean SIP as a possible ranking statistic.
The main advantage of the MSIP is its independence from the usual
summary statistics; its main drawback is that it depends on the set
of methods being compared and it is not transferable to comparisons
out of its definition set. Conflicts of the MSIP with the MUE reveal
disparities in the errors distribution.

\subsection{Ranking Probability Matrix}

The ranking probability matrix $\mathbf{P}_{r}$ provides a diagnostic
on the robustness of the ranking by any statistic. Our tests of MUE,
$Q_{95}$ and MSIP rankings show that the dataset size and the number
of methods influence notably the ranking uncertainty. Without any
surprise, the closer the performances of a group of methods, the more
uncertain their ranking. Depending on the datasets, the MUE and $Q_{95}$
rankings might conflict and present different levels of robustness
(\emph{cf.} case THA2015). We would advise to publish systematically
both of them, as they provide complementary information. 

In the various cases treated above, the rankings provided by the MSIP
are most often conform to the MUE rankings and are as sensitive as
the other rankings to sampling uncertainty. When ranking conflicts
for the first places occur with the MUE, as was observed in case PER2018,
one gets alerted that the method with the lowest MUE is not the one
providing the largest proportion of small absolute errors. Due to
the non-normality of error distributions, such scenarii are to be
expected, as for inversions in MUE and $Q_{95}$ rankings.

\subsection{Extension to composite datasets}

We considered here only datasets based on a single property. Many
modern benchmarks are based on composite datasets, involving weighting
schemes to incorporate data with different units \citep{Goerigk2017}.
The applicability of the SIP to such datasets is straightforward,
but the mean gain and mean loss statistics, having dimensions, should
become multivariate. 

The estimation of $P_{inv}$ and ranking probability matrices for
composite statistics (\emph{e.g.}, WTMAD \citep{Goerigk2017}) can
use directly the pair-based bootstrap sampling algorithms described
in the present article, although care should be taken to avoid imbalance
between the various components of a dataset by using the so-called
\emph{stratified} bootstrap \citep{Hesterberg2015}, preserving the
cardinal number of each component in the generated sample.

\section{Conclusion}

In Paper\,I \citep{Pernot2020}, we proposed several tools to test
the robustness of rankings or comparisons of methods based on error
statistics for non-exhaustive, limited size datasets. In order to
avoid hypotheses on the errors distributions, bootstrap-based methods
were used for the estimation of statistics uncertainty, $p$-values
and ranking uncertainty. In this paper, we illustrated and validated
these methods on nine datasets covering a representative panel of
properties and sizes. 

Most of these tools take into account the correlation between error
sets or their statistics, and we illustrated repeatedly that large
correlations occur that cannot be neglected. Moreover, we have seen
that the error sets correlation matrix can be useful to appreciate
the quality of a benchmark dataset, notably when experimental reference
data are used. To our knowledge, this topic has not previously been
discussed, and benchmarking studies do not presently make use nor
report such correlation matrices. 

The systematic improvement probability (SIP) is based on the system-wise
difference of absolute errors between two methods and, in conjunction
with the mean gain (MG) and mean loss (ML) statistics, it quantifies
the risk taken by a user when passing from a method to another. We
have seen in the applications that choosing a method with a lower
MUE might imply a non-negligible risk to produce large errors. Moreover,
only two of the showcased examples revealed a method which provides
a (nearly) full systematic improvement over one of its concurrents.
Even when comparing an elaborate composite method such as G4MP2 to
DFAs one observes partial SIP values (case NAR2019). A pedagogical
virtue of the SIP is to clearly show that computational chemistry
is a science of compromises.

We based the comparison between values of a statistic for two methods
on the inversion probability $P_{inv}$, which is simply linked to
the $p$-value for the test of the equality of those statistics ($p_{g}\simeq2P_{inv})$.
It is thus an important tool to assess if a difference between two
values is a real effect or if it might be due to the choice of dataset.
For ranking statistics, we suggest to report $P_{inv}$ with respect
to the method with the smallest value in results table.

The ranking probability matrix $\mathbf{P}_{r}$ for a chosen statistic
provides a clear diagnostic on the robustness of the corresponding
ranking. The impact of dataset size and number of compared methods
can be thoroughly tested, as shown in the examples above. It appeared
in these examples that the intermediate ranks are often weakly defined.
The robustness of the ranking might also depend on the ranking statistic,
and the statistic providing the most robust ranking depends on the
dataset. As we suggested earlier, one should therefore not rely on
the MUE alone to rank methods. We encourage benchmark authors to provide
ranking probability matrices for several statistics (at least the
MUE and $Q_{95}$), which can be obtained with a negligible overcharge
in computer time. 

We considered here for simplicity raw error sets, from which no care
has been taken to remove systematic trends. When this is possible,
such trend corrections, often simply linear, will provide much better
generalizability of the summary statistics derived from these error
sets. Besides, this is a necessary step if one wishes to estimate
the prediction uncertainty of any method \citep{Lejaeghere2014,Pernot2015,Proppe2017},
notably when dealing with non-uniform reference data uncertainties.
\begin{acknowledgments}
The authors are grateful to Pr. O.\,A.\,von~Lilienfeld for providing
the datasets of case ZAS2019, and to Pr. S.\,Grimme for providing
a corrected copy of the Supplementary Information for case CAL2019.
\end{acknowledgments}

\section*{Supplementary Information}

The data that support the findings of this study are openly available
in Zenodo at\linebreak{}
 \href{http://doi.org/10.5281/zenodo.3678481}{http://doi.org/10.5281/zenodo.3678481}\citep{SIPdata2020}.
Application \texttt{ErrView} implementing the methods described in
this article is archived in Zenodo at \href{http://doi.org/10.5281/zenodo.3628489}{http://doi.org/10.5281/zenodo.3628489});
a test web interface is freely accessible at \url{http://upsa.shinyapps.io/ErrView}.

\bibliographystyle{unsrturlPP}
\phantomsection\addcontentsline{toc}{section}{\refname}\bibliography{packages,NN}

\end{document}